\ifdefined\DataTraceProceedings
  \RequirePackage{xr-hyper}
\fi
\documentclass[sigconf]{acmart}
\AtBeginDocument{}

\usepackage{amssymb}
\usepackage{xspace}
\usepackage{multirow}
\usepackage{tikz}
\usepackage{enumitem}
\usetikzlibrary{positioning,arrows.meta,decorations.pathreplacing}
\usepackage{etoolbox}
\AtBeginEnvironment{table}{\small}
\AtBeginEnvironment{table*}{\small}
\usepackage{adjustbox}
\makeatletter
\newcommand{\clearExternalBookkeeping}{%
  \@for\dtlabel:=TotPages,tocindent-1,tocindent0,tocindent1,tocindent2,tocindent3,tocindent4,tocindent5\do{%
    \expandafter\global\expandafter\let\csname r@\dtlabel\endcsname\@undefined}}
\makeatother
\newcommand{\method}{\textsc{DataTrace}\xspace}

\newcommand{\alert}[1]{\textbf{#1}}
\newcommand{\artifacturl}{https://hkustdial.github.io/datatrace/}

\acmConference[KDD '27]{ACM SIGKDD Conference on Knowledge Discovery and Data Mining}{August 1--5, 2027}{San Jose, CA, USA}
\acmYear{2027}
\copyrightyear{2027}

\setcopyright{none}
\acmISBN{}
\acmDOI{}

\ifdefined\DataTraceProceedings
  \clearExternalBookkeeping
\fi

\begin{document}
\title{DataTrace: Benchmarking Data Agents for Evidence-Grounded Root-Cause Localization}
% Author order, affiliations, and email fields retained from the accepted source.
\author{Tung Sum Thomas Kwok}
\affiliation{%
  \institution{UCLA}
  \country{USA}
}
\email{}
\author{Xiaotian Lin}
\affiliation{%
  \institution{HKUST(GZ)}
  \country{China}
}
\author{Yuyu Luo}
\affiliation{%
  \institution{HKUST(GZ)}
  \country{China}
}
% \email{yuyuluo@hkust-gz.edu.cn}
\renewcommand{\shortauthors}{Kwok et al.}

\begin{abstract}
Database agents diagnose pipeline failures, but a correct root cause is trustworthy only when it is supported by evidence the agent inspected. Existing benchmarks usually grade the final query or answer but fail to distinguish an evidence-based diagnosis from an unsupported correct answer.  We introduce \method, a benchmark for evidence-grounded root-cause localization over typed graphs of database artifacts. \method uses controlled fault injection to derive root causes and evidence paths within 83 runnable environments without human annotations, and scores root-cause correctness and coverage of the supporting path separately. We also introduce \textsc{CoSE}, a generator-solver protocol in which an external verifier admits reproducible faults and checks both the submitted answer and the inspected evidence. Our evaluation shows that seven proprietary models achieve average root-cause accuracy $48\%$, with a mean per-model conditional unsupported rate of $18\%$.  Fine-grained analysis shows that this gap varies sharply with fault mechanism, dependency distance, and evidence modality. After in-loop training on a matched held-out batch, the share of grounded correct diagnoses rises from $83.7\%$ to $94.6\%$, while root-cause accuracy rises from $29.2\%$ to $36.7\%$.  These results show that while final-answer accuracy can hide unsupported diagnoses, evidence coverage provides an actionable signal to improve database agents. Code and data are available at \url{\artifacturl}.
\end{abstract}

\maketitle

\section{Introduction}
\label{sec:intro}

Large language model (LLM)-based data agents are autonomous systems that use natural language reasoning, external tools, and data-system interfaces to perform end-to-end data tasks~\cite{DBLP:journals/corr/abs-2510-23587, DBLP:conf/sigmod/LiPXLZMLL26}. They generate SQL, navigate warehouse schemas, diagnose and repair data pipelines, and modify data-engineering repositories~\cite{bird,spider2,eltbench,dsbench,dabench,beaver}. 

\begin{figure*}[t]
  \centering
  \includegraphics[width=\textwidth]{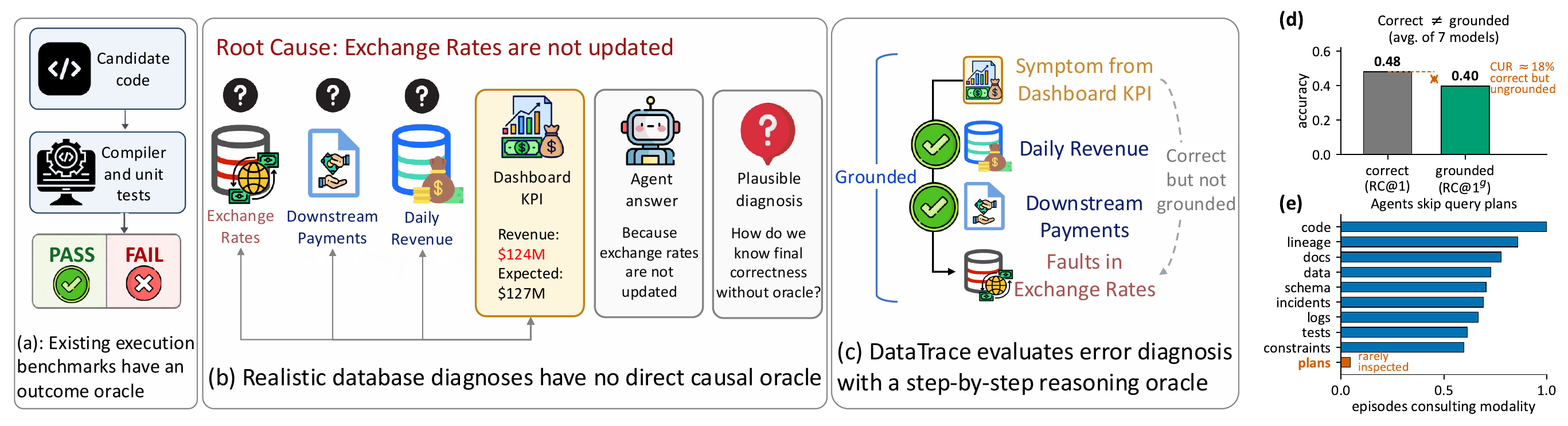}
  \vspace{-5mm}
  \caption{\textbf{A correct answer is not necessarily a grounded diagnosis.} \method simulates a realistic database diagnosis setting without an oracle by evaluating answer correctness and supporting evidence separately. \emph{(a) - (c):} Code submission can be validated by compiler or unit tests~(a), but a database diagnosis has no direct outcome oracle~(b), e.g. the dashboard is wrong (\$124M against the expected \$127M) without error raised because the exchange rates from multiple steps upstream did not update. \method grades the answer against an accepted evidence path~(c), scoring answer correctness and supporting evidence separately. \emph{(d):} Over seven proprietary models, we show that final answer accuracy ($\mathrm{RC@1}$) overstates grounded accuracy ($\mathrm{RC@1}^{g}$) with a mean per-model conditional unsupported rate ($\mathrm{CUR}$) of $18\%$. \emph{(e):} Agents occasionally ignore informative modalities such as query plans, which may miss useful evidence in diagnosis. We use the same example in Fig.~\ref{fig:pipeline} to ease referencing.}
  \label{fig:groundinggap}
  \Description{A motivating revenue discrepancy propagates from stale exchange rates through upstream dependencies. The diagram contrasts answer correctness with inspected evidence, then summarizes the grounding gap and modality consultation across seven models.}
\end{figure*}

However, as data agents evolve from generating queries and analytics results toward diagnosing failures in complex data systems, a new evaluation challenge emerges.
Unlike a program that either compiles or does not~(Fig.~\ref{fig:groundinggap}(a)), a data agent's diagnosis carries no built-in oracle. A pipeline can run to completion and still be wrong, \textit{e.g.,} an unupdated exchange rate may propagate silently through downstream payments into daily revenue and surface later as a dashboard figure off by a few percent without error~(Fig.~\ref{fig:groundinggap}(b)). Even when the error is explicit, it does not mark the upstream object that causes it. As there is no structured way to automatically check the validity of a submitted reasoning chain, a correct-looking answer can be reached either by genuinely tracing the fault upstream or by naming a plausible cause without establishing it. Therefore, the final answer is difficult to trust in isolation, because knowing whether a diagnosis is right requires verifying the reasoning trajectory that produced it, not just the answer string.

Existing evaluations do not isolate this requirement.  Text-to-SQL and data-agent benchmarks primarily grade the final query, answer, or repaired repository state~\cite{bird,spider2,eltbench,birdcritic}. SQL-error benchmarks move closer to diagnosis but still score whether the submitted error type, object, or fix is correct~\cite{nl2sqlbugs,birdcritic,sqlgovernor}.  In short, they measure whether an agent gets the answer right, not whether it got it for the right reasons, i.e. whether the answer is \emph{grounded} in evidence the agent actually inspected. An agent can submit the correct root cause without opening a single supporting object, and an outcome-only score cannot tell this apart from a genuine multi-object investigation~(Fig.~\ref{fig:groundinggap}(b)).  Adding human annotation to every case does not solve this, as it is expensive to scale, hard to keep consistent, and vulnerable to annotation errors that can change benchmark conclusions~\cite{annoterrors,agenticbestpractices}.  Data agents therefore need \emph{grounded} evaluation that scores an answer together with the evidence supporting it.

We study \emph{evidence-grounded root-cause localization for database agents}.  Given a workspace and an observed problem, an agent must identify the upstream object responsible and cite the artifacts it inspected, grading the answer and its evidential support separately. We organize the paper around three research questions: 
\begin{itemize}[leftmargin=*, topsep=3pt, itemsep=1pt, parsep=0pt]
  \item \textbf{RQ1:} Do agents ground their diagnoses?
  \item \textbf{RQ2:} What makes grounding fail? 
  \item \textbf{RQ3:} Can the grounding signal improve agents? 
\end{itemize} 

Answering these questions requires four design properties.  First, the benchmark must cover realistic, multi-artifact workspaces rather than isolated queries. Second, it must expose fine-grained dimensions of difficulty, including fault mechanism, dependency distance, or evidence modality. Third, reference causes and evidence paths must be reproducible without a handwritten rationale. Finally, evaluation should support the generation of new verified cases without relying on a permanently fixed task set.

We observe that data pipelines already contain the necessary structure for this evaluation, as database provenance records relationships among data, transformations, and outputs~\cite{whyWhereProvenance,provenanceSurvey}. Representing database and engineering artifacts as a typed graph makes these dependencies explicit, so we can inject a controlled fault into the graph for a known root cause and a reference path of supporting evidence. We therefore introduce \method~(Fig.~\ref{fig:groundinggap}(c)), an automatically graded provenance-graph benchmark spanning $83$ runnable environments (e.g., finance and software) across five source families (e.g., BIRD-SQL) covering 21 different fault operators (e.g., renaming upstream columns or corrupting source data). Each solver trajectory is graded separately for root-cause correctness and coverage of the accepted evidence path. We further introduce \textsc{CoSE}, a generator-solver protocol in which an external verifier admits only reproducible tasks and scores both the submitted answer and its evidential support.

The evaluation answers the research questions in turn. First, \textbf{accuracy does not ensure grounding}, as there is a persistent gap between producing a correct answer and establishing it from inspected evidence. Seven proprietary models reach root-cause accuracy of $48\%$ on average, but their mean per-model unsupported rate is $18\%$~(Fig.~\ref{fig:groundinggap}(d)).  Second, \textbf{grounding failure is structured, driven largely by fault family and lineage distance}, with common ignorance of informative artifacts such as query plans~(Fig.~\ref{fig:groundinggap}(e)). Third, the \textbf{grounding signal is actionable}, as our in-loop training study on a matched held-out batch shows the share of correct diagnoses that are also grounded and root-cause accuracy rising from $83.7\%$ to $94.6\%$ and from $29.2\%$ to $36.7\%$ respectively. This paper makes three contributions.
\begin{itemize}[leftmargin=*, topsep=3pt, itemsep=1pt, parsep=0pt]
  \item \textbf{Problem definition.} We formalize evidence-grounded root-cause
        localization for data systems to score correctness and inspected evidence separately (Sec.~\ref{sec:problem}).
  \item \textbf{Benchmark.} We introduce \method, a provenance-graph benchmark with mechanically derived root causes and evidence paths, together with the \textsc{CoSE}  generator-solver protocol that converts grounding signal into a training objective (Sec.~\ref{sec:method}, Sec.~\ref{sec:trainable}).
  \item \textbf{Empirical findings.} Our evaluation shows that final-answer accuracy overstates grounded diagnosis with structured localization failure patterns, which can be improved via training with the grounding signal (Sec.~\ref{sec:experiments}).
\end{itemize}

\section{Problem Formulation}
\label{sec:problem}
\label{sec:problem:objects}

We study \emph{evidence-grounded root-cause localization} in data systems. The database agent receives an observed symptom, investigates a workspace through typed tools, and returns an upstream root-cause object and inspected evidence. We score answer correctness and evidence coverage separately; a diagnosis is credited as grounded when it is correct and meets the coverage threshold defined in Sec.~\ref{sec:method:verify}. This section defines the graph substrate, fault, symptom, and evaluation task. Sec.~\ref{sec:method} explains how \method constructs reference labels at scale without human annotation.

{\textbf{Database as a provenance graph.}}
Each task is a typed provenance graph $\mathcal{G}=(\mathcal{V},\mathcal{E})$. Nodes $\mathcal{V}$ cover database and engineering modalities such as tables, columns, views, statements, constraints, plan operators, indexes, tests, logs, and incidents. Edges $\mathcal{E}$ record recoverable relations across these modalities, e.g. a view \emph{depends on} its inputs, a statement \emph{produces} a relation, or a constraint \emph{references} columns. We follow database provenance traditions~\cite{provenanceSurvey,whyWhereProvenance} and extract the edges mechanically from catalog metadata and parsed SQL definitions. The graph serves as the agent's explorable state space and the substrate for
evidence scoring.

{\textbf{Fault.}} A task contains a single fault $f=(v^\star,\tau_f)$, where $v^\star\in\mathcal{V}$ is the defective object and $\tau_f$ denotes its fault type, distinct from the grounding threshold $\tau_g$ defined below. We distinguish six benchmark families:
\begin{itemize}[leftmargin=*, topsep=3pt, itemsep=1pt, parsep=0pt]
  \item \emph{view-definition faults} that contain incorrectly derived relations caused by erroneous SQL definitions, such as a wrong join key, join type, aggregation function, or temporal filter,
  \item \emph{constraint faults} with missing or weak integrity rule such as uniqueness constraints, orphan rows, or duplicate business keys,
  \item \emph{schema faults} where an upstream schema changes without compatible propagation, e.g., a column rename or type change that dependent views still interpret under the old schema,
  \item \emph{source-data faults} with corrupt, incomplete or malformed incoming records such as duplicate primary key records, corrupted encoding or late-arriving records, 
    \item \emph{plan or configuration faults}, where execution settings select an incorrect access path or materialization, and
  \item \emph{requirement misspecification}, where the database state and query execution mismatch the stated analytical intent.
\end{itemize}

{\textbf{Symptom.}} The symptom $s$ is what the agent observes at a node $v_s$, usually downstream of $v^\star$ with distance $d_{\mathcal{G}}(v_s,v^\star)\ge 1$. We distinguish \emph{explicit failure} symptoms, such as raised errors, failing tests, or build failures, from \emph{silent drift} symptoms, where a downstream value, stale result, or analytical answer is wrong without an error being raised. \method focuses on silent drift because the agent cannot read the cause from a stack trace or test name, hence requiring the agent to follow the dependency structure and confirm the defect at its source. We reduce the boundary case $d_{\mathcal{G}}=0$ to local single-query bug detection in our setting.

{\textbf{Evaluation task.}} The solver acts through typed actions $\mathcal{A}$, including database navigation, schema and constraint inspection, table profiling and row sampling, incident and documentation reading, SQL execution and plan explanation, pipeline or test runs, lineage tracing, file editing, and submission. Each action is parameterized by a graph-node identifier to form a trajectory $T=((a_1,v_1),\ldots,(a_n,v_n))$. We write $V(T)\subseteq\mathcal{V}$ for the set of graph nodes inspected during $T$, as recorded by the tools. After investigating, the solver emits a predicted root-cause object $\hat{v}$, an evidence set $\hat{\pi}$, and an optional repair $\hat{r}$. These outputs are compared with the mechanically derived ground-truth tuple $(v^\star,\pi^\star,\mathcal{C}^\star)$. We further define the
metrics that score answer correctness and evidence coverage in  Sec.~\ref{sec:method:verify}.

\section{The \method Benchmark}
\label{sec:method}

\method extends Sec.~\ref{sec:problem} as an automated  corpus of executable tasks. Each task is built on a typed provenance graph, containing a mechanically derived root cause, evidence path and reasoning chain without human annotation. The score is computed over trajectories diagnosed from the graph. We present the design goals, the co-evolutionary construction pipeline, and the grounded metrics. \vspace{-2mm}

\subsection{Design goals and scope}
\label{sec:method:goals}

\method is guided by four design goals.  \textbf{G1: Wide coverage} requires tasks to span over all modalities within a database system including but not limited to data, schema, code, or plans, tests, logs, and incidents.  \textbf{G2: Fine-grained diagnosis} requires the benchmark to vary fault mechanism, graph distance, symptom type, modality span, and decoy evidence. \textbf{G3: Scalable reference construction} requires root causes and accepted evidence paths to follow from executable task structure without human annotation. \textbf{G4: Reproducible extension} requires every case to pass an executable admission gate to enable new benchmark versions without accepting unverifiable tasks. The benchmark evaluates whether an agent can localize upstream defective objects and support that diagnosis with inspected artifacts. While the current grader recognizes one mechanically accepted evidence path, we defer alternative valid paths as future work (Sec.~\ref{sec:limitations}). \vspace{-2mm}

\subsection{Co-evolutionary benchmark construction}
\label{sec:method:tasks}
\label{sec:method:coevoeval}

\begin{figure*}[t]
  \centering
  \includegraphics[width=\textwidth]{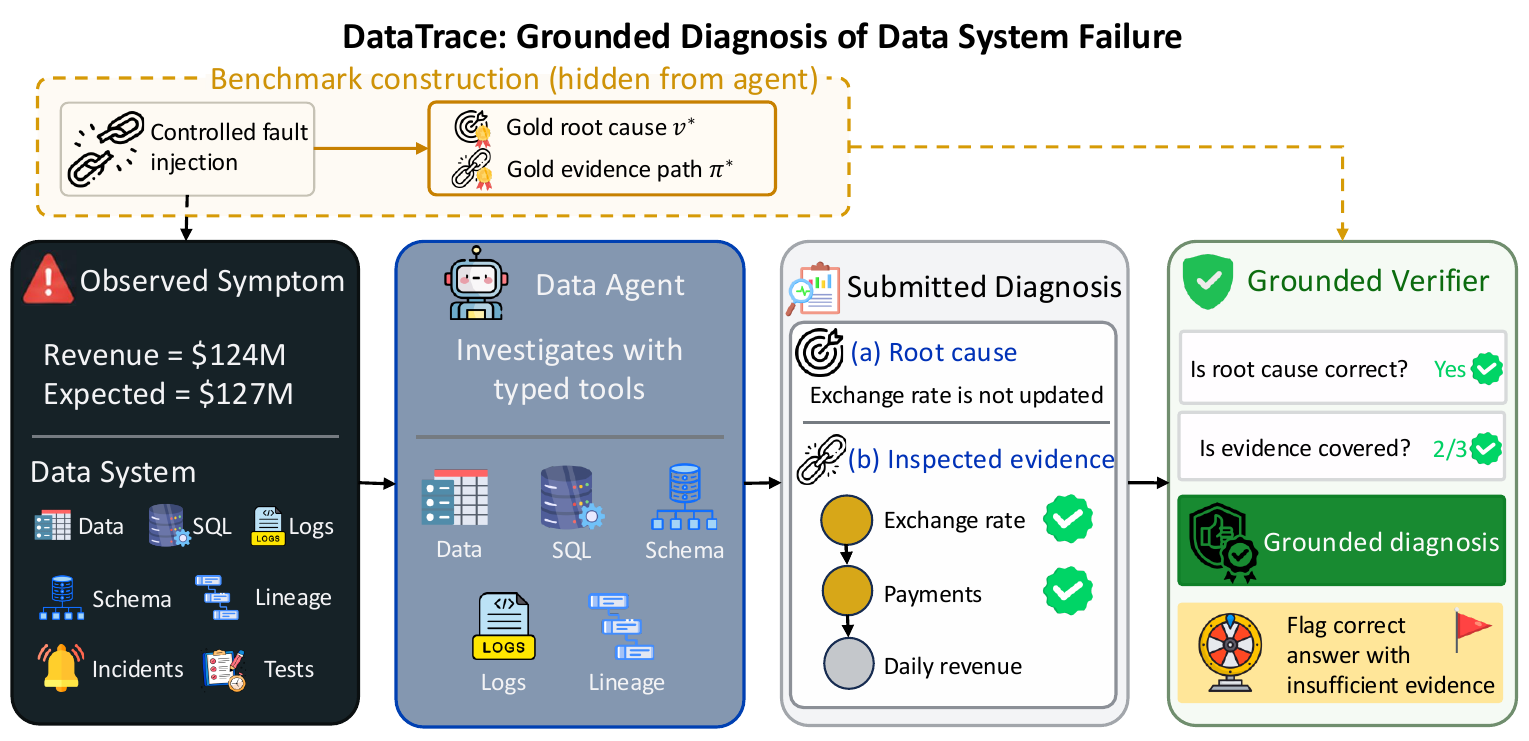}
  \vspace{-3mm}
  \caption{\textbf{The \method pipeline.} A controlled fault injection yields a
  gold root cause $v^\star$ and gold evidence path $\pi^\star$, hidden from the
  agent (top). The agent observes only a materialized symptom (here a daily-revenue
  KPI off by \$3M) within a data system that exposes many artifact modalities,
  investigates with typed tools, and submits a root cause together with the
  evidence it inspected. A grounded verifier scores answer correctness and evidence
  coverage ($\mathrm{ER}$) separately, crediting a \emph{grounded} diagnosis only
  when both hold and flagging a correct answer with too little inspected
  evidence ($\mathrm{ER}<\tfrac12$) as an unsupported lucky guess.}
  \label{fig:pipeline}
  \Description{A fault generator creates a reproducible database symptom and a hidden reference cause and evidence path. A solver investigates through typed tools. A verifier scores answer correctness and inspected-path coverage separately.}
\end{figure*}

\begin{figure}[t]
  \centering
  \includegraphics[width=0.95\columnwidth]{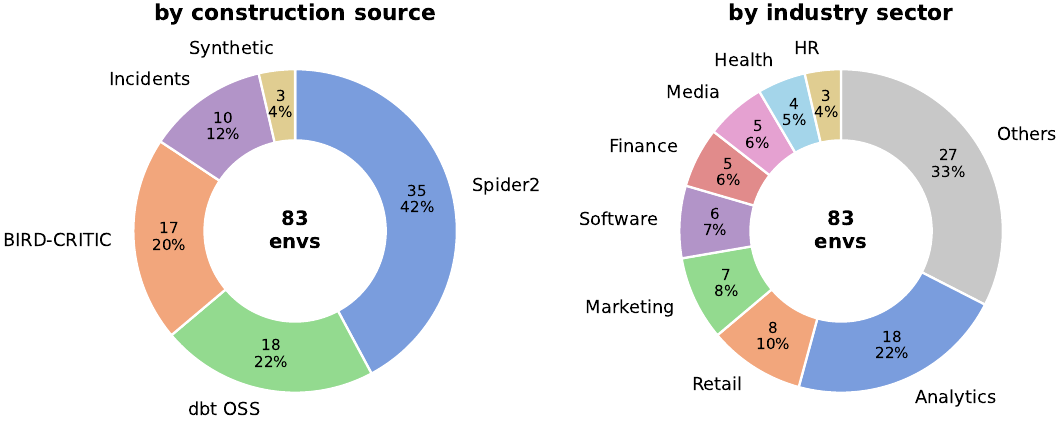}
  \caption{\textbf{\method spans 83 buildable environments across nine industry
  sectors} (slice size is the environment count), drawn from five construction
  sources: BIRD-CRITIC, Spider2, dbt repositories, mined incidents, and synthetic
  pipelines.}
  \label{fig:composition}
  \Description{A donut chart groups 83 buildable environments by nine industry sectors. The environments come from five construction sources listed in the caption.}
  \vspace{-5mm}
\end{figure}

\method assembles 83 data-engineering environments from five source families, namely BIRD-CRITIC~\cite{birdcritic}, Spider2~\cite{spider2}, dbt repositories, mined production incidents, and synthetic scenarios, spanning nine industry sectors (Fig.~\ref{fig:composition}). Each containerized environment is parsed into a provenance graph. We include corpus profile, node and edge types, actions and admission filters in Appendix~\ref{app:envspec}.

The generator creates tasks through three channels. First, it injects one of 21 typed fault operators into a working pipeline, re-executes the pipeline, and harvests the resulting symptom. The injection record and augmented pipeline determine root cause, evidence path and reasoning chain without human annotation. These operators cover source data, code, schema, constraint, and plan/configuration mechanisms. Second, it mines real incidents by retaining public-repository commits whose pre-fix state reproduces the symptom and whose post-fix state clears it. Third, an optional exploration partition reverse-generates questions from graph walks. These open-ended tasks lack mechanical witness paths, use agreement and checklist verifiers, and are excluded from the frozen leaderboard (App.~\ref{app:formal}).

\textsc{CoSE} is an optional generator-solver protocol supporting G3 and G4. The generator proposes a fault pattern and decoys; admission requires a reproducible symptom, an oracle-solvable cause, and a non-trivial evidence path. An external verifier then scores the solver's answer and inspected evidence. Mechanical references support G3, while executable admission gates support G4. Static evaluation uses the frozen board without adaptive generation. In RQ3, the same verifier selects training trajectories and guides adaptive task generation, testing whether the grounding signal can improve investigation behavior (Sec.~\ref{sec:trainable}).

\subsection{Metrics and evidence-grounded rating}
\label{sec:method:actions}
\label{sec:method:verify}
\label{sec:method:goldtraj}

\method reports four metric groups. The \emph{outcome metric} records whether the solver localizes the injected root cause: $\mathrm{RC}@1=\mathbf{1}[\hat v=v^\star]$, where $\mathbf{1}[P]$ equals 1 if condition $P$ holds and 0 otherwise. \emph{Evidence metrics} compare $V(T)$ with the accepted path $\pi^\star$ using evidence recall (ER), precision (EP), and F1 ($\mathrm{EF}_{1}$):
\begin{equation*}
  \mathrm{ER}(T)=\frac{|V(T)\cap\pi^\star|}{|\pi^\star|},
  \mathrm{EP}(T)=\frac{|V(T)\cap\pi^\star|}{|V(T)|},
  \mathrm{EF_1}(T)=\frac{2\,\mathrm{ER}\cdot\mathrm{EP}}{\mathrm{ER}+\mathrm{EP}}.
\end{equation*}
The \emph{reasoning metric} measures the proportion of gold reasoning steps that
were performed and justified (RR):
\begin{equation*}
  \mathrm{RR}(T)=\frac{1}{|\mathcal{C}^\star|}
    \sum_{i=1}^{|\mathcal{C}^\star|}
    \mathbf{1}\big[\text{step } i \text{ performed and justified}\big].
\end{equation*}
\emph{Efficiency metrics} report cost and step count, including expensive SQL queries and pipeline runs. With grounding threshold $\tau_g=0.5$, we flag an \emph{unsupported correct answer}, or lucky guess (LG), when a correct answer covers less than half of the accepted evidence path:
\begin{equation}
  \mathrm{LG}(T) \;=\; \mathbf{1}\big[\,\hat v=v^\star \;\wedge\;
  \mathrm{ER}(T)<\tau_g\,\big],
  \label{eq:luckyguess}
\end{equation}

We aggregate over a task cell for the conditional unsupported rate
$\mathrm{CUR}=\Pr[\mathrm{ER}<\tau_g\mid\mathrm{RC}@1=1]$.
Grounded accuracy, $\mathrm{RC@1}^g$, counts correct answers with
$\mathrm{ER}\ge\tau_g$. ER measures coverage of one accepted path, not causal
dependence on its nodes; valid alternatives can be under-credited. EP and
$\mathrm{EF}_1$ penalize broad scanning, and RR additionally checks whether
steps were performed and justified. App.~\ref{app:formal} formalizes the grader
as an Investigation Reward Machine. For the separate exploration partition,
agreement verifiers assess consistency across runs, and checklist verifiers
assess satisfaction of rubric criteria; their uncertainty does not affect the
leaderboard.
Table~\ref{tab:worked} works the running stale-exchange-rate example end to end,
showing how the \emph{same} correct answer scores as grounded or as a lucky guess
depending on the evidence actually inspected.

\begin{table}[t]
\centering\small
\setlength{\tabcolsep}{4pt}
\caption{A worked \method case: the stale-exchange-rate running example of
Figs.~\ref{fig:groundinggap} and~\ref{fig:pipeline}. Two solvers submit the
\emph{same} correct root cause, but only the one whose inspected evidence covers
the gold path is credited as grounded, while the other is flagged as an unsupported
lucky guess (Eq.~\ref{eq:luckyguess}).}
\label{tab:worked}
\begin{tabular}{@{}p{0.30\columnwidth}p{0.63\columnwidth}@{}}
\toprule
Case element & Value \\
\midrule
Symptom (observed) & \texttt{kpi\_daily\_revenue} reads \$124M vs.\ \$127M
  expected, with no failing test and no raised error \\
Root cause $v^\star$ & \texttt{stg\_fx\_rates}, operator
  \texttt{off\_by\_one\_temporal}: a shifted effective-date serves the prior
  day's exchange rates \\
Lineage ($v^\star\!\to$ symptom) & \texttt{stg\_fx\_rates} $\to$
  \texttt{stg\_payments} (USD conversion) $\to$ \texttt{fct\_revenue} $\to$
  \texttt{kpi\_daily\_revenue} \\
Gold evidence path $\pi^\star$ & $\{$\texttt{stg\_fx\_rates},
  \texttt{stg\_payments}, \texttt{fct\_revenue}$\}$ \\
Planted decoys $E^\star$ & a sibling pass-through view of payments and a tracker
  issue blaming the dashboard query \\
\midrule
\multicolumn{2}{@{}p{0.94\columnwidth}}{\emph{Both solvers submit the correct
  node} \texttt{stg\_fx\_rates}:} \\
\;grounded ($\mathrm{RC@1}^g$) & inspects the rates and payments and traces
  lineage: $\mathrm{ER}=1.0\ge\tfrac12$ \\
\;lucky guess ($\mathrm{LG}$) & reads only the KPI and build log, then guesses:
  $\mathrm{ER}=\tfrac13<\tfrac12$ \\
\bottomrule
\end{tabular}
\vspace{-3mm}
\end{table}

To quantify relative progress in self-play, we adapt Elo ratings used in LLM evaluation~\cite{3692070.3692401} to denote relative performance between two roles. In our Evidence-Grounded Elo (EG-Elo), the solver wins a rollout only when it is both correct and grounded ($\mathrm{RC}@1=1$ and $\mathrm{ER}\ge\tau_g$), or the generator wins otherwise. We adopt standard $1{,}500$ initial points and Glicko-2 updates to track rating uncertainty~\cite{glickman2001dynamic}. If EG-Elo${}<1{,}500$, the current solver is underperforming against the current generator. App.~\ref{app:egelo} gives details regarding the standard Elo definition, our Glicko-2 construction, and leaderboard interpretations. We incorporate a cost-aware reward for co-evolving training-pool selection,
\begin{equation}
  R(T) \;=\; \alpha \cdot \mathbf{1}[\text{outcome correct}]
        \;+\; \beta \cdot \mathrm{EF_1}(T)
        \;-\; \lambda \cdot c(T),
  \label{eq:reward}
\end{equation}
where $\alpha$ and $\beta$ weight correctness and evidence F1, $\lambda$ weights the cost penalty, and $c(T)$ is trajectory cost. Process-supervised variants replace $\mathrm{EF_1}$ with $\mathrm{RR}$. Sec.~\ref{sec:trainable} and Sec.~\ref{sec:exp:selfplayparity} respectively use the grounded-win signal as a training selector and as an evaluator of whether solvers
keep pace with adaptive task generation. \vspace{-2mm}

\subsection{Grounded trajectory pool}
\label{sec:method:flywheel}
Grounded-correct trajectories enter the training pool. Multiple rollouts on the same task form contrastive pairs for preference optimization. The experiments use these trajectories for pool selection and in-loop optimization, which adapts operator difficulty to solver success, on the condition that the proposed task must pass the admission gate for reproducible symptom, cause solvable by oracle and non-trivial evidence path. Full operational settings (the system prompt, decoding and serving configuration, GRPO hyperparameters, and the evaluation protocol) are given in Appendix~\ref{app:impl}.

\section{Grounding as a Trainable Objective}
\label{sec:trainable}
After formulating database diagnosis as an evidence-grounded evaluation problem in Sec.~\ref{sec:method}, we next ask whether this signal is useful beyond evaluation. \textsc{CoSE} identifies a solver weakness that outcome accuracy misses, while the grounding signal can select shortcut-prone training tasks to improve solver behavior in the loop.

\subsection{The signal identifies shortcut-prone tasks}
\label{sec:train:signal}
\begin{table}[t]
\centering\small
\setlength{\tabcolsep}{5pt}
\caption{\textsc{CoSE} self-play performance shows how the agent evolves as a solver and generator. These runs use the adaptive self-play batch, not the frozen $128$-case board of Table~\ref{tab:hero}, so absolute accuracies are not directly comparable across the two tables. EG-Elo is initialized at $1,500$ with the final self-play gap $\Delta$ denoting the generator's win rate (lower is better). Older models show a low accuracy in solving the problems whereas the newer models showcase a more balanced EG-Elo ratings between two roles.}
\label{tab:coevoeval}
\resizebox{0.95\linewidth}{!}{
\begin{tabular}{lccccc}
\toprule
Model & $\mathrm{RC@1}$ & $\mathrm{RC@1}^g$ & $\Delta$\,(final)$\downarrow$ & $R_s/R_g$ & $\Delta$ traj. \\
\midrule
\multicolumn{6}{l}{\emph{Proprietary}}\\
\texttt{gpt-4o-mini}   & 0.38 & 0.17 & 0.89 & 1308/1692 & $0.78\!\to\!0.89$ \\
\texttt{gpt-5}         & 0.83 & 0.38 & \textbf{0.42} & 1530/1470 & $0.93\!\to\!0.42$ \\
\midrule
\multicolumn{6}{l}{\emph{Open-sourced}}\\
\texttt{qwen3.6-27b}   & 0.89 & \textbf{0.56} & 0.45 & 1519/1481 & $0.19\!\to\!0.45$ \\
\texttt{qwen3.5-9b}    & 0.44 & 0.44 & 0.56 & 1479/1521 & $0.81\!\to\!0.56$ \\
\texttt{qwen3.6-35b}   & 0.56 & 0.33 & 0.67 & 1437/1563 & $0.62\!\to\!0.67$ \\
\texttt{qwen3-32b}     & 0.00 & 0.00 & 0.99 & \phantom{0}1048/1952$^\dagger$ & $0.99\!\to\!0.99$ \\
\texttt{qwen3-30b}     & 0.00 & 0.00 & 0.99 & \phantom{0}1048/1952$^\dagger$ & $0.99\!\to\!0.99$ \\
\bottomrule
\end{tabular}}
\end{table}

Using CUR as defined in Sec.~\ref{sec:method:verify}, Table~\ref{tab:coevoeval}
shows that a generator can surface tasks on which the solver's apparent success is under-grounded. We test the signal's utility for task generation with two generators that rank the same fault cells, one using CUR and the other the standard outcome gap $1-\mathrm{RC@1}$, to be solved by the same solver. Table~\ref{tab:motivation:coevo} shows that CUR selects tasks with a higher conditional grounding gap and final correct rate, suggesting that the CUR signal can identify shortcut-prone tasks that are not captured as well by outcome difficulty. The difference is directional rather than conclusive at this sample size (two-proportion $z=1.74$, two-sided $p=0.082$, one-sided $p=0.041$), which motivates but does not by itself establish its use as the task-selection signal. App.~\ref{app:coevo:setup} provides the full experimental setup and a difficulty-matching probe. \textbf{Finding~1: CUR-based selection tends to surface more ungrounded agent behavior than outcome-error selection, though the effect is marginal at the current sample size. }

\begin{table}[t]
\centering
\small
\setlength{\tabcolsep}{2.5pt}
\caption{CUR selects more correct-but-ungrounded behavior than the outcome-based
control with two-proportion $z=1.74$ (one-sided $p=0.041$, two-sided $p=0.082$).}
\label{tab:motivation:coevo}
% \resizebox{0.75\linewidth}{!}{
\small
\begin{tabular}{lcccc}
\toprule
Signal & RC@1 & ER & CUR & Lucky\\
\midrule
CUR (ours) & 0.309 & 0.842 & \textbf{0.238} & 0.074 \\
Outcome & 0.260 & 0.842 & 0.113 & 0.029 \\
\bottomrule
\end{tabular}
\vspace{-3mm}
\end{table}

\subsection{In-loop optimization improves grounding}
\label{sec:train:inloop}
By selecting tasks that elicit unsupported correct answers, we train the solver to prefer evidence-supported derivations. Because all arms in Table~\ref{tab:motivation:coevo3} share the same held-out batch, this table also serves as the attribution control: the difference across arms isolates the effect of the training objective from the task population. It shows that offline approaches such as SFT on gold paths and static preference tuning over correct answers inflate the grounding gap because the answer label is identical whether or not the trajectory is grounded (Table~\ref{tab:motivation:coevo2} in App.~\ref{app:controls}). By sampling multiple rollouts per task, we incorporate group-relative policy optimization (GRPO)~\cite{shao2024deepseekmathpushinglimitsmathematical} to optimize the evidence reward in-loop using evidence recall on correct answers in the advantage. After two GRPO rounds, CUR already falls below the base value with higher RC@1, which is consistent with GRPO's behavior to amplify grounded behavior when a rollout group contains both evidence-following solutions and near-deterministic answer shortcuts. Apps.~\ref{app:coevo:setup} and \ref{app:trainfindings} provide reward details and generalization analyses. \textbf{Finding~2: on the matched held-out batch, in-loop evidence optimization improves both
support and final root-cause accuracy.} We include a transferability study in Appendix~\ref{app:n9}. 

\begin{table}[t]
\centering
\small
\setlength{\tabcolsep}{3pt}
\caption{Evaluation on a matched held-out batch ($n{\approx}500$) shows that offline SFT / RWR inflate the conditional grounding gap while GRPO reduces it below base while improving RC@1.}
\label{tab:motivation:coevo3}
\begin{tabular}{lcccc}
\toprule
Arm & RC@1 & ER & CUR & Lucky \\
\midrule
Base ($30$B) & 0.292 & 0.824 & 0.163 & 0.048 \\
\midrule
Offline SFT & 0.191 & 0.870 & 0.521 & 0.100 \\
Offline RWR & 0.227 & 0.852 & 0.465 & 0.105 \\
\midrule
\textbf{In-loop GRPO} & \textbf{0.367} & 0.869 & \textbf{0.054} & 0.020 \\
\bottomrule
\end{tabular}
\vspace{-3mm}
\end{table}

\subsection{End-to-end co-evolution loop}
\label{sec:train:loop}
Finally, we incorporate co-evolution into training by iteratively letting the generator reselect batches against the current solver to train the solver on those batches. We conduct the co-evolution loop for six rounds to track progress with solver EG-Elo. Table~\ref{tab:motivation:coevoloop} shows a trajectory that peaks at round 2 and then fluctuates across later rounds, consistent with the run-to-run variance characteristic of reinforcement-learning training~\cite{henderson2018deeprl}. Reading at the endpoint of round 6, we observe improvement over the untrained base on RC@1, evidence recall, CUR, and solver EG-Elo, though the intermediate rounds are noisy rather than monotone, suggesting a net shift from ungrounded shortcuts toward grounded exploration. Details of the reward, input controls, and rating setup appear in Appendix~\ref{app:coevo:setup}.

\begin{table}[t]
\centering
\small
\setlength{\tabcolsep}{3pt}
\caption{A co-evolution loop using \texttt{Qwen3-30B-A3B} over six rounds, reporting means across three seeds. Each round trains a LoRA on a batch reselected using CUR. The round-6 endpoint improves over the base on all four metrics, but intermediate rounds fluctuate and the best values occur at round~2.}
\label{tab:motivation:coevoloop}
% \resizebox{0.95\linewidth}{!}{
\begin{tabular}{lcccc}
\toprule
Round & RC@1 & ER & CUR & solver EG-Elo \\
\midrule
1 (base) & 0.896 & 0.463$\pm$.018 & 0.507$\pm$.113 & 1438$\pm$77 \\
2 & 0.972 & \textbf{0.613}$\pm$.034 & \textbf{0.185}$\pm$.112 & \textbf{1790}$\pm$110 \\
3 & 0.986 & 0.482$\pm$.090 & 0.482$\pm$.238 & 1514$\pm$240 \\
4 & 0.979 & 0.521$\pm$.013 & 0.347$\pm$.023 & 1638$\pm$20 \\
5 & 0.986 & 0.566$\pm$.013 & 0.302$\pm$.079 & 1686$\pm$74 \\
6 & 0.993 & 0.554$\pm$.047 & 0.328$\pm$.166 & 1666$\pm$161 \\
\midrule
$\Delta$ (6$-$1) & $+0.097$ & $+0.091$ & $\mathbf{-0.179}$ & $+228$ \\
\bottomrule
\end{tabular}
% }
\end{table}

\section{Experiments}
\label{sec:experiments}
We organize the evaluation around the three questions from Sec.~\ref{sec:intro}. RQ1 measures the gap between final-answer accuracy and evidence-grounded diagnosis. RQ2 locates that gap across fine-grained dimensions of difficulty. RQ3 evaluates how the evidence signal supports adaptive solver improvement. Concretely, we have RQ1, RQ2 and RQ3 addressed by Finding~3, Finding~4, and Findings~1 and~2 respectively, with Finding~5 a self-play corollary. Table~\ref{tab:hero} reports the overall leaderboard, with analyses explaining where the aggregate scores come from. We include additional controls, transfer results, and statistical tests in Appendix~\ref{app:validation} and~\ref{app:findings}.

\subsection{Experimental setup}
\label{sec:exp:setup}
Table~\ref{tab:hero} uses a frozen board of 64 tasks with two repeats ($128$ episodes per model), covering nine operators: null injection, duplicate primary keys, incorrect join key, off-by-one temporal filter, unenforced uniqueness, dropped join predicate, and three column-rename variants. It spans 13 environments: three local DuckDB projects and ten BIRD-derived databases~\cite{bird}. Under the API budget, all frozen-board comparisons, trained and untrained, use the same task assignment; the released library of 83 environments and 21 operators describes available coverage, not the scope of the ranking. We evaluate proprietary and open-weight models available by early July 2026~\cite{openai2026gpt56,yang2025qwen3technicalreport,qwenteam2026qwen35omnitechnicalreport,qwen3.6-27b,qwen3.6-35b-a3b}, alongside supplementary scaffold and published-agent studies~\cite{yang2024sweagentagentcomputerinterfacesenable,yao2023react,spider2}. The seven-model proprietary board contains $896$ episodes. The separate RL rows use a matched held-out batch (Sec.~\ref{sec:train:inloop}). Environment-clustered intervals separate capability bands but overlap for adjacent models (App.~\ref{app:unified:leaderboard}); synthetic--organic comparisons preserve ordering among three backbones while showing different fault distributions (App.~\ref{app:unified:organic}). Apps.~\ref{app:envspec},~\ref{app:impl}, and~\ref{app:validation} give the specifications, protocols, and supplementary evaluations.

\subsection{Capability does not remove grounding gap}
\label{sec:exp:hero}
\label{sec:exp:paradox}

\newcommand{\best}[1]{\textbf{#1}}

% Inline RC@1^g -> RC@1 "dumbbell" for the rightmost column (folds former Fig.~3
% into the table, per-row, on a shared 0--70(%) scale). #1 = grounded RC@1^g (%),
% #2 = outcome RC@1 (%). Blue circle = grounded, orange square = outcome; the
% segment between them is the correct-but-unsupported (inflation) gap. Markers are
% sized in mm (absolute) so every row's axis lines up.
\newcommand{\gapbar}[2]{%
  \begin{tikzpicture}[x=0.30mm,baseline=-0.6ex]
    \draw[black!12,line width=0.4pt] (0,0) -- (70,0);
    \draw[very thick,black!45] (#1,0) -- (#2,0);
    \node[circle,fill=blue!65,draw=blue!80!black,inner sep=0pt,minimum size=1.6mm] at (#1,0){};
    \node[rectangle,fill=orange!80,draw=orange!80!black,inner sep=0pt,minimum size=1.5mm] at (#2,0){};
  \end{tikzpicture}%
}

\begin{table*}[t]
\centering\small
\setlength{\tabcolsep}{5pt}
\caption{\textbf{Outcome accuracy exceeds grounded accuracy across model tiers.}
Frozen-board rows share a $64$-task$\times2$-repeat assignment; the trained
Qwen3.6-27B row has $127$ completed episodes. The bottom RL rows use a separate
matched held-out batch ($n\approx500$, Sec.~\ref{sec:train:inloop}) and are not
directly comparable to frozen-board rows. Blue circles denote $\mathrm{RC@1}^g$
and orange squares denote $\mathrm{RC@1}$ on a shared $0$--$70\%$ scale.
Their separation is LG, the unconditional unsupported-correct rate; CUR is
LG divided by RC@1. Arrows indicate whether higher or lower is better.
Apps.~\ref{app:n9} and~\ref{app:trainfindings} give training comparisons.}
\label{tab:hero}
\begin{tabular}{llccccccc}
\toprule
 & & \multicolumn{1}{c}{\emph{Outcome}} & \multicolumn{4}{c}{\emph{Grounded evidence}} & \multicolumn{1}{c}{\emph{Rating}} & \multicolumn{1}{c}{\emph{Grounding gap}} \\
\cmidrule(lr){3-3}\cmidrule(lr){4-7}\cmidrule(lr){8-8}\cmidrule(lr){9-9}
Tier & Model
 & $\mathrm{RC@1}\!\uparrow$
 & $\mathrm{RC@1}^g\!\uparrow$
 & $\mathrm{ER}\!\uparrow$
 & $\mathrm{LG}\!\downarrow$
 & $\mathrm{CUR}\!\downarrow$
 & EG-Elo\,$\uparrow$
 & {\scriptsize $\mathrm{RC@1}^g\!\rightarrow\!\mathrm{RC@1}$} \\
\midrule
% ---- Proprietary. Top block: OpenRouter frozen-corpus run (2026-07-18), SAME
%      128-ep frozen corpus as the open-weight rows (identical tasks + tool parser)
%      -> DIRECTLY COMPARABLE and bold-eligible. gpt-5.6-sol best overall (takes the
%      global RC@1^g/EG-Elo bolds from Qwen3.6-27B, which stays best-open); CUR band
%      0.15-0.23 across all tiers. ----
\multirow{7}{*}{\rotatebox{90}{\scriptsize Proprietary}}
  & \texttt{gpt-5.6-sol}       & \best{0.688} & \best{0.586} & 0.816 & 0.102 & 0.148 & \best{1588} & \gapbar{58.6}{68.8} \\
  & \texttt{kimi-k3}           & 0.641 & 0.531 & 0.741 & 0.109 & 0.171 & 1532 & \gapbar{53.1}{64.1} \\
  & \texttt{claude-sonnet-5}   & 0.633 & 0.523 & 0.686 & 0.109 & 0.173 & 1524 & \gapbar{52.3}{63.3} \\
  & \texttt{glm-5.2}           & 0.492 & 0.398 & 0.804 & 0.094 & 0.190 & 1396 & \gapbar{39.8}{49.2} \\
  & \texttt{deepseek-v4-flash} & 0.414 & 0.344 & 0.841 & 0.070 & 0.170 & 1340 & \gapbar{34.4}{41.4} \\
  & \texttt{grok-4.5}          & 0.305 & 0.234 & 0.821 & 0.070 & 0.231 & 1229 & \gapbar{23.4}{30.5} \\
  & \texttt{gemini-3.5-flash}  & 0.195 & 0.164 & 0.840 & 0.031 & 0.160 & 1157 & \gapbar{16.4}{19.5} \\
\midrule
% ---- Open, small (REAL RC@1/RC@1^g/ER/LG/Infl from 8 x 128 evaluations;
%      CUR/EG-Elo still pending difficulty_study/rating aggregation) ----
\multirow{12}{*}{\rotatebox{90}{\scriptsize Open source models}}
  & \texttt{qwen2.5-7b}   & 0.047 & 0.039 & 0.502 & 0.008 & 0.167 & 1029 & \gapbar{3.9}{4.7} \\
  & \texttt{qwen2.5-14b}  & 0.273 & 0.203 & 0.535 & 0.070 & 0.257 & 1197 & \gapbar{20.3}{27.3} \\
  & \texttt{qwen3-4b}     & 0.141 & 0.109 & 0.685 & 0.031 & 0.222 & 1101 & \gapbar{10.9}{14.1} \\
  & \texttt{qwen3-8b}     & 0.062 & 0.023 & 0.665 & 0.039 & 0.625 & 1013 & \gapbar{2.3}{6.2} \\
  & \texttt{qwen3-30b}    & 0.039 & 0.016 & 0.682 & 0.023 & 0.600 & 1005 & \gapbar{1.6}{3.9} \\
  & \textbf{+ co-evolution} & 0.039 & \textbf{0.031} & \textbf{0.706} & \textbf{0.008} & \textbf{0.200} & N/A & \gapbar{3.1}{3.9} \\
  & \texttt{qwen3-32b}    & 0.195 & 0.148 & 0.628 & 0.047 & 0.240 & 1141 & \gapbar{14.8}{19.5} \\
  & \texttt{qwen3.5-9b}   & 0.276 & 0.260 & 0.764 & 0.016 & 0.057 & 1255 & \gapbar{26.0}{27.6} \\
  & \texttt{qwen3.5-35b}  & 0.242 & 0.219 & 0.774 & 0.023 & 0.097 & 1213 & \gapbar{21.9}{24.2} \\
  & Qwen3.6-27B       & 0.586 & 0.508 & 0.832 & 0.078 & 0.133 & 1508 & \gapbar{50.8}{58.6} \\
  & \textbf{\;+ co-evolution} & \textbf{0.606} & \textbf{0.528} & 0.820 & 0.078 & \textbf{0.130} & N/A & \gapbar{52.8}{60.6} \\
  & Qwen3.6-35B-A3B   & 0.328 & 0.281 & 0.801 & 0.047 & 0.143 & 1277 & \gapbar{28.1}{32.8} \\
\midrule
% ---- Ours (frozen): co-evolved checkpoint on the SAME frozen 128-ep corpus as the
%      open-weight rows (identical tasks, hermes parser) -> DIRECTLY COMPARABLE.
%      Same outcome accuracy (0.039), grounding improves (CUR 0.60->0.20). Base row
%      duplicates the open-weight qwen3-30b line intentionally. Directional; small-n
%      significance deferred to App.~\ref{app:n9}. ----

% ---- Ours (batch): in-loop GRPO on the matched held-out batch (easier population
%      WITH headroom; NOT the frozen corpus). Larger closure. § quarantines it. ----
  \multirow{2}{*}{\rotatebox{90}{\scriptsize RL}}
  & \texttt{qwen3-30b} base           & 0.292 & 0.244 & 0.824 & 0.048 & 0.163 & N/A & \gapbar{24.4}{29.2} \\
  & \textbf{\;+ in-loop GRPO} & \textbf{0.367} & \textbf{0.347} & 0.869 & \textbf{0.020} & \textbf{0.054} & N/A & \gapbar{34.7}{36.7} \\
\bottomrule
\end{tabular}
\end{table*}

Table~\ref{tab:hero} reports outcome accuracy and grounded accuracy side by side. Across model tiers, RC@1 is consistently higher than grounded accuracy RC@1$^g$, suggesting that a nontrivial share of correct root-cause answers lack sufficient evidence-path coverage. Despite final-answer accuracy rising steadily from $0.195$ to $0.688$ with capability, evidence gating lowers every score visualized in dumbbells of Table~\ref{tab:hero}. Among correct answers, the unsupported rate stays consistently between $0.148$ and $0.231$. In the trained rows, we observe that optimizing the evidence signal reduces CUR while preserving or improving outcome accuracy (Sec.~\ref{sec:trainable}), suggesting that the signal serves both as a diagnostic metric and as a criterion for selecting training data. \textbf{Finding~3: Capability improves outcomes without eliminating unsupported success.} We further include conditional analysis clustered on environment in Appendix~\ref{app:unified:conditional}.

\subsection{Evidence grounding fails structurally}
\label{sec:exp:where}

\begin{figure*}[t]
  \centering
  \includegraphics[width=0.9\textwidth]{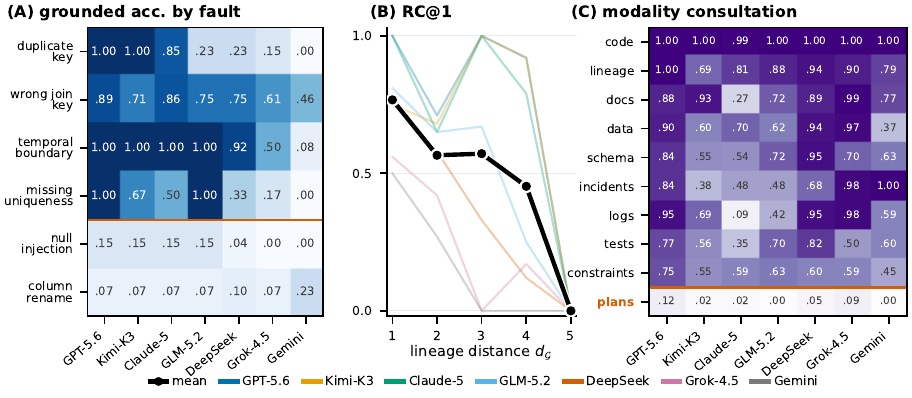} \vspace{-5mm}
  \caption{\textbf{Evidence grounding varies with fault mechanism, lineage distance,
  and artifact modality.} Seven proprietary models use the frozen board of
  $128$ episodes each. \textbf{(A)} Grounded accuracy is higher for duplicate-key,
  join-key, and temporal-boundary faults than for null injection and column
  renaming. \textbf{(B)} Root-cause accuracy declines with lineage distance;
  the five-hop estimate uses few cases (App.~\ref{app:findings}).
  \textbf{(C)} Query plans are inspected in at most $12\%$ of episodes for any
  model. Consultation rates measure behavior, not exposure during pretraining.}
  \label{fig:rq2}
  \Description{Three panels show grounded accuracy by fault family, outcome accuracy by lineage distance, and consultation rates by artifact modality for seven proprietary models. Null injection and column renaming remain difficult, accuracy falls with distance, and query plans are rarely consulted.}
  \vspace{-2mm}
\end{figure*}

Specific fault operators lead to grounding failures that aggregate accuracy hides. Fig.~\ref{fig:rq2}(A) shows that duplicate keys, incorrect joins, and temporal boundaries become more tractable for stronger models, while null injections and column renaming remain difficult across models. Fig.~\ref{fig:rq2}(B) shows that root-cause accuracy declines toward zero at five hops from the symptom, although few cases reach that depth (App.~\ref{app:findings}). Fig.~\ref{fig:rq2}(C) shows that query plans are rarely inspected (at most $12\%$ for GPT-5.6), whereas code and lineage are consulted more often. Limited pretraining exposure is one possible explanation for this modality imbalance; these experiments do not measure it directly.

Symptom access has more direct experimental support. All $75$ unsupported correct answers on the unified board occur among $326$ correct explicit-failure diagnoses ($\mathrm{CUR}=0.230$); all $105$ correct silent-drift diagnoses meet the grounding threshold (App.~\ref{app:unified:heatmap}). Keeping the precise symptom while withholding logs leaves localization nearly unchanged (App.~\ref{app:controls:logblind}); additionally making the symptom generic reduces both accuracy and lucky guessing (App.~\ref{sec:exp:shortcut}). These controls support symptom-text shortcuts in the tested settings, without establishing a pretraining cause. \textbf{Finding~4: Grounding failures are structured by graph distance and specific fault types.}

\subsection{Most solvers fall behind generators}
\label{sec:exp:selfplayparity}
While the preceding results concern a fixed task distribution, we next ask whether a solver can keep pace when the generator adapts its fault to the solver's current grounding failures. We observe from Table~\ref{tab:coevoeval} that, except for \texttt{gpt-5} and the state-of-the-art open source \texttt{qwen3.6-27b} finishing in the solver's favor on EG-Elo (App.~\ref{app:egelo}), most other models finish in the generator's favor. Because the self-play batch differs from the frozen board and covers only seven models, we read this as an observation rather than a causal claim: in our runs, only the strongest proprietary and most recent open source models keep pace with an adaptive generator. \textbf{Finding~5: in our self-play setting, only the strongest recent models keep pace with an adaptive generator, consistent with (but not isolating) stronger out-of-distribution adaptability. }

\subsection{Similar negligence from agent frameworks}
\label{sec:exp:shipped}
Aside from the ReAct-DS architecture~\cite{yao2023react}, we also evaluate various LLM database agents including SWE-agent~\cite{yang2024sweagentagentcomputerinterfacesenable}, spider-agent~\cite{spider2}, and Alpha-SQL~\cite{li2025alphasql}, all using \texttt{gpt-4o-mini} due to API budget constraints. Table~\ref{tab:motivation:realagents} shows that all three agents fail to localize fault from source data. Specifically, agents with access to incident tracker do not consult it during runs even when it explicitly contains relevant user report. This suggests that existing database agentic frameworks are not sufficiently exposed to multiple database modalities, hence relying on model's own capability to explore modalities itself.

\subsection{Case studies}
\label{sec:exp:case}
We conduct a case study on two trajectories to observe how the evidence layer separates a correct answer from a grounded one. 

First, agents can use error-text clues to name the root cause without tracing its dependencies. In the pilot's \texttt{column\_rename} cases, the downstream build error exposes the offending column; \texttt{gpt-5} reaches RC@1 of $1.0$, mean ER of $0.37$, and LG of $0.92$ (App.~\ref{sec:exp:shortcut}). These are operator-level averages, illustrating the shortcut seen in individual trajectories. The generic-symptom ablation, rather than log removal alone, tests dependence on this clue.

In the second trajectory, \texttt{gpt-5} correctly names the fault type but localizes it to the wrong table. It never opens the incident report, which explicitly states ``the link to \texttt{account} is not populated''. This illustrates neglect of relevant evidence available in the workspace; it does not identify the source of that behavior in training data. Details appear in Appendix~\ref{sec:exp:crossmodal}.

\begin{table}[t]
\centering
\scriptsize
\setlength{\tabcolsep}{2.5pt}
\caption{Performance of published agents on BIRD-derived cases shows that the test runs commonly ignore faults from source data and seldom use the incident tracker. ER refers to a lower bound inferred from native actions. For Alpha-SQL, Fix and incident-read are not applicable.}
\label{tab:motivation:realagents}
\vspace{-3mm}
\resizebox{0.75\linewidth}{!}{
\begin{tabular}{llcccc}
\toprule
System & Fault & RC@1 & Fix & ER & Inc. \\
\midrule
\multirow{2}{*}{SWE-agent} & null & 0.00 & 0.12 & 0.23 & 0\% \\
 & join-key &\textbf{0.43} & 0.14 & 0.48 & 0\% \\
\midrule
\multirow{2}{*}{Spider2} & null & 0.00 & 0.00 & 0.23 & 0\% \\
 & join-key & 0.10 & 0.20 & 0.51 & 0\% \\
\midrule
\multirow{2}{*}{Alpha-SQL} & null & 0.00 & N/A & 0.16 & N/A \\
 & join-key & 0.00 & N/A & 0.29 & N/A \\
\bottomrule
\end{tabular}}
\vspace{-3mm}
\end{table}

\section{Related Work}
\label{sec:related}

\method intersects with SQL-error and data-agent benchmarks, root-cause analysis, process-level supervision, and data provenance, with the most directly related works highlighted here. App.~\ref{app:related} gives the extended comparison and Table~\ref{tab:related}.

\subsection{SQL-error and data-agent benchmarks}
The nearest benchmark in topic, \emph{NL2SQL-Bugs}~\cite{nl2sqlbugs}, detects and \emph{types} semantic errors in a single generated query. Its localization is categorical (which kind of bug) rather than positional, and models detect that a query is wrong at about 75$\%$ far more reliably than they pinpoint why. Data-pipeline and agentic benchmarks, i.e. Spider~2.0~\cite{spider2}, ELT-Bench~\cite{eltbench}, BEAVER~\cite{beaver}, and DSBench / DABench~\cite{dsbench,dabench}, grade the \emph{outcome} (final SQL, produced tables, or answer) and do not supervise which evidence the agent consulted. \method lifts the question from the single query to the system level and grades the root-cause \emph{object} and the evidence path to it, explicitly flagging unsupported correct answers (Eq.~\ref{eq:luckyguess}).

\subsection{Process supervision and root-cause analysis}
A growing line rewards the reasoning \emph{process} rather than the answer. \emph{OpenRCA~2.0}~\cite{openrca2} adopts the same separation, supplementing outcome labels with causal process supervision and reporting that many correct identifications are ungrounded, but over microservice telemetry instead of a typed data-provenance graph and without an in-loop training signal, while \emph{AgentLens}~\cite{agentlens} independently documents our lucky-guess phenomenon (the ``Lucky Pass'') in SWE-agents. Building on step-level process rewards that outperform outcome-only supervision~\cite{lightman2024verify,mathshepherd}, \method conducts process supervision for data systems with the gold reasoning chain derived mechanically from lineage and separates grounding into path coverage and reasoning recall with signal optimized in-loop (Sec.~\ref{sec:trainable}).

\subsection{Data provenance and grounded attribution}
\method scores over a typed provenance graph, building on classical database provenance~\cite{provenanceSurvey,whyWhereProvenance} and on outlier-explanation methods that trace a data anomaly to its source, such as \emph{Scorpion}~\cite{scorpion} and view-conditioned causality~\cite{meliou2011}. Its two graded axes have independent precedent: evidence recall and precision are the pipeline-diagnosis analog of citation-grounding evaluation~\cite{alce}, asking whether a submitted root cause rests on inspected artifacts. The need to separate the answer from its support follows from evidence that a model's stated reasoning is often unfaithful to the true cause of its answer~\cite{turpin2023}, where a correct label can sit atop an ungrounded trajectory, which is the lucky guess \method flags.

\section{Discussion}
\label{sec:discussion}
\label{sec:limitations}

\textbf{Limitations.}
The grounded flag measures coverage of one accepted witness, not causal evidence use. Counterfactual masking provides a small pilot, and the expert-audit instrument has only been tested with simulated raters (Apps.~\ref{app:unified:intervention}--\ref{app:unified:audit}). A graph reconstruction identifies $7/64$ tasks where following a single shortest route can fall below $\tau_g$, including cases whose accepted witness combines several symptom routes (App.~\ref{app:unified:altpaths}). EP, evidence F1, and RR complement recall, whose per-operator difficulty ordering is relatively stable across thresholds $0.1$--$0.9$ ($20{,}026$ episodes; mean pairwise Spearman $\rho=0.843$, App.~\ref{app:val:tau}). Generalization claims are limited to the evaluated 13 environments and nine operators; synthetic faults broaden coverage without matching natural incident frequencies. Finally, short self-play schedules do not establish convergence. Appendix~\ref{app:limitations} details these boundaries.

\textbf{Research opportunities.}
We observe four directions for next steps. Grounding can be \emph{optimized as an investigation behavior}, e.g. through contrastive trajectories that reach the same cause with and without inspecting relevant dependencies. Agents need \emph{provenance-aware cross-modality routing} to further integrate use of modalities. Evaluation should \emph{allow multiple valid paths} and test counterfactual evidence use for further causal grounding. Lastly, dynamic benchmarks need \emph{transparent versioning} with a frozen batch for reproducible ranking. Details are covered in Appendix~\ref{app:opportunities}.

\section{Conclusion}
\label{sec:conclusion}
We introduced \method, a benchmark that grades whether a database agent both
identifies the correct root cause and inspects the evidence needed to support it,
using typed provenance graphs to recover root causes and accepted evidence paths
mechanically. The benchmark spans $83$ buildable environments across five source
families with a frozen $128$-case board, and every metric is recomputable from
saved trajectories, so grounding can be measured and improved without human annotation. Across the unified board, stronger models solve more cases but a
conditional grounding gap persists, concentrated in particular fault mechanisms,
growing with dependency distance, and marked by systematic neglect of informative
artifacts. The same signal supports adaptive task generation and, in the matched
in-loop study, improves both evidential support and root-cause accuracy. More
broadly, whether a diagnosis is trustworthy depends on the investigation behind it,
which grounded evaluation makes measurable and, in turn, optimizable.

\ifdefined\DataTraceProceedings
\noindent\textbf{Supplementary material.} Detailed protocols, controls, and
validity analyses appear in the companion supplement (Appendices A--O).
Project materials are linked from \url{\artifacturl}.
\fi

\bibliographystyle{ACM-Reference-Format}
\bibliography{references}
\ifdefined\DataTraceProceedings
\else
  \clearpage
  % Keep the accepted appendix lettering A--O in all companion builds.
\appendix
\setcounter{table}{0}
\setcounter{figure}{0}
\setcounter{equation}{0}
\renewcommand{\thetable}{S\arabic{table}}
\renewcommand{\thefigure}{S\arabic{figure}}
\renewcommand{\theequation}{S\arabic{equation}}
\section{Environment and Action-Space Specification}
\label{app:envspec}
This appendix gives the full type system, fault-operator library, and action set
summarized in Sec.~\ref{sec:method}.

\paragraph{Node types ($\mathcal{T}_V$, 13).}
\{\texttt{table}, \texttt{column}, \texttt{view}, \texttt{cte},
\texttt{constraint}, \texttt{plan}, \texttt{statement}, \texttt{test},
\texttt{config}, \texttt{doc}, \texttt{incident}, \texttt{log}, \texttt{run}\}:
a materialized relation and its columns, a dbt model compiled to a view/table and
the CTEs within its SQL, declared or enforced constraints, \texttt{EXPLAIN} query
plans and individual SQL statements, data tests, pipeline configuration,
catalogue/README/data-dictionary docs, incident-tracker reports, build/run/test
logs, and pipeline-run records.

\paragraph{Edge types ($\mathcal{T}_E$, 6).}
\{\texttt{depends}, \texttt{references}, \texttt{derives}, \texttt{produces},
\texttt{validates}, \texttt{mentions}\}: model-level dependency
(\texttt{ref()}/\texttt{source()}), foreign-key or cross-artifact reference,
column-level provenance recovered by static SQL analysis (an output column
\texttt{derives} from input column(s)), a run/statement that \texttt{produces} a table,
a test that \texttt{validates} a table/column/constraint, and an incident or doc
that \texttt{mentions} an artifact. Model- and view-level dependencies come from
the compiled dbt DAG, column-level \texttt{derives} edges from sqlglot, constraint
and \texttt{plan} nodes from the catalogue and \texttt{EXPLAIN}.

\paragraph{Fault-operator library ($\mathcal{I}$, 21).}
The injection operators span the five mechanism classes of
Table~\ref{tab:datatrace:profile} (source-data 6, code 6, schema 5, constraint 3,
plan 1):
\begin{itemize}
  \item \emph{source-data faults}: null injection, duplicate primary keys,
        type drift, timezone shift, character-encoding corruption,
        late-arriving records
  \item \emph{query / view-definition faults}: incorrect join key or join type,
        off-by-one temporal filter, wrong aggregation function, dropped
        \texttt{WHERE} predicate in a CTE
  \item \emph{constraint faults}: a missing or unenforced foreign key or
        uniqueness constraint that admits orphans or duplicates
  \item \emph{schema faults}: upstream column rename without downstream
        propagation
  \item \emph{config faults}: incorrect incremental-load cursor field.
\end{itemize}
\emph{Silent} operators (value drift with no failing test, detectable only by
statistical checks) instantiate the no-test regime.

\paragraph{Action space (14).}
Every action is parameterized by provenance-graph node identifiers. A trajectory is
therefore a walk on $\mathcal{G}$ (Sec.~\ref{sec:method:actions}).
\begin{center}\small
\begin{adjustbox}{max width=0.95\linewidth}
\begin{tabular}{ll}
\toprule
Action & Effect \\
\midrule
\texttt{inspect\_schema}     & return a table/view schema \\
\texttt{inspect\_constraint} & list declared/enforced constraints \\
\texttt{profile\_table}      & column stats, null and distinct counts \\
\texttt{sample\_rows}        & sample rows from a relation \\
\texttt{read\_incident}      & read an incident-tracker report \\
\texttt{read\_doc}           & read catalogue / README / dictionary \\
\texttt{run\_sql}            & execute an ad-hoc query \\
\texttt{explain\_plan}       & return the \texttt{EXPLAIN} plan \\
\texttt{run\_pipeline}       & (re)build the dbt pipeline \\
\texttt{read\_log}           & read build/run/test logs \\
\texttt{trace\_lineage}      & walk model/column lineage on $\mathcal{G}$ \\
\texttt{edit\_file}          & edit a model or config file \\
\texttt{run\_test}           & run the data-test suite \\
\texttt{submit}              & submit $(\hat{v},\text{fix})$ with evidence pointers \\
\bottomrule
\end{tabular}
\end{adjustbox}
\end{center}

\paragraph{Benchmark composition.}
\emph{Fault mechanism classes} ($21$ operators): source-data (6),
code (6), schema (5), constraint (3), plan (1). \emph{Evidence modalities:} build
logs, schema/catalogue, query plans, data tests, declared constraints, incident
text. \emph{Environments vs.\ instances:} each environment is sampled behind a
self-verifying admission gate (reproducible symptom, reachable cause, non-empty
gold path), enabling fresh instance generation and reducing dependence on a
static, potentially contaminated task list. The released
\textsc{DataTrace-Eval-v1} split is frozen, seed-pinned, and content-hashed.
Spider2 counts the green-bridging pool ($35$ of $70$ example pipelines). The
remaining examples were not included in this evaluation.

\begin{table}[t]
\centering
\small
\caption{\method composition: \textbf{83 buildable environments across five
source families}, spanning hand-authored synthetic pipelines,
real enterprise dbt pipelines, and real open-source dbt repositories. Each
environment is a \emph{fault generator}: a procedural injector samples admitted,
difficulty-indexed instances (operator $\times$ decoys $\times$ seed) with
mechanically-derived ground truth ($v^\star,\pi^\star,\mathcal{C}^\star$) and no
  per-instance human annotation. The frozen leaderboard split
  \textsc{DataTrace-Eval-v1} samples
a balanced core subset ($13$ environments, $128$ cases). The full suite plus
unlimited fresh instances support training and stress-testing.}
\label{tab:datatrace:profile}
\begin{adjustbox}{max width=0.95\linewidth}
\begin{tabular}{llc}
\toprule
Source family & Provenance & Envs \\
\midrule
Native synthetic            & hand-authored dbt (ecom, retails, warehouse)          & 3 \\
BIRD-CRITIC warehouse-grade & real schemas, synthesized warehouse                   & 17 \\
Spider2 enterprise pipelines & real SaaS/enterprise dbt (salesforce, xero, \ldots)   & 35 \\
Real dbt OSS repositories   & dbt-utils, dbt-project-evaluator, \ldots (real PRs)    & 18 \\
Generator-B mined incidents & real BIRD-CRITIC user incidents                        & 10 \\
\midrule
\textbf{Total} & \emph{five source families} & \textbf{83} \\
\bottomrule
\end{tabular}
\end{adjustbox}
\end{table}

\subsection{Corpus Views and Difficulty Axes}
\label{sec:method:characteristics}
We separate three views of the corpus. The \emph{construction library} contains the full set of $83$ buildable environments across five source families (Fig.~\ref{fig:pipeline}(B)), while the \emph{validated core pool} admits $448$ of $544$ generated tasks across $14$ core environments. The \emph{frozen evaluation board} (\textsc{DataTrace-Eval-v1}) is a versioned $128$-case subset of $13$ environments. Environment count describes the available construction substrate, whereas the board is a fixed subset used for same-corpus model comparison. Across the validated core pool, accepted tasks vary in gold-path length ($2$--$7$ nodes, mean $5.12$), modality span ($1$--$4$, mean $3.09$), lineage distance ($1$--$5$ hops), and planted decoys ($0$--$2$). These axes form the benchmark's fine-grained difficulty taxonomy and support the analyses in Sec.~\ref{sec:exp:where}.

\section{Implementation Details: Prompts, Decoding, and Hyperparameters}
\label{app:impl}
This appendix consolidates the operational settings needed to reproduce the solver
rollouts, the serving stack, GRPO training, and the evaluation protocol. The
grounding reward and the co-evolution selection rule are specified separately in
App.~\ref{app:coevo:setup}. The EG-Elo rating is defined in App.~\ref{app:egelo}.

\subsection{Reproducibility configuration}
\label{app:impl:reproconfig}
The leaderboard used the ReAct-DS scaffold, two repeats per injection, a
22-tool-step budget, and temperature $0.3$ where the endpoint exposed that
parameter. Reasoning endpoints that only accept their provider default used that
default. All models received the same task prompt, tool schema, frozen manifest,
and parser. We retained per-step actions and observations, so RC@1, ER, EP,
EF$_1$, LG, and CUR can be recomputed without calling the model again. The
submission artifact includes the exact prompts, model identifiers, command-line
configuration, and content hash. Glicko-2 priors and the grounding threshold are
specified in App.~\ref{app:egelo}.

\subsection{Solver scaffold and system prompt}
\label{app:impl:prompt}
Every episode runs a single tool-calling agent under the \textsc{ReAct-DS}
scaffold, which exposes the full data-system action set:
\begin{itemize}
  \item schema/profile/sample introspection (\texttt{inspect\_schema},
        \texttt{profile\_table}, \texttt{sample\_rows}),
  \item documentation and an incident tracker (\texttt{list\_docs},
        \texttt{read\_doc}, \texttt{list\_issues}, \texttt{read\_issue}),
  \item code navigation (\texttt{list\_models}, \texttt{read\_file},
        \texttt{search\_code}),
  \item read-only SQL and build logs (\texttt{run\_sql}, \texttt{read\_log}),
  \item dependency and integrity inspection (\texttt{trace\_lineage},
        \texttt{explain\_plan}, \texttt{inspect\_constraint}, \texttt{run\_test}),
  \item repair and rebuild (\texttt{edit\_model}, \texttt{run\_pipeline}), and
  \item a terminal \texttt{submit} action with operator, node, and evidence arguments.
\end{itemize}
The two external-framework proxies
restrict this set: \textsc{swe} is code-centric (file view/edit/search plus
build/test, no database introspection or lineage) and \textsc{spider} is
database-centric (schema/sample/profile plus SQL, no lineage or repository
navigation). Affordance ablations (\textsc{-nolineage}, \textsc{-noprofile}) and a
salience control (\textsc{-salient}, which advertises the tracker and mandates
upstream tracing) drop or reweight individual primitives. The verbatim system
prompt, with per-instance fields shown in braces, is:

{\footnotesize
\begin{verbatim}
[Scaffold: react_ds] You have the full
data-system toolset, including trace_lineage.

You are a data-engineering diagnosis agent
working on a dbt-style ELT pipeline
materialized in DuckDB. A fault has been
planted somewhere in the pipeline and is now
causing a symptom. Your job:
  1. Find the ROOT CAUSE: the single offending
     node and the kind of fault.
  2. (Optional) Repair it so the pipeline
     returns to green.
  3. Submit your diagnosis with evidence
     citations.

Available relations (sources+models): {tables}
Available data tests: {tests}
Fault operator vocabulary (pick one): {ops}

Observed symptom(s):
  - {symptom; generic under the log-blind ctrl}

Use the typed actions to investigate. Be
efficient but make sure your diagnosis is
grounded in artifacts you actually inspected.
When confident, call
submit(operator, node, evidence).
\end{verbatim}
}
The braced fields are filled per instance from the materialized workspace. The
log-blind control (App.~\ref{app:controls}) replaces the symptom text with a
generic one.

\subsection{Rollout and decoding}
\label{app:impl:decode}
Rollouts use temperature $0.3$ under \textsc{ReAct-DS}, capped at $16$ tool-calling
steps and a per-turn generation budget of $768$ tokens for the $30$B co-evolution
loop and $1024$ for the frozen-board leaderboard. GRPO harvesting samples at
temperature $1.0$ for within-group diversity. Reasoning models that reject a custom
temperature (e.g.\ GPT-5) use their provider default. Open (vLLM, OpenAI-compatible)
and proprietary (Anthropic) models run behind one dispatch layer, so all backbones
see the identical action space and scoring.

\subsection{Serving configuration}
\label{app:impl:serve}
Open-weight models are served with vLLM in bfloat16 and \texttt{--enforce-eager} on
a single NVIDIA H200 NVL ($143$~GB). The Qwen3.5/3.6 family uses the
\texttt{qwen3\_xml} tool-call parser, \texttt{--max-model-len 32768}, and
\texttt{--gpu-memory-utilization 0.85}. The older \texttt{Qwen3-30B-A3B} uses the
\texttt{hermes} parser and \texttt{--max-model-len 16384}. Trained checkpoints are
served as base~$+$~LoRA adapter via \texttt{--enable-lora} rather than a merged
model: this avoids a merge-time config incompatibility on the multimodal wrapper of
\texttt{Qwen3.6-27B} and lets one served base host several adapters.

\subsection{GRPO training hyperparameters}
\label{app:impl:grpo}
Table~\ref{tab:impl:hp} lists the training and harvest settings. Each round samples
$G{=}12$ rollouts per CUR-selected task, scores them with the mechanical reward of
App.~\ref{app:coevo:setup}, and takes an advantage-weighted policy-gradient step on
the assistant tokens only, with a token-level KL to the frozen base (LoRA disabled)
anchoring against collapse. Zero-variance groups carry no learning signal and are
dropped. Adaptation is attention-only LoRA, so the recipe applies unchanged to dense
and sparse-MoE backbones. For multi-round co-evolution on \texttt{Qwen3.6-27B} the
best adapter is \emph{continued} each round (loaded trainable, no merge) rather than
re-initialized from the base.

\begin{table}[t]
\centering\small
\setlength{\tabcolsep}{4pt}
\caption{GRPO training and rollout hyperparameters. The KL weight $\beta$ and
learning rate are the tuned operating point ($\beta{=}0$ degenerates). All other
values are shared across backbones.}
\label{tab:impl:hp}
\begin{adjustbox}{max width=0.95\linewidth}
\begin{tabular}{ll}
\toprule
Setting & Value \\
\midrule
LoRA rank $r$ / $\alpha$ / dropout & $16$ / $32$ / $0.05$ \\
LoRA target modules & attention proj.\ ($q,k,v,o$) only \\
Optimizer & AdamW \\
Learning rate & $5\times10^{-6}$ \\
KL-to-reference weight $\beta$ & $0.05$ \\
Advantage clip & $\pm 3.0$ \\
Epochs / max sequence length & $1$ / $2048$ \\
LR schedule & cosine, $3\%$ warmup \\
Advantage & group-relative $(R_i{-}\mu_g)/(\sigma_g{+}\epsilon)$ \\
Rollouts per group $G$ (harvest) & $12$ \\
Harvest temperature / reward & $1.0$ / process \\
Precision / attention impl.\ & bfloat16 / eager \\
\bottomrule
\end{tabular}
\end{adjustbox}
\end{table}

\subsection{Evaluation protocol}
\label{app:impl:eval}
The frozen leaderboard corpus is $128$ episodes: $64$ unique tasks $\times\,2$
repeats over $13$ environments (\texttt{ecom}, \texttt{retails}, \texttt{warehouse},
and ten BIRD-derived \texttt{birdwh\_*} databases) and nine fault operators.
The construction pipeline supports planted decoys, including sibling
pass-through models and distractor tracker issues, but all $896$ saved episodes
of the unified proprietary board report $n_{\text{decoys}}=0$. Available decoy
support should therefore not be read as a property of those evaluated episodes. A
correct answer counts as grounded ($\mathrm{RC@1}^g$) only if evidence recall clears
$\tau_g{=}0.5$. The same threshold defines
$\mathrm{CUR}=\Pr[\mathrm{ER}{<}\tau_g\mid\text{correct}]$. Leaderboard evaluation is
serial and checkpointed (resumable), so every model traverses the identical task
order.

\section{Scope Comparisons}
\label{app:scope}

\begin{table}[t]
  \centering\scriptsize
  \setlength{\tabcolsep}{2.5pt}
  \begin{adjustbox}{max width=0.95\linewidth}
\begin{tabular}{p{0.18\columnwidth}p{0.40\columnwidth}p{0.34\columnwidth}}
    \toprule
    Dimension & \textbf{In scope (\method)} & Out of scope \\
    \midrule
    Unit & multi-object database, $d_{\mathcal{G}}\!\ge\!1$ & single isolated query \\
    Goal & localize root-cause object and justify it & generate SQL, classify bug type \\
    Signal & schema, data, query, constraint, plan, test, log, incident & schema and query text only \\
    Symptom & silent wrong result and hard failure & explicit error or failing test only \\
    Interaction & tool investigation over the provenance graph & one static forward pass \\
    Ground truth & mechanically derived from dependencies & human-annotated gold SQL \\
    Grading & answer, evidence path, reasoning & end-state correctness only \\
    \bottomrule
  \end{tabular}
\end{adjustbox}
  \caption{Scope of evidence-grounded root-cause localization in databases. The
  right column is the regime covered by prior single-query bug detection and
  test-gated repair.}
  \label{tab:scope}
\end{table}

\begin{table}[t]
\centering\small
\setlength{\tabcolsep}{4pt}
\caption{\textbf{\textsc{CoSE} versus prior evaluation paradigms.} \textsc{CoSE}
is the only paradigm in the comparison that jointly evaluates task generation and
solution under evidence-conditioned scoring. This makes the game harder to guess,
less tied to a static board, and less exposed to benchmark contamination.
Yes~= has the property; $\sim$~= partial; $\times$~= lacks it.}
\label{tab:selfplay:positioning}
\resizebox{0.95\linewidth}{!}{
\begin{tabular}{lccccc}
\toprule
Paradigm & Unguess. & Non-sat. & Dual & Contam.-res. & No humans \\
\midrule
Static accuracy (SWE-bench, BIRD)      & $\times$ & $\times$ & $\times$   & $\times$   & Yes \\
Preference arena (Chatbot Arena)       & $\times$ & $\sim$   & $\times$   & $\sim$     & $\times$   \\
Red-teaming / adversarial suites       & $\sim$   & $\times$ & $\times$   & $\sim$     & $\times$   \\
Self-instruct / generative evals       & $\times$ & $\sim$   & $\sim$     & Yes & Yes \\
Training self-play (SPIN, self-reward) & $\times$ & Yes & Yes & Yes & Yes \\
\midrule
\textbf{\textsc{CoSE} (ours)}          & Yes & Yes & Yes & Yes & Yes \\
\bottomrule
\end{tabular}}
\end{table}

\section{Additional Controls and Negative Results}
\label{app:controls}

\subsection{Log-substitution control}
\label{app:controls:logblind}
A natural objection is that agents skip the incident tracker only because the
run/build log offers a more convenient route to the same information. We test
this with a matched control on the same 10 enriched cases
(Table~\ref{tab:motivation:logblind}): we withhold the build/test log and the
changelog but \emph{keep} the precise failing-test symptom in the prompt and
leave the incident tracker fully readable, isolating ``the log'' as the single
removed variable. The results do not support this explanation. Removing the log
does \emph{not} redirect agents to the incident report: incident-consult remains near zero
(pooled $1.2\%\!\rightarrow\!2.8\%$, \texttt{gpt-5} $3.5\%\!\rightarrow\!5.9\%$,
\texttt{gpt-5-mini} unchanged at $0\%$), and localization and repair are
essentially flat (\texttt{gpt-5} $\mathrm{RC@1}$ $0.43\!\rightarrow\!0.42$,
fix-rate $0.58\!\rightarrow\!0.62$). Agents fall back on lineage, data, and code
to localize and rarely consult the incident reports. Two conclusions follow: under neutral
prompting the cross-modal neglect is \emph{not} a log-substitution artifact
(removing the log does not redirect agents to the tracker), and the verbose
build/test log is \emph{not load-bearing} once the precise symptom is retained.
The neglect is nonetheless a property of the \emph{default} scaffold rather than
an intrinsic limit: explicitly advertising the tracker recovers it for the
stronger backbone (Sec.~\ref{sec:motivation:robustness}). This also refines the
blind ablation of
Table~\ref{tab:motivation:ablation}: that condition collapses RC@1 because it
additionally makes the symptom generic. Holding the symptom fixed and
removing only the log leaves localization intact.

\begin{table*}[t]
\centering
\caption{Log-blind control on the 10 enriched BIRD cases (ReAct-DS, two-mode,
$n{=}86$ baseline / $n{\geq}81$ log-blind per backbone). \textsc{log-blind}
withholds the run/build log and changelog but keeps the precise failing-test
symptom and the readable incident tracker, isolating the log as the single removed
variable. Withholding the log neither increases incident-consult nor materially
changes localization or repair. The neglect is therefore not a log-substitution
artifact, and the log is not load-bearing when the precise symptom remains
available (the tracker is recoverable through salience,
Sec.~\ref{sec:motivation:robustness}).}
\label{tab:motivation:logblind}
\begin{adjustbox}{max width=0.95\linewidth}
\begin{tabular}{llccccc}
\toprule
Backbone & Condition & RC@1 & Fix & EF$_1$ & RR & incident-read \\
\midrule
\multirow{2}{*}{\texttt{gpt-4o-mini}} & baseline  & 0.11 & 0.02 & 0.57 & 0.55 & 0\%   \\
                                      & log-blind & 0.09 & 0.01 & 0.58 & 0.57 & 2.3\% \\
\midrule
\multirow{2}{*}{\texttt{gpt-5-mini}}  & baseline  & 0.19 & 0.40 & 0.53 & 0.57 & 0\%   \\
                                      & log-blind & 0.25 & 0.33 & 0.51 & 0.60 & 0\%   \\
\midrule
\multirow{2}{*}{\texttt{gpt-5}}       & baseline  & 0.43 & 0.58 & 0.52 & 0.60 & 3.5\% \\
                                      & log-blind & 0.42 & 0.62 & 0.54 & 0.58 & 5.9\% \\
\bottomrule
\end{tabular}
\end{adjustbox}
\end{table*}

\subsection{Offline supervision inflates the grounding gap}
Identifying shortcut mass does not establish whether training on these examples
closes the grounding gap. We first evaluate imitation: from each generator, we
synthesize matched grounded gold
trajectories (which inspect the evidence path and rule out the decoys),
LoRA-fine-tune the same $30$B backbone on each, and evaluate $\mathrm{CUR}$ on a
held-out batch with seed-disjoint decoy instances. The result is negative
(Table~\ref{tab:motivation:coevo2}): grounded-gold
SFT \emph{raises} the conditional grounding gap rather than shrinking it (base
$0.098 \rightarrow$ CUR $0.393$, $z{=}3.28$, $p{=}0.001$, with OG $0.295$), even as it
changes the surface behaviors: outcome accuracy rises (RC@1
$0.228\rightarrow0.34$) and decoy-blame falls to near zero
($0.072\rightarrow0.000/0.011$). One interpretation is that imitating short
grounded demonstrations encourages decisive, decoy-avoiding answers that improve
accuracy, while newly correct episodes on held-out cells still reach the correct
cause \emph{without} re-deriving its evidence. The result is more lucky guesses,
not fewer. \emph{Selecting} shortcut-laden tasks does not, by
itself, teach grounding. An imitation objective is blind to the distinction
between grounded- and lucky-correct because both share the same label.

This isolates the missing ingredient and motivates the
reward-coupled formulation of Sec.~\ref{sec:problem}: the training
signal must \emph{prefer the grounded derivation over the lucky guess for the
same correct answer}. We therefore replace imitation with a preference objective
(DPO) over pairs that share an identical correct conclusion but differ in
process. The \emph{chosen} trajectory traverses the evidence path, whereas the
\emph{rejected} trajectory submits the same root cause directly. The gradient
therefore rewards grounding conditioned on correctness, the distinction measured
by CUR.

\paragraph{Neither imitation nor preference tuning closes the gap: the signal
must be in the loop.}
We test the preference objective at a larger scale, using $96$ pairs per
arm, three epochs, and a held-out eval of $n{\approx}180$
($\sim\!50$ correct episodes per arm, against $\sim\!25$ in a preliminary
underpowered run), and the closure does \emph{not} materialize
(Table~\ref{tab:motivation:coevo2}). Two effects emerge. First, the imitation
result replicates at the larger scale: grounded-gold SFT raises $\mathrm{CUR}$
from $0.098$ (base) to $0.393$, now highly significant ($z{=}3.28$, $p{=}0.001$).
Second, grounding-coupled DPO does not reduce it: on
top of the CUR-SFT arm $\mathrm{CUR}$ is $0.429$ vs.\ $0.393$ ($p{=}0.71$), on top
of OG-SFT it is $0.392$ vs.\ $0.295$ ($p{=}0.28$). Every trained model remains far
above base (CUR$+$DPO vs.\ base $z{=}3.49$, $p<0.001$). A directional reduction
seen in the small preliminary run does not persist in the larger experiment and
is consistent with sampling variation. The offline preference signal, grounded
derivation $\succ$ same-answer lucky guess, does not overcome the
shortcut: it is confounded by surface form (the grounded branch is simply longer
and tool-richer), and it cannot transfer grounding to the held-out operators a
small selected batch does not cover. In this experiment, \emph{outcome-correct
supervision does not teach grounding, and neither does a static preference over
correct answers.} This result motivates optimizing evidence recall \emph{within}
the investigation loop rather than from a fixed dataset, as in the online reward of
Sec.~\ref{sec:problem} (reward the grounded derivation as it is
produced). The generator-side result remains independent: CUR is still a distinct
\emph{selection} signal (Table~\ref{tab:motivation:coevo}). What this experiment
shows is that, under the tested conditions, selection benefits from an
\emph{in-loop} rather than offline grounding objective.

\begin{table*}[t]
\centering
\caption{Loop-closure attempt, held-out decoys (seed-disjoint, $n{\approx}180$,
$\sim\!50$ correct/arm). Grounded-gold SFT \emph{raises} the conditional grounding
gap $\mathrm{CUR}$ far above base (CUR $0.098\!\rightarrow\!0.393$, $z{=}3.28$,
$p{=}0.001$). It improves answer selection without improving grounding. Adding a grounding-coupled
preference objective (DPO: grounded derivation $\succ$ same-answer lucky guess)
does \emph{not} reduce it (CUR$+$DPO vs.\ CUR-SFT $p{=}0.71$, OG$+$DPO vs.\ OG-SFT
$p{=}0.28$). All trained models stay significantly above base ($p<0.001$). Offline
imitation \emph{and} preference supervision both fail to close the gap, motivating
an in-loop (RL) evidence reward.}
\label{tab:motivation:coevo2}
\begin{adjustbox}{max width=0.95\linewidth}
\begin{tabular}{lccccc}
\toprule
Arm & RC@1 & ER & $\mathrm{CUR}$ & Lucky & Decoy-blame \\
\midrule
Base ($30$B)         & 0.228 & 0.834 & \textbf{0.098} & 0.022 & 0.072 \\
\midrule
SFT, CUR-selected     & 0.341 & 0.840 & 0.393 & 0.134 & 0.000 \\
SFT, OG-selected     & 0.339 & 0.858 & 0.295 & 0.100 & 0.011 \\
\midrule
CUR-SFT $+$ DPO       & 0.275 & 0.860 & 0.429 & 0.118 & 0.022 \\
OG-SFT $+$ DPO       & 0.283 & 0.877 & 0.392 & 0.111 & 0.011 \\
\bottomrule
\end{tabular}
\end{adjustbox}
\end{table*}

\subsection{Generational robustness}
To test whether correct-but-ungrounded behavior is specific to one backbone, we
repeat the base evaluation and in-loop training
across two Qwen generations and four sizes
(Table~\ref{tab:motivation:gen}). The gap appears on \emph{every} model capable of
solving the shortcut operator, including the recent Qwen3-8B and Qwen3-30B, not
only Qwen2.5. Offline supervision increases it on the models that most clearly
exhibit it (Qwen3-8B: $\mathrm{CUR}\ 0.13\!\rightarrow\!0.28$, Qwen3-30B:
$0.16\!\rightarrow\!0.52$). Only the sub-capable Qwen2.5-7B, which cannot solve the
rename at all, shows no gap, consistent with boundary~(i). The
newest-generation Qwen3.5-35B-A3B, an attention/state-space \emph{hybrid} model
released after our toolchain and served through a forward-compatibility shim, solves
the rename well ($\mathrm{RC@1}=0.67$, $n{=}48$) but never grounds it
($\mathrm{CUR}=1.0$, $\mathrm{ER}\!\approx\!0.29$): the correct-but-ungrounded gap is
not receding in this model generation, and the newer model has a larger shortcut
rate than the $30$B. Within that generation, scale improves
accuracy but not grounding: the smaller Qwen3.5-9B solves the rename far less often
($\mathrm{RC@1}=0.35$) but is equally ungrounded ($\mathrm{CUR}=1.0$), so
the larger model is more accurate but not more grounded. By contrast, closure
tracks capability: the $30$B converts its grounded exploration into a grounded
policy, whereas the $8$B and $14$B, which solve the rename rarely and shortcut it
near-deterministically, exhibit the gap but do not close it under matched GRPO,
consistent with the exploration condition in boundary~(iii). The
correct-but-ungrounded gap therefore appears across the tested generations of
capable data agents. Closing it in-loop remains capability-dependent.

\begin{table*}[t]
\centering
\small
\caption{Generational robustness (\texttt{ecom}). The gap appears on both Qwen2.5
and the recent Qwen3, at every size above the capability floor, and offline
supervision inflates it. Only Qwen2.5-7B (which cannot solve the rename) shows no
gap. In-loop closure is demonstrated at $30$B (Tables~\ref{tab:motivation:coevo3},
\ref{tab:motivation:xdomain}). Smaller capable models exhibit the gap but do not
close it (boundary~(iii)).}
\label{tab:motivation:gen}
\begin{adjustbox}{max width=0.95\linewidth}
\begin{tabular}{llccc}
\toprule
Family & Size & RC@1 & $\mathrm{CUR}$ base & $\mathrm{CUR}$ offline \\
\midrule
Qwen2.5 & 7B  & 0.10 & --- & --- \\
Qwen2.5 & 14B & 0.44 & 0.28 & 0.27 \\
Qwen3   & 8B  & 0.24 & 0.13 & 0.28 \\
Qwen3   & 30B & 0.29 & 0.16 & 0.52 \\
\bottomrule
\end{tabular}
\end{adjustbox}
\end{table*}

\section{Constraint and Plan Modalities: A Probe}
\label{sec:motivation:constraint}

The reframing of Sec.~\ref{sec:problem} makes \texttt{constraint} and
\texttt{plan} first-class modalities. We probe both with a constraint operator,
\texttt{unenforced\_unique}: the \texttt{order\_id} uniqueness invariant on
\texttt{stg\_orders} is declared but its guard test is dropped, so a single
duplicate order is admitted and silently fans out through the payment join,
inflating downstream aggregates (\texttt{fct\_orders} rows $52\!\to\!53$,
\texttt{dim\_customers} $\sum$\texttt{n\_orders} $52\!\to\!53$) with \emph{no
failing test}. The root cause is the unenforced \emph{constraint object}, not
the data, and the symptom is silent, squarely the no-test regime of
Sec.~\ref{sec:problem}. Two tools make the new modalities observable:
\texttt{inspect\_constraint} (lists declared constraints, their enforcement
status, and current violation counts) and \texttt{explain\_plan}. The instance
is well-posed: an oracle that inspects the constraint localizes it
($\mathrm{RC@1}{=}1$, $\mathrm{ER}{=}0.75$) while a null agent scores zero.

Table~\ref{tab:motivation:constraint} reports $10$ rollouts per backbone under
the \texttt{react\_ds} scaffold on \emph{three} projects: the synthetic
e-commerce ELT, the real-schema, synthetic-pipeline TPC-H \textsc{retails}
project, and the cross-modal
\textsc{warehouse} project, whose tracker carries an incident
(\texttt{DATA-452}) that explicitly names the inflated-rollup symptom. Three
patterns appear across all six cells. \emph{First, the new modalities are not
consulted when they would be informative}: \texttt{explain\_plan} is called in $0\%$ of
episodes, and the incident tracker in $0\%$, even on \textsc{warehouse}, where
the ticket names the symptom, reproducing the cross-modal neglect of
Sec.~\ref{sec:exp:crossmodal} in the constraint setting. \emph{Second,
agents see but cannot localize}: \texttt{gpt-4o-mini} opens the constraint in
\emph{every} episode on all three projects ($100\%$ \texttt{inspect\_constraint}),
reads ``\texttt{[NOT ENFORCED]; violating rows now: 1},'' and \emph{never}
localizes it ($\mathrm{RC@1}{=}0$ throughout), blaming the downstream symptom
tables instead. \texttt{gpt-5-mini} reaches high evidence recall
($\mathrm{ER}{=}0.76$--$0.93$) but localizes the constraint only $0.1$--$0.5$ of
the time, reading the fault as an ordinary duplicate-\emph{data} problem and
submitting the constrained staging model or the raw source table rather than the
unenforced invariant. The constraint fault is consistently among the
lowest-$\mathrm{RC@1}$ operators, the single lowest of all six on e-commerce.
\emph{Third, no run is a lucky guess} ($\mathrm{LG}{=}0$): when agents name the
constraint, they have grounded it. We report this as a three-project existence
result showing that the \texttt{constraint} and \texttt{plan} modalities are in
scope but underused. A multi-database constraint suite and a dedicated
plan-localization fault are left to the full benchmark.

\begin{table*}[t]
  \centering\small
  \caption{Constraint- and plan-modality localization across three projects
  (synthetic e-commerce, real-schema TPC-H \textsc{retails}, cross-modal
  \textsc{warehouse}) under \texttt{react\_ds}, $10$ rollouts/backbone. Agents reach
  high evidence recall (ER) but rarely localize the unenforced-uniqueness
  constraint. The query \emph{plan} and the incident tracker are never consulted
  ($0\%$), even on \textsc{warehouse} where the ticket names the cause. \texttt{c.insp}
  $=$ constraint-inspect rate, \texttt{incid} $=$ incident-read rate.}
  \label{tab:motivation:constraint}
  \begin{adjustbox}{max width=0.95\linewidth}
\begin{tabular}{llcccccccc}
    \toprule
    Project & Backbone & RC@1 & ER & EF$_1$ & LG & \texttt{c.insp} &
      \texttt{plan} & \texttt{incid} \\
    \midrule
    \multirow{2}{*}{e-commerce} & \texttt{gpt-4o-mini} & 0.00 & 0.75 & 0.48 & 0 & 100\% & 0\% & 0\% \\
    & \texttt{gpt-5-mini}  & 0.40 & 0.88 & 0.56 & 0 & 60\%  & 0\% & 0\% \\
    \midrule
    \multirow{2}{*}{\textsc{retails}} & \texttt{gpt-4o-mini} & 0.00 & 0.65 & 0.51 & 0 & 100\% & 0\% & 0\% \\
    & \texttt{gpt-5-mini}  & 0.10 & 0.93 & 0.58 & 0 & 70\%  & 0\% & 0\% \\
    \midrule
    \multirow{2}{*}{\textsc{warehouse}} & \texttt{gpt-4o-mini} & 0.00 & 0.44 & 0.45 & 0 & 100\% & 0\% & 0\% \\
    & \texttt{gpt-5-mini}  & 0.50 & 0.76 & 0.51 & 0 & 80\%  & 0\% & 0\% \\
    \bottomrule
  \end{tabular}
\end{adjustbox}
\end{table*}

\section{Robustness Checks}
\label{sec:motivation:robustness}

We evaluate the sensitivity of the main claims to sample size, the lucky-guess
threshold, reasoning-judge choice, prompt salience, and the reliability of the
mechanically derived gold.
All bootstrap intervals use $10^4$ resamples.

\paragraph{Confidence intervals: the grounding gap widens with capability.}
On a larger controlled run ($n\!\approx\!72$ per backbone), the lucky-guess rate
rises $0.04\,[0,0.10]\to0.16\,[0.07,0.24]\to0.25\,[0.15,0.35]$ across
\texttt{gpt-4o-mini}/\texttt{gpt-5-mini}/\texttt{gpt-5} ($95\%$ bootstrap CIs).
The paired bootstrap difference is $+0.21$ with $95\%$ CI $[+0.10,+0.32]$,
\emph{excluding zero}. The rise is partly within-task: it persists after
conditioning on the operator for $2$ of $6$ operators and is therefore not solely
a composition effect. The widening is most pronounced on controlled projects.
On the hardest real-schema cases the lucky-guess rate is near zero (agents fail
rather than shortcut), so we scope the claim to settings where outcome success is
attainable.

\paragraph{Threshold sensitivity.}
This effect is not an artifact of the $\mathrm{ER}<\tfrac12$ cutoff: sweeping the
threshold over $[0.1,0.9]$, the lucky-guess rate rises with capability at every
$\tau_g\ge0.4$. A full $20{,}026$-episode sweep (App.~\ref{app:validation},
Table~\ref{tab:val:tau}) confirms that \emph{which} operators are shortcuttable is
essentially independent of $\tau_g$ (per-operator rank stability
$\rho=0.843$).

\paragraph{Abstention.}
$\mathrm{RC@1}$ can conflate localization skill with willingness to answer. At an
adequate step budget ($35$ steps) abstention is low ($\le4\%$) and $\mathrm{RC@1}$
conditioned on submission matches unconditional $\mathrm{RC@1}$. The high
abstention seen at a tighter budget was a step-budget artifact, not an answering
policy.

\paragraph{Scripted-strategy floor.}
To separate task difficulty from agent strategy, we run non-LLM scripted
baselines (symptom-blame, lineage-first, incident-first, random walk) on every
task. All reach $\le0.11$ node-localization. The LLM results therefore cannot be
explained by these simple heuristics, although this comparison does not establish
the intrinsic difficulty of every task.

\paragraph{Reasoning-judge reliability.}
Grading the same trajectories with a second judge model, aggregate $\mathrm{RR}$
is stable (judge means $0.679$ vs.\ $0.687$) but per-episode agreement is only
fair (Cohen's $\kappa=0.33$). We therefore rest reasoning claims on aggregate
$\mathrm{RR}$ and avoid per-episode interpretation.

\paragraph{Is incident-neglect a salience artifact?}
A natural worry is that agents ignore the incident tracker only because a neutral
ReAct prompt under-advertises it. We test this with a scaffold that explicitly
advertises the tracker and requires upstream tracing before answering
(Table~\ref{tab:motivation:salience}). Salience has a substantial effect:
\texttt{gpt-5-mini} goes from $0\%$ to $100\%$ incident-read and $\mathrm{RC@1}$
$0.46\to0.71$, so the neglect is a property of \emph{default} prompting, not an
intrinsic limit. The smaller backbone benefits less ($0\%\to24\%$,
$\mathrm{RC@1}\,0.0\to0.14$): the capability gap in exploiting cross-modal evidence
persists even when the tool is advertised.

\begin{table*}[t]
  \centering\small
  \caption{Salience ablation on \textsc{warehouse}: advertising the incident
  tracker and requiring upstream tracing. The neglect is recoverable for the
  stronger backbone, but the weak one barely benefits.}
  \label{tab:motivation:salience}
  \begin{adjustbox}{max width=0.95\linewidth}
\begin{tabular}{llccc}
    \toprule
    Backbone & Scaffold & incident-read & RC@1 & ER \\
    \midrule
    \multirow{2}{*}{\texttt{gpt-4o-mini}} & \texttt{react\_ds} & 0\% & 0.00 & 0.49 \\
     & \texttt{+salient} & 24\% & 0.14 & 0.52 \\
    \midrule
    \multirow{2}{*}{\texttt{gpt-5-mini}} & \texttt{react\_ds} & 0\% & 0.46 & 0.73 \\
     & \texttt{+salient} & 100\% & 0.71 & 0.85 \\
    \bottomrule
  \end{tabular}
\end{adjustbox}
\end{table*}

\paragraph{Plan modality.}
On a fault whose signature is uniquely in the query plan, a dropped join
predicate producing a cartesian product, \texttt{explain\_plan} is still
consulted in only ${\sim}4\%$ of episodes. Agents localize via the failing test
instead. The plan modality is therefore available and occasionally diagnostic but
rarely used, matching the pattern observed for the incident modality.

\paragraph{Validity of the mechanical gold.}
Finally, we audit the mechanically derived gold, which uses no per-instance human
annotation. When we cross-check the
sqlglot-recovered lineage against an independent reference, model-level edge
extraction is exact (precision $=$ recall $=1.0$ over $147$ edges across $12$
synthetic and $9$ real-schema lifted pipelines), and $19/19$ gold paths are
well-formed (the injected cause lies on a connected lineage path from the
symptom). Lineage for real dbt projects is ingested from the dbt manifest. The
remaining extraction risk previously identified, column-level
provenance on arbitrary SQL, is now \emph{measured} rather than assumed: against
an independent differential DuckDB oracle over $13$ projects, the
column-level \texttt{derives} edges reach precision $1.000$, recall $0.944$,
$\mathrm{F}_1 0.971$ (App.~\ref{app:validation},
Table~\ref{tab:val:provenance}). This precision-favoring asymmetry is appropriate
for a gold-standard extractor.

\section{EG-Elo: Formal Definition and Interpretation}
\label{app:egelo}

\paragraph{Standard Elo.}
Elo represents each competitor $i$ by a scalar rating $R_i$. For a match between
$i$ and $j$, the standard logistic expected score and update are
\begin{equation}
  p_{i>j}=\frac{1}{1+10^{(R_j-R_i)/400}},\qquad
  R_i' = R_i + K(y_i-p_{i>j}),
  \label{eq:elo:standard}
\end{equation}
where $y_i\in\{0,1\}$ is the observed loss/win outcome and $K$ controls the
update magnitude. Equal ratings imply $p_{i>j}=0.5$. Consequently, 1500 is a
conventional initialization and parity point, not an absolute level of skill.

\paragraph{Grounded match outcome.}
We instantiate a match between a solver and a generated task. For rollout $e$,
the solver outcome is
\begin{equation}
  y_e^{(s)}=\mathbf{1}\!\left[
    \mathrm{RC@1}_e=1 \wedge \mathrm{ER}_e\ge\tau_g
  \right],\qquad
  y_e^{(g)}=1-y_e^{(s)},
  \label{eq:egelo:outcome}
\end{equation}
with grounding threshold $\tau_g=0.5$. Thus the solver wins only by producing a
grounded-correct diagnosis. A wrong diagnosis and a correct-but-ungrounded
diagnosis are both generator wins. The empirical solver win rate is exactly
$\mathrm{RC@1}^{g}$ and, when $\mathrm{RC@1}>0$, satisfies
\begin{equation}
  \mathrm{RC@1}^{g}
  =\mathrm{RC@1}\bigl(1-\mathrm{CUR}\bigr).
  \label{eq:egelo:currelation}
\end{equation}
CUR remains a conditional diagnostic among correct answers. EG-Elo instead rates
the complete grounded-win event over all episodes.

\paragraph{Glicko-2 update used by EG-Elo.}
We implement the paired update with Glicko-2 because it associates each rating
with a rating deviation and volatility and is therefore explicit about
uncertainty in small or non-stationary match sets~\cite{glickman2001dynamic}.
Each player has state $(r,\mathrm{RD},\sigma)$, initialized to
$(1500,350,0.06)$. We transform to the internal scale
\begin{equation*}
  \mu=\frac{r-1500}{173.7178},\qquad
  \phi=\frac{\mathrm{RD}}{173.7178},
\end{equation*}
and, for opponent $j$, compute
\begin{align*}
  g(\phi_j)&=\left(1+\frac{3\phi_j^2}{\pi^2}\right)^{-1/2},\\
  E_j&=\frac{1}{1+\exp[-g(\phi_j)(\mu-\mu_j)]}.
\end{align*}
For one rating period with outcomes $y_j$, the estimated variance and improvement
are
\begin{align*}
  v&=\left[\sum_j g(\phi_j)^2E_j(1-E_j)\right]^{-1},\\
  \Delta&=v\sum_j g(\phi_j)(y_j-E_j).
\end{align*}
Let $a=\log\sigma^2$. The updated volatility $\sigma'$ is obtained by solving
for $x$ in the standard Glicko-2 equation
\begin{equation*}
 f(x)=\frac{e^x(\Delta^2-\phi^2-v-e^x)}
 {2(\phi^2+v+e^x)^2}-\frac{x-a}{\tau_v^2}=0,
\end{equation*}
where our volatility constraint is $\tau_v=0.5$, and setting
$\sigma'=e^{x/2}$. The remaining update is
\begin{align*}
  \phi^* &= \sqrt{\phi^2+\sigma'^2}, \\
  \phi' &= \left[(\phi^*)^{-2}+v^{-1}\right]^{-1/2},\\
  \mu' &= \mu+(\phi')^2\sum_jg(\phi_j)(y_j-E_j), \\
  r' &= 173.7178\mu'+1500,\\
  \mathrm{RD}' &= 173.7178\phi'.
\end{align*}

\paragraph{Static leaderboard interpretation.}
For the hero leaderboard, each solver and task cell enters the rating period at
1500, and the solver is updated from its grounded-win outcomes against the common
frozen task pool. A solver EG-Elo below 1500 therefore means that its observed
grounded-win performance pulls it below parity against that reference pool. It
does not mean that an adaptive generator learned faster, because this calculation
contains no longitudinal opponent. Cross-model ratings are directly comparable
only when the task population, tools, prompt, budget, and rating prior are shared.

\paragraph{Dynamic self-play interpretation.}
In \textsc{CoSE}, solver and generator states persist across rounds and are updated
symmetrically from their pre-round snapshots. We report
\begin{equation}
  \Delta_t=P(g>s)
  =\frac{1}{1+\exp[-g(\phi_s)(\mu_g-\mu_s)]},
  \label{eq:egelo:selfplaygap}
\end{equation}
the expected generator win probability after round $t$. The parity boundary is
$R_s=R_g$ and $\Delta_t=0.5$. Hence $R_s<R_g$ (equivalently,
$\Delta_t>0.5$) indicates a generator-favored endpoint, whereas $R_s>R_g$
indicates a solver-favored endpoint. For frozen-weight evaluation this is a
relative competition result, not evidence of parameter learning. A claim about
which role improves faster would additionally require rating trajectories and a
fixed-plan replay control that separates solver change from generator difficulty
drift.

\section{Formal Specification: The Investigation Reward Machine}
\label{app:formal}

The main experiments grade path coverage against one mechanically derived
witness path $\pi^\star=\mathrm{path}_{\mathcal{G}}(s\to v^\star)$ and use the
threshold lucky guess of Eq.~\ref{eq:luckyguess}. The label \emph{grounded}
therefore denotes path coverage, not a claim that the answer causally depended on
every visited artifact. This single-witness surrogate
emphasizes recall: a solver may legitimately inspect decoys to rule them
out, so evidence precision can be lower even for sound investigations. For
open-ended exploration tasks that lack a mechanical witness path, agreement
verifiers assess consistency across repeated runs and checklist verifiers assess
satisfaction of rubric criteria. This small, optional partition is reported
separately as exploratory and is excluded from the frozen leaderboard, whose
root causes and evidence paths are mechanically derived. Uncertainty in these
verifiers therefore does not affect leaderboard scores. A fuller
automaton view treats valid investigations as an accepted language over tool
observations, with an evidence cut-set $E^\star$ and a reward machine that can
both monitor evaluation runs and shape training rewards. This appendix gives the
full formalization: the agent/specification split, the evidence cut-set, the
reward machine and its product with the environment POMDP, gold as an accepted
language, the monitor characterization of the metrics, the reward-shaping and
flywheel use, and an encode--decode (autoencoder) reading of the construction.

\paragraph{Agent vs.\ specification.}
We use two complementary formalisms. The
\emph{agent}'s decision problem is a partially observable Markov decision
process (POMDP): the fault $v^\star$ is hidden, observations arrive only through
tool calls, and the optimal next action depends on the entire history of
evidence gathered so far, the standard model for multi-turn tool-using
agents~\cite{agentpomdp}. We therefore do \emph{not} model the agent's policy as
finite-state. Doing so would fold the accumulated-evidence set into the state and
incur a $2^{|\mathcal{V}|}$ blow-up. Instead, the task \emph{specification},
grader, and reward are a single finite-state machine over the labeled
action--observation trace, a \emph{reward machine} in the sense of Toro Icarte
et al.~\cite{rewardmachines}, which supplies non-Markovian, regular-language
reward on top of the POMDP without constraining the policy.

\paragraph{Evidence cut-set and labeling.}
Let $D\subset\mathcal{V}$ be the set of \emph{decoys}: a priori plausible but
incorrect causes of the symptom $s$ (sibling views, alternative join keys,
correlated source tables). The \emph{evidence cut-set} $E^\star$ is the \emph{smallest} set of observations
that still suffices to tell the true cause apart from every decoy. Formally,
\begin{equation}
  \begin{aligned}
  E^\star &\;=\; \arg\min_{E\subseteq\mathcal{V}} |E|\\
  \text{s.t.}\quad &\mathrm{obs}(E)\text{ distinguishes }v^\star\text{ from every }d\in D,
  \end{aligned}
  \label{eq:cutset}
\end{equation}
equivalently a minimal separator between $v^\star$ and $D$ in the dependency
graph. $E^\star$ turns ``required evidence'' into a graph-theoretic object with a
minimality guarantee rather than a hand-drawn list. A labeling function
$L:(\mathcal{A}\times\mathcal{V})^{*}\rightarrow 2^{\mathcal{P}}$ maps each step
$(a_t,v_t)$ (a tool applied at a node, Sec.~\ref{sec:method:actions}) to
high-level propositions $\mathcal{P}=\{\,\texttt{seen}(e):e\in
E^\star\,\}\cup\{\texttt{read\_incident},\,\texttt{confirmed}(v^\star),\,
\texttt{answered}\}$.

\paragraph{The reward machine.}
The investigation automaton is a reward machine
$\mathcal{R}=\langle U,u_0,F,\delta_u,\delta_r\rangle$ over inputs
$2^{\mathcal{P}}$: states $U$ track which cut-set propositions have been
satisfied (and whether the incident was read), the transition function
$\delta_u:U\times 2^{\mathcal{P}}\rightarrow U$ advances on observed evidence,
and the accepting set $F$ requires that the cause be confirmed \emph{after} the
cut-set is covered, an ordering constraint expressible as a regular-language or
$\mathrm{LTL}_f$ property. Learning takes place on the product of the environment
POMDP and $\mathcal{R}$,
\begin{equation}
  \mathcal{P} \;=\; \mathrm{POMDP}(\mathcal{G}) \,\times\, \mathcal{R},
  \label{eq:product}
\end{equation}
the standard reward-machine construction~\cite{rewardmachines}: a state $(b,u)$
pairs the agent's belief over $\mathcal{G}$ with the automaton's progress $u$.

\paragraph{Gold as an accepted language.}
The gold is \emph{not} a single canonical path. It is the language
\begin{equation}
  \mathcal{L}(\mathcal{R}) \;=\;
  \{\, T \;:\; L(T)\ \text{drives}\ \mathcal{R}\ \text{to a state in}\ F \,\},
  \label{eq:goldlang}
\end{equation}
i.e.\ \emph{every} trace that covers the cut-set and confirms the cause in an
admissible order. A task may admit many valid investigations, making exact
matching to one annotated path unreliable~\cite{trace}. For curriculum and difficulty control
we still compute a minimal witness, the shortest accepting run
\begin{equation}
  \pi^\star \;=\; \arg\min_{T\in\mathcal{L}(\mathcal{R})}\ \mathrm{len}(T),
  \label{eq:goldwalk}
\end{equation}
so the \emph{specification} grades by membership in $\mathcal{L}(\mathcal{R})$
rather than by distance to any single path. \textbf{What we compute in this
paper} is a tractable instantiation of that
specification: we grade against the single accepted witness $\pi^\star$,
equivalently the lineage path $\mathrm{path}_{\mathcal{G}}(s\to v^\star)$, by
path coverage, and leave full language-membership grading over
$\mathcal{L}(\mathcal{R})$ to future work. Generators~A and B become the cases in
which the run is seeded at a \texttt{test} node (planted symptom) or an
\texttt{incident} node (mined incident) of the same construction.

\paragraph{Grading as runtime verification.}
An agent run, labeled by $L$, is a string over $2^{\mathcal{P}}$. Grading is then
a form of runtime verification: does the monitor $\mathcal{R}$ accept it? Related
work checks LLM-agent trajectories against LTL-derived B\"uchi
monitors~\cite{ltlmonitor} or uses product-automaton
reachability~\cite{ltlfmas}. Our metrics become properties of this run:
$\mathrm{RC}@1$ is acceptance at a correct cause state, evidence recall
$\mathrm{ER}$ is the fraction of cut-set propositions ever satisfied (automaton
progress), and the \emph{lucky guess} of Eq.~\ref{eq:luckyguess} acquires an
exact monitor characterization,
\begin{equation}
\begin{aligned}
  \mathrm{LG}(T)=1 \iff{}&
  \big(\texttt{answered}\wedge\texttt{confirmed}(v^\star)\big)\ \text{fires}\\
  &\text{while}\ \mathcal{R}\notin U_{\mathrm{cov}},
\end{aligned}
  \label{eq:lg_monitor}
\end{equation}
where $U_{\mathrm{cov}}\subseteq U$ are the states reachable only after the
cut-set is covered. A lucky guess is thus a trace that asserts the right answer
\emph{before the automaton has entered an evidence-covered state}, a regular
property of the trace. These are the \emph{specification-level} forms of the
metrics. The numbers reported in this paper use their operational instantiation
(Sec.~\ref{sec:method:verify}): $\mathrm{RC}@1$ by exact node match, $\mathrm{ER}$
as coverage of $\pi^\star$, and the \emph{threshold} lucky guess of
Eq.~\ref{eq:luckyguess}. Eq.~\ref{eq:lg_monitor} is the equivalent automaton-state
view, which the threshold instantiates when the covered-state boundary
$U_{\mathrm{cov}}$ is set at half cut-set coverage.

\paragraph{One object, three uses, and the flywheel.}
Because $\mathcal{R}$ also carries a state-reward function $\delta_r$, the same
automaton that defines gold and grades runs can shape reinforcement reward.
$\delta_r$ rewards each newly covered cut-set element and then the confirmed
cause, instantiating Eq.~\ref{eq:reward}. This yields a dense, non-Markovian signal
and supports automated reward shaping and counterfactual experience
replay~\cite{rewardmachines}. Thus, at
evaluation $\mathcal{R}$ is a monitor. In training it is a reward machine, and
the trajectory flywheel (Sec.~\ref{sec:method:flywheel}) admits to the SFT pool
the grounded-correct runs, the operational $\mathcal{L}(\mathcal{R})$ surrogate
(root cause $\wedge\ \mathrm{ER}\ge\tau_g$), grounded by construction and
excluding lucky guesses (Eq.~\ref{eq:luckyguess}).

\paragraph{An encode--decode view.}
The construction can be viewed as a lossy, task-specific \emph{autoencoder}
of the data environment (Fig.~\ref{fig:autoencoder}). The \textbf{encoder} is the
mechanical compilation of Sec.~\ref{sec:method:goldtraj}: it maps a full database
$\mathcal{G}$, rows, columns, and SQL, to the compact finite-state code
$\mathcal{R}$, discarding everything that does not bear on the diagnosis and
keeping only the causal skeleton (the cut-set $E^\star$, the path order, and the
\textsc{confirm} checks). The \textbf{bottleneck} is $E^\star$ itself: by
Eq.~\ref{eq:cutset} it is the smallest code that incurs no diagnostic
``reconstruction error,'' i.e.\ one that still separates $v^\star$ from every
decoy. The \textbf{decoders} are the three uses of $\mathcal{R}$: instantiation
expands the code into a concrete task (with gold
$\mathcal{C}^\star=\mathcal{L}(\mathcal{R})$), monitoring decodes an agent run
into a verdict, and $\delta_r$ decodes it into a learning signal. Under this
interpretation, a grounded answer must be decoded
\emph{through} the bottleneck, and a lucky guess (Eq.~\ref{eq:lg_monitor}) is
exactly a reconstruction of the right answer that never passed through the code.
This analogy has two limits. The encoder is \emph{structural and deterministic},
not a network trained with a reconstruction loss. The encoding is also
\emph{lossy by design}: it reconstructs the diagnosis rather than the full
environment. Its formal counterpart in our POMDP is
reward-relevant state abstraction~\cite{stateabstraction} and bisimulation /
model minimization~\cite{bisimulation,deepmdp}, which formalize ``compress the
environment while preserving decision-relevant structure.'' The autoencoder is
an interpretation of this construction rather than its formal definition.

\begin{figure*}[t]
  \centering
  \begin{tikzpicture}[
    node distance=12mm and 18mm,
    box/.style={draw, rounded corners, align=center, inner sep=4pt,
                font=\small, minimum height=9mm},
    code/.style={draw, rounded corners, align=center, inner sep=4pt,
                 font=\small, fill=black!5, minimum height=12mm},
    arr/.style={-{Stealth[length=2mm]}, thick},
    lbl/.style={font=\footnotesize\itshape, align=center}
  ]
    \node[box, text width=24mm] (env)
      {Real data\\environment $\mathcal{G}$\\
       {\scriptsize schema $\cdot$ data $\cdot$ SQL $\cdot$ constraints}};
    \node[code, right=of env, text width=27mm] (rm)
      {Reward machine $\mathcal{R}$\\(finite-state \emph{code})\\
       {\scriptsize bottleneck $E^\star$ (Eq.~\ref{eq:cutset})}};
    \node[box, right=24mm of rm, yshift=13mm, text width=28mm] (task)
      {Task $+$ gold\\$\mathcal{C}^\star=\mathcal{L}(\mathcal{R})$};
    \node[box, right=24mm of rm, text width=28mm] (grade)
      {Verdict\\{\scriptsize runtime verification}};
    \node[box, right=24mm of rm, yshift=-13mm, text width=28mm] (rew)
      {Reward\\{\scriptsize RL / flywheel}};
    \draw[arr] (env) -- node[lbl, above]{encode}
      node[lbl, below]{\scriptsize compile (\S\ref{sec:method:goldtraj})} (rm);
    \draw[arr] (rm.east) to[out=35,in=180] node[lbl, above]{decode} (task.west);
    \draw[arr] (rm.east) -- node[lbl, above]{decode} (grade.west);
    \draw[arr] (rm.east) to[out=-35,in=180] node[lbl, below]{decode} (rew.west);
  \end{tikzpicture}
  \caption{The construction as a lossy, task-specific autoencoder. A mechanical
  \emph{encoder} compresses the data environment into the finite-state code
  $\mathcal{R}$ whose bottleneck is the evidence cut-set $E^\star$
  (Eq.~\ref{eq:cutset}). Three \emph{decoders} expand the code into a task (with
  gold $\mathcal{C}^\star$), a grading verdict, and a learning reward. A grounded
  answer is decoded through the bottleneck. A lucky guess bypasses it.}
  \label{fig:autoencoder}
  \Description{The formal construction maps a database environment into a reward machine, then uses that representation to construct a task, grade an investigation, and supply a training reward.}
\end{figure*}
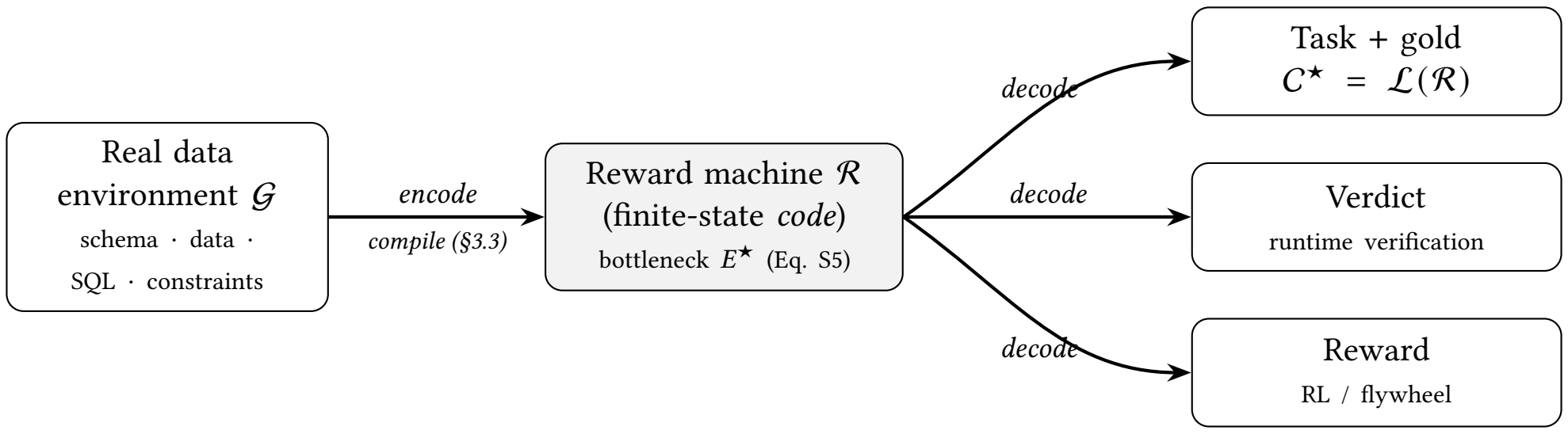

\paragraph{Derived, not authored.}
Prior automata- and LTL-based specifications for agents are written by
hand~\cite{ltlmonitor,ltlfmas}. Here $\mathcal{R}$ is \emph{synthesized
mechanically} from the provenance graph: its propositions come from the cut-set
$E^\star$ (a graph separator between $v^\star$ and its decoys), its ordering
constraints from the dependency edges, and its shaped reward from the
distance-to-cause $d_{\mathcal{G}}$. Compiling the reward machine from data
lineage avoids per-instance manual specification of its states and transitions
(Sec.~\ref{sec:method:verify}).

\paragraph{Controllable difficulty.}
This makes difficulty graph-theoretic and explicit: the walk length (dependency
distance $d_{\mathcal{G}}$), the number of \emph{modality switches} an accepting
run must make (a
$\texttt{data}\!\to\!\texttt{test}\!\to\!\texttt{query}\!\to\!\texttt{data}$
chain is harder than $\texttt{data}\!\to\!\texttt{data}$), and the decoy count
$|D|$, which sets the size of the separating cut-set $E^\star$. Sampling
$(\mathcal{G},v^\star,\tau_f)$ and compiling $\mathcal{R}$ yields a fully graded,
difficulty-indexed instance, enabling a curriculum at scale.

\section{Benchmark Validity and Robustness Checks}
\label{app:validation}

We collect validation experiments that evaluate the main claims at larger scale or against an
independent oracle: an eight-model open-weight leaderboard
(Sec.~\ref{app:val:leaderboard}), the cross-backbone stability of
grounding difficulty (Sec.~\ref{app:val:difficulty}), a full lucky-guess
threshold sweep (Sec.~\ref{app:val:tau}), a decomposition of the grounding
gap into localization vs.\ classification (Sec.~\ref{app:val:decomp}),
corpus diversity (Sec.~\ref{app:val:diversity}), the gameability of
outcome-\textsc{Fix} under blind baselines (Sec.~\ref{app:val:fix}), and a
column-level provenance audit against a differential engine oracle
(Sec.~\ref{app:val:provenance}), and a full-corpus evaluation of the co-evolved
checkpoint, which makes the trained \emph{Ours} row directly comparable to the
open-weight rows (Sec.~\ref{app:n9}).

\subsection{Open-weight leaderboard (eight models)}
\label{app:val:leaderboard}
We evaluate eight open-weight models spanning $4$B--$35$B, including dense and
MoE architectures. Each model is evaluated on the same $128$ frozen episodes,
for $1024$ model--episode evaluations in aggregate, and ranked by grounded
accuracy $\mathrm{RC@1}^g$ (the
open-weight rows of the consolidated leaderboard, Table~\ref{tab:hero}). The
ordering exhibits the same \emph{grounding gap} in the open-weight regime: the evidence-recall-heavy \texttt{qwen3.5}
pair leads, whereas \texttt{qwen2.5-14b}, competitive on outcome accuracy,
carries the \emph{largest inflation} between naive and grounded scoring
($0.070 = \text{naive}-\text{grounded}$), i.e.\ its apparent accuracy is most
inflated by ungrounded correct answers. Grounded ranking on a diverse
open-weight ladder therefore exhibits the same gap. (The eight-model
$\mathrm{RC@1}^g$ figures are the open-weight rows of
Table~\ref{tab:hero}. The standalone table is folded into that consolidated
leaderboard.)

\subsection{Full-corpus evaluation of the trained checkpoint}
\label{app:n9}
The \emph{Ours\,(frozen)} block of Table~\ref{tab:hero} evaluates the co-evolved
checkpoint on the \emph{same} frozen leaderboard corpus and tool parser as the
open-weight rows, so the base$\to$trained delta is directly comparable rather than
cross-population. Because the tasks are identical, the comparison is \emph{paired}
(Table~\ref{tab:n9}). Outcome accuracy is unchanged ($\mathrm{RC@1}=0.039$, $5/128$
correct for both), but among those correct answers the co-evolved model is grounded
more often ($4/5$ versus $2/5$): $\mathrm{RC@1}^g$ rises $0.016\!\to\!0.031$, the
conditional ungrounded rate falls from $\mathrm{CUR}=0.60$ to $0.20$, and lucky-guess
mass halves ($0.023\!\to\!0.008$). The direction matches the larger matched-batch
closure (\emph{Ours\,(batch)}). The two evaluations therefore show the same
directional pattern at different base strengths.

\paragraph{Significance and power.} For the $30$B base this effect is
\emph{directional but not statistically significant}, and we report it as such. The
higher-headroom rerun below replicates the direction with more power.
The base is a near-floor solver: only $5/128$ episodes are correct, so the
grounded-accuracy change rests on two episodes converting from lucky to grounded. A
paired McNemar exact test on grounded-correct outcomes gives $p=0.63$
(outcome-correct $p=1.00$), and the $95\%$ bootstrap intervals on $\mathrm{RC@1}^g$
overlap ($[0.00,0.04]$ versus $[0.01,0.06]$). The best-powered signal, a Wilcoxon
signed-rank test on \emph{per-episode} evidence recall over all paired episodes,
is likewise only suggestive ($\Delta\overline{\mathrm{ER}}=+0.023$, $p=0.12$). The
small number of correct base-model episodes limits statistical power, so we
rerun the identical co-evolution recipe on a higher-headroom base,
\texttt{Qwen3.6-27B} (frozen $\mathrm{RC@1}^g=0.508$, roughly two orders of
magnitude more grounded episodes, bottom block of Table~\ref{tab:n9}). At this
scale the same directional pattern holds and is no longer floor-limited: over the
$127$ paired episodes, $\mathrm{RC@1}^g$ rises $0.508\!\to\!0.528$ and
$\mathrm{RC@1}$ $0.586\!\to\!0.606$ (both $+0.020$, grounded-correct
$65/128\!\to\!67/127$), while $\mathrm{CUR}$ does not increase
($0.133\!\to\!0.130$) and no lucky-guess mass is added ($\mathrm{LG}=0.078$ in
both). The closure is smaller in magnitude than the near-floor $30$B case but lands
on the same grounded metric without trading it for ungrounded correctness,
indicating that the grounding reward remains an informative, non-degenerate
training signal on a state-of-the-art open model rather than an artifact of a
near-floor base.

\begin{table}[t]
\centering\small
\caption{\textbf{Trained checkpoint on the frozen corpus (N9), paired over the
identical episodes the open-weight rows used.} For the near-floor \texttt{qwen3-30b}
base ($128$ episodes) outcome accuracy is unchanged and grounding improves. On the
higher-headroom \texttt{Qwen3.6-27B} base ($127$ paired episodes) both outcome and
grounded accuracy improve by $+0.020$ with $\mathrm{CUR}$ non-increasing. The shift
is directional. See the text for the small-sample significance analysis.}
\label{tab:n9}
\begin{adjustbox}{max width=0.95\linewidth}
\begin{tabular}{lcccccc}
\toprule
 & $\mathrm{RC@1}$ & $\mathrm{RC@1}^g\!\uparrow$ & $\mathrm{CUR}\!\downarrow$ & $\mathrm{ER}\!\uparrow$ & $\mathrm{LG}\!\downarrow$ & corr./gr. \\
\midrule
\texttt{qwen3-30b} base   & 0.039 & 0.016 & 0.60 & 0.682 & 0.023 & $5/2$ \\
\;+ co-evolution          & 0.039 & \textbf{0.031} & \textbf{0.20} & \textbf{0.706} & \textbf{0.008} & $5/4$ \\
\midrule
\texttt{Qwen3.6-27B} base & 0.586 & 0.508 & 0.133 & 0.832 & 0.078 & $75/65$ \\
\;+ co-evolution          & \textbf{0.606} & \textbf{0.528} & \textbf{0.130} & 0.820 & 0.078 & $77/67$ \\
\bottomrule
\end{tabular}
\end{adjustbox}
\end{table}

\subsection{Grounding difficulty is a model-robust property}
\label{app:val:difficulty}
Is grounding difficulty a stable, curriculum-usable property of a task, or an
artifact of one backbone? Across eight backbones we measure the rank-stability of
per-operator \emph{grounding difficulty} ($\mathrm{GD}=\mathrm{CUR}$) and of
outcome difficulty ($\mathrm{OD}$) as the mean pairwise Spearman correlation of
operator rankings between backbones (Table~\ref{tab:val:difficulty}). Grounding
difficulty has higher rank stability ($\rho=0.839$) than outcome difficulty
($\rho=0.286$): within these models, the operators that are hard to \emph{ground}
are more consistent than those that are hard to \emph{solve}.
Per operator, \texttt{column\_rename} is the most shortcuttable
($\mathrm{GD}=0.73$), followed by \texttt{null\_injection} ($0.333$) and
\texttt{incorrect\_join\_key} ($0.057$). The rest are near zero. Grounding
These results support using grounding difficulty as a curriculum signal across
the tested model set.

\begin{table}[t]
\centering\small
\caption{Cross-backbone ($n{=}8$) rank-stability of difficulty, mean pairwise
Spearman $\rho$ of per-operator rankings. Grounding difficulty
($\mathrm{GD}=\mathrm{CUR}$) is more stable than outcome difficulty
($\mathrm{OD}$). Right: per-operator $\mathrm{GD}$.}
\label{tab:val:difficulty}
\begin{adjustbox}{max width=0.95\linewidth}
\begin{tabular}{lc}
\toprule
Quantity & Spearman $\rho$ \\
\midrule
Grounding difficulty ($\mathrm{GD}=\mathrm{CUR}$) & \textbf{0.839} \\
Outcome difficulty ($\mathrm{OD}$)                & 0.286 \\
\midrule
\multicolumn{2}{l}{\emph{Per-operator} $\mathrm{GD}$} \\
\quad \texttt{column\_rename}      & 0.730 \\
\quad \texttt{null\_injection}     & 0.333 \\
\quad \texttt{incorrect\_join\_key}& 0.057 \\
\quad others                       & $\approx 0$ \\
\bottomrule
\end{tabular}
\end{adjustbox}
\end{table}

\subsection{E1: lucky-guess threshold sweep}
\label{app:val:tau}
The lucky-guess threshold $\tau_g$ (evidence-recall cutoff below which a correct
answer is scored a lucky guess) is a free parameter. The paper uses $\tau_g=0.5$.
Over $20{,}026$ episodes we sweep $\tau_g\in[0.1,0.9]$ (Table~\ref{tab:val:tau}).
While the absolute rates move with $\tau_g$ (by construction), \emph{which}
operators are shortcuttable does not: the per-operator grounding-difficulty
ranking is essentially invariant (mean pairwise Spearman $\rho=0.843$). A
lucky-guess mass persists for thresholds from $0.2$ through $0.9$, and the
operator ordering is stable across the sweep. The reported pattern is therefore
not specific to the default $\tau_g=0.5$.

\begin{table}[t]
\centering\small
\caption{Lucky-guess threshold sweep over $20{,}026$ episodes. Per-operator
grounding-difficulty rank stability across $\tau_g$ is $\rho=0.843$. Paper default
is $\tau_g=0.5$ ($\mathrm{CUR}=0.411$).}
\label{tab:val:tau}
\begin{adjustbox}{max width=0.95\linewidth}
\begin{tabular}{lccc}
\toprule
$\tau_g$ & $\mathrm{RC@1}^g$ & LG rate & $\mathrm{CUR}$ \\
\midrule
0.1 & 0.592 & 0.000 & 0.000 \\
0.2 & 0.589 & 0.003 & 0.005 \\
0.3 & 0.410 & 0.182 & 0.308 \\
0.4 & 0.400 & 0.192 & 0.325 \\
0.5 & 0.349 & 0.243 & \textbf{0.411} \\
0.6 & 0.180 & 0.412 & 0.697 \\
0.7 & 0.177 & 0.415 & 0.701 \\
0.8 & 0.127 & 0.465 & 0.786 \\
0.9 & 0.123 & 0.469 & 0.792 \\
\bottomrule
\end{tabular}
\end{adjustbox}
\end{table}

\subsection{E2: the grounding gap is a localization failure}
\label{app:val:decomp}
$\mathrm{RC@1}$ requires \emph{both} the gold fault-type (operator) and the gold
node. Decomposing it (Table~\ref{tab:val:decomp}) shows the two failure modes are
very unequal: type-hit is high everywhere (\texttt{incorrect\_join\_key}
$0.886$, \texttt{column\_rename} $0.944$, \texttt{null\_injection} $0.931$), but
node-hit collapses on the shortcut-prone operators (\texttt{null\_injection}
$0.058$, \texttt{duplicate\_pk} $0.129$), and the right-type/wrong-node rate
reaches $0.878$ for \texttt{null\_injection}. Models name the right \emph{kind}
of fault but blame the wrong \emph{object}. In this decomposition, localization
is the dominant source of the grounding gap, which outcome-accuracy and
fault-type metrics do not expose.

\begin{table}[t]
\centering\small
\caption{$\mathrm{RC@1}$ decomposition pooled by operator. High type-hit with
lower node-hit (and high right-type/wrong-node) indicates that localization is
the dominant failure mode.}
\label{tab:val:decomp}
\begin{adjustbox}{max width=0.95\linewidth}
\begin{tabular}{lcccc}
\toprule
Operator & $\mathrm{RC@1}$ & type-hit & node-hit & r.t./w.n. \\
\midrule
\texttt{incorrect\_join\_key} & 0.715 & 0.886 & 0.807 & 0.171 \\
\texttt{column\_rename}       & 0.712 & 0.944 & 0.755 & 0.231 \\
\texttt{duplicate\_pk}        & 0.076 & 0.590 & 0.129 & 0.514 \\
\texttt{null\_injection}      & 0.054 & 0.931 & 0.058 & 0.878 \\
\bottomrule
\end{tabular}
\end{adjustbox}
\end{table}

\subsection{E3: corpus diversity}
\label{app:val:diversity}
The admitted corpus is $448/544$ instances (yield $0.824$) spanning the $14$
frozen core environments (the \textsc{DataTrace-Eval-v1} subset of the
$83$-environment suite, Table~\ref{tab:datatrace:profile}) $\times$ $7$ base
operators, populating $54$ distinct operator$\times$environment cells
(Table~\ref{tab:val:diversity}). Structural
difficulty varies widely: gold-path length $2$--$7$ (mean $5.12$), modality span
$1$--$4$ (mean $3.09$), lineage distance $1$--$5$, and $0$--$2$ decoys per
instance. Diversity is carried by the product of environments, operators, and
difficulty/decoy/modality settings, not by raw instance count.

\begin{table}[t]
\centering\small
\caption{Corpus diversity. $448/544$ admitted (yield $0.824$), $14$ environments
$\times$ $7$ operators $= 54$ operator$\times$environment cells.}
\label{tab:val:diversity}
\begin{adjustbox}{max width=0.95\linewidth}
\begin{tabular}{lcccc}
\toprule
Axis & min & mean & max & $n$ \\
\midrule
Gold-path length   & 2 & 5.12 & 7 & 448 \\
Modality span      & 1 & 3.09 & 4 & 448 \\
Lineage distance $d_{\mathcal{G}}$ & 1 & 3.00 & 5 & 448 \\
\# decoys          & 0 & 1.00 & 2 & 448 \\
\bottomrule
\end{tabular}
\end{adjustbox}
\end{table}

\subsection{E4: outcome-\textsc{Fix} gameability, and why grounded metrics
matter}
\label{app:val:fix}
We test whether an outcome-\textsc{Fix} check (does a repair recover the
golden output?) can be satisfied by blind ``self-healing'' patches that never
diagnose the fault. Over the full native$+$BIRD corpus
($64$ tasks: $53$ fault-visible to the coarse rows$+$sums fingerprint, $11$
invisible), a diagnosis-free blind NULL/dedup patch self-heals
$\mathbf{0.509}$ of the fault-visible tasks
(Table~\ref{tab:val:fix}), concentrated in \texttt{null\_injection} and
\texttt{incorrect\_join\_key} via \texttt{COALESCE} ($0.769$, $0.636$) and
\texttt{duplicate\_pk} via dedup ($0.769$). Thus, outcome-\textsc{Fix} alone is
sensitive to diagnosis-free repairs. The same blind patches
\emph{localize} only $0.057$ (they patch the symptom rather than the cause),
whereas a single-node oracle revert fixes $0.943$. This is precisely why
\method\ ranks on \emph{grounded} evidence metrics ($\mathrm{RC@1}^g$,
$\mathrm{ER}$, $\mathrm{LG}$), which score these blind baselines $\approx 0$,
rather than on outcome-\textsc{Fix}. Two caveats are that (a) $11/64$ faults are
invisible to the coarse rows$+$sums fingerprint (a fingerprint-coverage
limitation, not a self-heal), and (b) \texttt{unenforced\_unique} is not fixable
by re-enforcing the dropped guard alone, the admitted duplicate row must also be
removed, so even the oracle revert is $0.943$ rather than $1.0$.

\begin{table}[t]
\centering\small
\caption{Non-LLM fix baselines on the golden-fingerprint \texttt{validate\_fix}
check, full corpus ($53$ fault-visible tasks). A blind diagnosis-free patch
self-heals $0.509$ by \emph{outcome} but localizes only $0.057$. Grounded metrics
($\mathrm{RC@1}^g$/$\mathrm{ER}$/$\mathrm{LG}$) score all blind baselines
$\approx 0$.}
\label{tab:val:fix}
\begin{adjustbox}{max width=0.95\linewidth}
\begin{tabular}{lcc}
\toprule
Baseline & \textsc{Fix} rate & localizes (node-RC) \\
\midrule
\texttt{blind\_dropna}   & 0.000 & 0.057 \\
\texttt{blind\_distinct} & 0.189 & 0.057 \\
\texttt{blind\_coalesce} & 0.321 & 0.057 \\
\texttt{blind\_best}     & \textbf{0.509} & 0.057 \\
\texttt{oracle\_revert}  & 0.943 & --- \\
\bottomrule
\end{tabular}
\end{adjustbox}
\end{table}

\subsection{E5: column-level provenance audit against an engine oracle}
\label{app:val:provenance}
The mechanical-gold construction depends on the fidelity of the sqlglot column
lineage (\texttt{derives} edges). We validate it against an \emph{independent}
differential DuckDB oracle: perturb one source column, re-run the model's real
SQL, and observe which output columns move (engine truth, native to CTEs, window
functions, and nested subqueries). Over $13$ projects ($3$ native $+$ $10$
warehouse-grade BIRD), the value-derivation edges reach precision
$\mathbf{1.000}$, recall $\mathbf{0.944}$, $\mathrm{F}_1 \mathbf{0.971}$ ($904$
TP, $0$ FP, $54$ FN). Precision is $1.000$ in \emph{every} project and every
construct (Table~\ref{tab:val:provenance}). Within this sample, the audit finds
no fabricated dependencies and misses a small fraction of value-derivation edges.
Separately, $225$ key-influence edges (join/group keys moving aggregate measures)
are bucketed as a documented limitation, $0$ are captured by static
value-lineage by design (they are carried on the model-level lineage edge, not
the column \texttt{derives} edge). $28$ perturbations were inconclusive (engine
type error, skipped). These results quantify the fidelity and known coverage
limitation of the mechanically derived provenance used in this study.

\begin{table}[t]
\centering\small
\caption{Column-level provenance audit: sqlglot \texttt{derives} edges vs.\ a
differential DuckDB oracle over $13$ projects. Value-derivation precision
$1.000$, recall $0.944$, $\mathrm{F}_1 0.971$ ($904$ TP, $0$ FP, $54$ FN).
Key-influence edges ($225$) and inconclusive perturbations ($28$) are bucketed
separately.}
\label{tab:val:provenance}
\begin{adjustbox}{max width=0.95\linewidth}
\begin{tabular}{lcccc}
\toprule
Construct & edges & precision & recall & $\mathrm{F}_1$ \\
\midrule
simple    & 877 & 1.000 & 0.938 & 0.968 \\
join      & 56  & 1.000 & 1.000 & 1.000 \\
aggregate & 25  & 1.000 & 1.000 & 1.000 \\
\midrule
\textbf{overall} & 904 & \textbf{1.000} & \textbf{0.944} & \textbf{0.971} \\
\bottomrule
\end{tabular}
\end{adjustbox}
\end{table}

\subsection{Pilot backbone and scaffold tables (folded from the body)}
\label{app:pilottables}
These give the full-column detail behind the proprietary tier of the consolidated
leaderboard (Table~\ref{tab:hero}): the pilot backbone breakdown
(Table~\ref{tab:motivation:main}) and the GPT-5-family scaffold ablation
(Table~\ref{tab:motivation:scaffold}). Their
$\mathrm{RC@1}$/$\mathrm{RC@1}^g$/ER/LG summary appears in the evidence-grounded leaderboard. The
outcome/cost columns (Fix, EP, EF$_1$, Cost) not carried there are retained here.

\begin{table}[t]
\centering\small
\caption{Pilot main results ($M{=}180$ episodes), pooled over scaffolds and
projects, by backbone. Outcome metrics (RC@1, Fix) vs.\ evidence metrics
(ER, EP, EF$_1$) and the lucky-guess rate LG. $\mathrm{RC@1}^{g}$ is the
evidence-gated accuracy (outcome correct \emph{and} $\mathrm{ER}\ge 0.5$).
The gap $\mathrm{RC@1}-\mathrm{RC@1}^{g}$ widens with capability. Summarized in the
proprietary tier ($\dagger$) of Table~\ref{tab:hero}.}
\label{tab:motivation:main}
\begin{adjustbox}{max width=0.95\linewidth}
\begin{tabular}{lcccccccc}
\toprule
Backbone & RC@1 & $\mathrm{RC@1}^{g}$ & Fix & ER & EP & EF$_1$ & LG$\downarrow$ & Cost \\
\midrule
\texttt{gpt-4o-mini} & 0.18 & 0.13 & 0.10 & 0.65 & 0.51 & 0.52 & \textbf{0.05} & \textbf{5.1k} \\
\texttt{gpt-5-mini}  & 0.50 & 0.33 & 0.37 & \textbf{0.81} & 0.46 & 0.55 & 0.17 & 9.5k \\
\texttt{gpt-5}       & \textbf{0.82} & \textbf{0.55} & \textbf{0.63} & 0.72 & 0.47 & 0.53 & 0.27 & 18.0k \\
\bottomrule
\end{tabular}
\end{adjustbox}
\end{table}

\begin{table}[t]
\centering\small
\caption{Scaffold ablation, GPT-5 family pooled ($n{=}40$ each). RC@1 narrows
across scaffolds but evidence precision and cost separate them. The lineage-less
Spider-style scaffold pays the most to reach evidence. Summarized as the scaffold
sub-rows ($\ddagger$) under \texttt{gpt-5} in Table~\ref{tab:hero}.}
\label{tab:motivation:scaffold}
\begin{adjustbox}{max width=0.95\linewidth}
\begin{tabular}{lcccccc}
\toprule
Scaffold & RC@1 & ER & EP & EF$_1$ & LG$\downarrow$ & Cost \\
\midrule
ReAct-DS            & 0.60 & 0.76 & \textbf{0.50} & 0.56 & 0.23 & 13.9k \\
SWE-Agent-style     & \textbf{0.73} & 0.70 & 0.43 & 0.50 & 0.23 & \textbf{6.1k} \\
Spider-Agent-style  & 0.65 & \textbf{0.83} & 0.47 & \textbf{0.57} & \textbf{0.20} & 21.1k \\
\bottomrule
\end{tabular}
\end{adjustbox}
\end{table}

\section{Extended Related Work}
\label{app:related}

\method lies at the intersection of SQL error analysis, data-pipeline and agent
benchmarks, system and database root-cause analysis, data provenance, and
process-level supervision. We position prior work along three axes: the
\emph{unit} of analysis, the \emph{context} available to the model, and
\emph{what is graded}. Sec.~\ref{sec:method} targets multi-object database
diagnosis in which both the final root cause and the inspected evidence are
evaluated mechanically. Table~\ref{tab:related} summarizes the comparison.

\subsection{Single-query SQL bug detection and repair}

The work nearest to ours in topic is \emph{NL2SQL-Bugs}~\cite{nl2sqlbugs}, a
benchmark of $2{,}018$ BIRD-derived instances for detecting and \emph{typing}
semantic errors in generated SQL. Given a triple $(\mathrm{NL},\mathrm{DB},
\mathrm{SQL})$ it asks a model to decide whether the SQL is semantically
correct and, if not, to classify the error into one of $31$ subcategories
across a $9$-class taxonomy. Its localization is \emph{categorical}
(which \emph{kind} of bug), not positional: there are no clause spans, AST
references, or pipelines, and the bug and its observation occur in the same
query string. The model receives only static context
(schema, NL, an evidence hint, the SQL) and emits a single forward-pass
verdict, with no tools, no execution, and no evidence path. Their headline empirical
result is directly relevant to us: models detect that a query is wrong with
$\sim\!75\%$ accuracy but are markedly less reliable at pinpointing the
specific cause: they know \emph{something} is wrong but not \emph{why}.
\method lifts exactly this gap from the single-query level to the
\emph{system} level: the cause is upstream of the symptom, the evidence is
spread across constraints, query plans, logs, data, tests, and incident
reports, and the agent must \emph{investigate} rather than classify. Whereas
NL2SQL-Bugs grades the bug label, we grade the root-cause \emph{object} and
the evidence path to it, and we
explicitly flag unsupported correct answers, or lucky guesses (Eq.~\ref{eq:luckyguess}), a quantity that has
no analogue when there is no separate evidence path to recall.

\emph{SQLGovernor}~\cite{sqlgovernor} is a complementary method paper: an
LLM-powered toolkit that post-processes a single query through four
functions, syntax correction, rewriting, modification, and equivalence
verification, backed by a self-learning knowledge base mined from DBMS
execution logs. It shares our use of logs and schema as evidence, but its unit of
analysis is again a single query, recursively decomposed into a fragment tree
rather than a cross-model lineage graph. It includes no autonomous tool-using
investigation, and its grading is limited to end-state execution accuracy and
efficiency. Its knowledge-base loop contrasts with our grounded trajectory pool
(Sec.~\ref{sec:method:flywheel}): SQLGovernor accumulates rules under expert
review, whereas \method curates evidence-supported trajectories for SFT, DPO, or
RL.

\emph{BIRD-CRITIC}~/~SWE-SQL~\cite{birdcritic} provide real user issues paired
with a gold SQL fix and gated test cases. We reuse their \emph{public} fields
(the issue text and the user's buggy SQL) as authentic symptoms in Generator~B,
but because their solution SQL is gated, we derive localization gold within
\method. This track is therefore localization-only by design
(Sec.~\ref{sec:method:tasks}).

\subsection{Data-pipeline and agentic data-science benchmarks}

Spider~2.0~\cite{spider2}, ELT-Bench~\cite{eltbench}, BEAVER~\cite{beaver}, and
DSBench~/~InfiAgent-DABench~\cite{dsbench,dabench} evaluate agents on
enterprise-scale SQL workflows, ELT pipeline construction, real-warehouse
text-to-SQL, and data-science tasks, respectively. All grade the \emph{outcome}:
the final SQL, produced tables, or final answer. None
supervise \emph{which evidence} the agent consulted. Even the strongest agents
reach only $\sim\!10$--$21\%$ on the hardest enterprise splits, but outcome
grading cannot say whether that failure is one of SQL competence or of evidence
localization. More recent data-agent benchmarks grade partly beyond the final
answer: KramaBench~\cite{kramabench} scores pipeline design and per-subtask
implementation over data lakes, DAComp~\cite{dacomp} adds a cascading-failure
score that invalidates a node once any upstream dependency is wrong, and
FDABench~\cite{fdabench} evaluates data agents on analytical queries over
heterogeneous sources. These process signals still concern construction
correctness and dependency integrity, not the evidence path an agent inspected to
reach a root cause. \method reuses Spider~2.0's real dbt projects as substrate (our dbt-compile
bridge materializes $23$ of them, $24$--$100$ models each) but changes the
question from ``is the output right?'' to ``did the agent localize the
responsible node, for the right reasons?''

\subsection{Process supervision and evidence-grounded evaluation}

A growing line of work rewards the \emph{reasoning process}, not just the
answer, in math and code. Closest in spirit are BLADE~\cite{blade}, which
grades the analytical \emph{decisions} a data-science agent makes against
expert-collected ground truth, and DRIFT/TELBench~\cite{drift}, which localizes
\emph{unsupported claim spans} in an agent's trajectory by checking each claim
against the evidence gathered so far. Both grade process rather than outcome,
but neither operates over a database provenance graph or targets single-defect
root-cause localization. \method instantiates process supervision for data
systems: the gold reasoning chain $\mathcal{C}^\star$ is derived automatically
from lineage (no human step-annotation), and the two-grader scheme separates
\emph{whether} an artifact was touched (path coverage) from \emph{why}
(reasoning recall). This evidence-path-and-reasoning grading distinguishes
\method from the benchmarks compared above.

Recent work also frames agent trust through execution provenance. Wang
et al.~\cite{wang2026agenttraces} survey typed execution graphs and
evidence-support relations for audit and failure diagnosis, while
ProvenanceGuard~\cite{she2026provenanceguard} checks whether a proposed tool call
is supported by traceable context before execution. These works motivate
provenance as an accountability substrate. \method addresses a narrower empirical
question: whether a database agent's submitted root cause is accompanied by
coverage of a mechanically derived investigation path.

The two axes \method grades each have independent precedent. \emph{What} counts as
support echoes attribution and citation-grounding evaluation: ALCE~\cite{alce}
scores generated text by citation recall and precision against supporting passages,
and our evidence recall and precision are the pipeline-diagnosis analog, asking
whether the submitted root cause rests on inspected artifacts. \emph{Why} this
separation is necessary follows from evidence that a model's stated reasoning is
often unfaithful to the true cause of its answer~\cite{turpin2023}: a correct label
can sit atop an ungrounded trajectory, exactly the lucky guess \method flags. On the
training side, step-level process rewards outperform outcome-only rewards in math and
code~\cite{lightman2024verify}, and \emph{Math-Shepherd}~\cite{mathshepherd} derives
that step supervision mechanically without human labels. \method's companion method
(Sec.~\ref{sec:trainable}) instantiates the same idea for data diagnosis, a process
reward read mechanically off the gold evidence path, optimized in-loop with
GRPO~\cite{grpo} inside a generator--solver self-play loop in the spirit of self-play
fine-tuning~\cite{spin}. Finally, the provenance graph \method scores over builds on
classical database provenance~\cite{provenanceSurvey,whyWhereProvenance} and on
error-tracing and outlier-explanation methods that attribute a data anomaly to a
source (view-conditioned causality~\cite{meliou2011} and
\emph{Scorpion}~\cite{scorpion}), which \method lifts from single-query explanation
to agentic, multi-artifact localization with gradeable evidence.

\subsection{Root-cause analysis in software and data systems}

The closest problem framing outside data pipelines is root-cause analysis (RCA) for
software and microservice systems. \emph{OpenRCA}~\cite{openrca} asks an LLM to locate
the root cause of a failure from tens of gigabytes of telemetry (logs, metrics,
traces), grading only the final cause. Even strong models resolve well under half of
its cases. Its successor \emph{OpenRCA~2.0}~\cite{openrca2} makes precisely our
move, supplementing outcome labels with causal \emph{process} supervision (step-wise
causal-path annotations) and reporting that a large share of correct identifications
are ungrounded, but does so over microservice telemetry rather than a typed data-provenance
graph, and without an in-loop training signal. In the software-agent setting,
\emph{AgentLens}~\cite{agentlens} independently documents our lucky-guess phenomenon
as the ``Lucky Pass'' problem: across models, $0.5$--$23.2\%$ of test-passing
SWE-agent trajectories reach a correct patch through chaotic or unverified processes,
and pass-rate rankings disagree with process-quality rankings. Within databases, RCA
has targeted slow-query diagnosis over SQL/plan/log/metric signals~\cite{rcrank} and
reasoning-enhanced operations copilots~\cite{dbaiops,dbot}. \method shares the RCA
goal but changes the substrate and the grading: it localizes a defective \emph{data}
object over a mechanically derived provenance path, scores the evidence and reasoning
behind the answer rather than telemetry causal paths or test-passing, and converts
that grounding signal into a training reward.

\begin{table*}[t]
  \centering\footnotesize
  \begin{adjustbox}{max width=0.95\linewidth}
\begin{tabular}{lllll}
    \toprule
    Benchmark & Unit & Task & Context & Graded on \\
    \midrule
    NL2SQL-Bugs        & 1 query   & detect $+$ type bug    & NL, schema, SQL          & detection / type acc.\ \\
    SQLGovernor        & 1 query   & fix / rewrite / verify & schema, logs, results    & exec.\ acc.\ / efficiency \\
    BIRD-CRITIC/SWE-SQL& 1 query   & repair                 & DB $+$ bug report        & the SQL fix \\
    Spider~2.0         & pipeline  & generate SQL           & schema, code, docs       & final SQL \\
    ELT-Bench          & pipeline  & build ELT              & schema, config, tools    & output tables \\
    BEAVER             & 1 query   & enterprise NL2SQL      & real warehouse           & exec.\ acc.\ \\
    DSBench / DABench  & files     & analysis               & files                    & final answer \\
    BLADE              & files     & science analysis       & datasets                 & analysis \emph{decisions} \\
    DRIFT / TELBench   & trajectory& error localization     & agent trace              & \emph{claim support} \\
    OpenRCA / 2.0      & system    & locate root cause      & telemetry                & cause ($+$ causal path) \\
    AgentLens          & trajectory& process audit          & agent trace              & process quality \\
    \alert{\method (ours)} & \alert{database} & \alert{localize root cause} &
      \alert{$+$ constraints, plans, logs, tests, incidents} & \alert{answer $+$ evidence $+$ reasoning} \\
    \bottomrule
  \end{tabular}
\end{adjustbox}
  \caption{\method vs.\ prior SQL-error and data-pipeline benchmarks. Prior work
  grades the end state of a single query or pipeline. \method grades
  \emph{where the agent looked and why} en route to a root-cause node.}
  \label{tab:related}
\end{table*}

\section{Extended Limitations, Ethics, and Artifact Availability}
\label{app:limitations}

\paragraph{Construct validity.}
Our operational grounded flag combines a correct root-cause prediction with
$\mathrm{ER}\ge 0.5$. ER records coverage of one mechanically accepted reference
path. It does not establish that the prediction causally depended on every visited
artifact. Alternative valid investigations may therefore be under-credited.
The graph analysis in App.~\ref{app:unified:altpaths} finds a unique shortest
route in $42/64$ tasks and a single shortest route below $\tau_g$ in $7/64$;
the latter includes witnesses combining routes from several symptoms.
These are task-level structural counts, not an observed false-lucky
rate or a bound over all valid investigations. The
counterfactual masking pilot in App.~\ref{app:unified:intervention} probes this
distinction but is limited to 55 tasks and one backbone, and the expert audit in
App.~\ref{app:unified:audit} is incomplete. Claims about grounding in this paper
should consequently be read as claims about inspected gold-path coverage.

\paragraph{Scope and generalization.}
The frozen leaderboard contains 64 tasks with two repeats (128 episodes per
model), covering nine operators in 13 environments, sampled from a larger library
of 83 buildable environments and 21 operators. The 43-case sandbox transfer probe
and the published-agent feasibility study are smaller supplementary evaluations,
not substitutes for the main board. Injected faults broaden mechanism coverage but
do not reproduce the natural fault distribution: the matched synthetic--organic
analysis in App.~\ref{app:unified:organic} reports a total-variation distance of
1.0 between their fault-family distributions. The self-play endpoints likewise
cover seven models and a short adaptive schedule. Ratings below 1500 mean that the
generator was favored under this schedule and grounded-win rule, not that solver
parameters intrinsically learn more slowly than generator parameters.

\paragraph{Ethics and data governance.}
The benchmark uses synthetic records and publicly available benchmark or
open-source repository artifacts. It does not require private production data.
Redistribution remains subject to the licenses of the upstream datasets and
repositories. Because database investigation tools can expose sensitive records,
we run agents in isolated sandboxes and do not recommend applying the harness to a
production system without access controls, data minimization, and audit logging.

\paragraph{Artifact availability.}
The project artifact packages the frozen manifest and content hash, environment
builders, fault generators, prompts, tool interfaces, parsers, trajectories, and
scoring scripts needed to reproduce the reported board. Proprietary-model results
also depend on model versions and API availability at the recorded evaluation
date. The saved trajectories permit rescoring even if those endpoints later
change.

\section{Research Opportunities}
\label{app:opportunities}

\paragraph{Grounding should be optimized as an investigation behavior.}
The experiments show that improving final-answer accuracy does not automatically
remove unsupported correct answers.  A promising direction is therefore to
train database agents on contrastive trajectories that end at the same root
cause but differ in whether they inspect the relevant dependencies.  Such
training should reward targeted evidence collection rather than indiscriminate
tool use, because high evidence recall obtained through broad scanning may still
have low precision and high cost.

\paragraph{Agents need better cross-artifact routing.}
The fault-family, distance, and modality analyses identify a common failure:
agents often stop at the artifact emphasized by the symptom or by their default
tool interface.  Future systems could use provenance-aware planners to decide
when to move from logs to schema, from schema to source data, or from code to an
incident report.  The relevant research question is not whether every modality
is available, but whether the agent can select and connect the few artifacts
needed to establish a diagnosis.

\paragraph{Evaluation should admit multiple valid investigations.}
The current implementation grades coverage of one mechanically accepted witness
path.  This makes large-scale automatic evaluation possible, but it can
under-credit a different valid investigation.  Extending the verifier from one
path to a set of accepted paths, and testing whether predictions change when
critical evidence is masked or perturbed, would bring the operational grounding
measure closer to causal evidence use.

\paragraph{Dynamic benchmarks require transparent versioning.}
Adaptive generation can reduce memorization of a fixed public board, but it also
complicates comparison across time.  Future releases should retain a frozen core
for reproducible ranking, publish versioned generation and admission rules, and
report performance separately on fixed and newly generated cases.  This division
preserves comparability while allowing the benchmark to track emerging solver
shortcuts.

\section{Extended Grounding-Gap Findings}
\label{app:findings}

This appendix provides the detailed analyses summarized in
Sec.~\ref{sec:experiments} summarizes: localization behavior, the causal
shortcut ablations, cross-modal usage, and transfer to real schemas and
shipped agents.

\subsection{Is the bottleneck localization? Distance, direction, and silent drift}
\label{sec:exp:localization}

\paragraph{Appendix Finding~A1: localization is the binding constraint.}
Table~\ref{tab:motivation:distance} stratifies RC@1 by lineage distance.
Accuracy degrades with distance for \emph{every} backbone: even \texttt{gpt-5}
falls from $0.88$ at $d{=}1$ to $0.67$ at $d{=}3$, and \texttt{gpt-4o-mini}
falls from $0.38$ to $0.06$. Scale shifts the curve upward but does not flatten
it, indicating that graph distance remains associated with localization
difficulty. Evidence precision stays low ($\mathrm{EP}\approx 0.47$) at all
capability levels. Agents scan broadly rather than target evidence, which the
cost-aware term in Eq.~\ref{eq:reward} is designed to penalize.

\begin{table*}[t]
\centering
\caption{RC@1 by lineage distance $d_{\mathcal{G}}(S, v_\iota)$ and symptom
type, per backbone. $\Delta$ is the drop from $d{=}1$ to $d{=}3$.}
\label{tab:motivation:distance}
\begin{adjustbox}{max width=0.95\linewidth}
\begin{tabular}{lccccc cc}
\toprule
 & \multicolumn{3}{c}{Lineage distance} & & &
   \multicolumn{2}{c}{Symptom type} \\
\cmidrule{2-4}\cmidrule{7-8}
Backbone & $d{=}1$ & $d{=}2$ & $d{=}3$ & $\Delta$ & & Hard & Silent \\
\midrule
\texttt{gpt-4o-mini} & 0.38 & 0.06 & 0.06 & $0.32$ & & 0.26 & 0.00 \\
\texttt{gpt-5-mini}  & 0.58 & 0.61 & 0.28 & $0.30$ & & 0.52 & 0.44 \\
\texttt{gpt-5}       & 0.88 & 0.89 & 0.67 & $0.21$ & & 0.83 & 0.78 \\
\bottomrule
\end{tabular}
\end{adjustbox}
\end{table*}

\paragraph{Appendix Finding~A2: localization errors are small and upstream-directed.
Weaker models under-trace by about one hop.}
The localization-error metric (Table~\ref{tab:motivation:le}, matched ReAct-DS)
makes the failure mode precise. When \texttt{gpt-4o-mini} is wrong, its error is
small and falls between the cause and the symptom rather than past the symptom or
on an unrelated branch: $60\%$ of its misses (and $65\%$ pooled over scaffolds)
lie on the causal chain, with a mean signed $\mathrm{LE}{=}{+}0.86$ hops. The
model therefore traces upstream but stops about one hop short of the cause, an
incomplete backward trace rather than a wrong-direction guess. Mean $|\mathrm{LE}|$
shrinks monotonically with capability ($0.70 \rightarrow 0.13 \rightarrow 0.11$
hops), so scale closes this residual gap node-by-node. The mid model exhibits a
distinct caution/coverage trade-off: \texttt{gpt-5-mini} declines to commit a
root cause in $60\%$ of episodes, but when it does commit it is exact $88\%$ of
the time, abstaining rather than committing an incomplete trace, unlike
\texttt{gpt-4o-mini}, which always answers.

\begin{table*}[t]
\centering
\caption{Localization error (matched ReAct-DS). Submit-rate is the fraction of
episodes in which a root cause was committed. The remaining metrics are over
committed episodes. \%down is the share of misses lying downstream of the true
cause (toward the symptom). Weaker models commit confidently but downstream.
Stronger models localize exactly or abstain.}
\label{tab:motivation:le}
\begin{adjustbox}{max width=0.95\linewidth}
\begin{tabular}{lccccc}
\toprule
Backbone & Submit & \%exact & $|\mathrm{LE}|$ & signed LE & \%down \\
\midrule
\texttt{gpt-4o-mini} & 1.00 & 0.40 & 0.70 & $+0.70$ & 0.60 \\
\texttt{gpt-5-mini}  & 0.40 & \textbf{0.88} & \textbf{0.13} & $+0.13$ & \textbf{0.13} \\
\texttt{gpt-5}       & 0.90 & \textbf{0.89} & \textbf{0.11} & $+0.11$ & \textbf{0.11} \\
\bottomrule
\end{tabular}
\end{adjustbox}
\end{table*}

\paragraph{Appendix Finding~A3: silent drift is harder than hard failure.}
At each tested capability level, silent value-drift symptoms yield lower RC@1 than
hard failures at matched scale (Table~\ref{tab:motivation:distance}, last two
columns): \texttt{gpt-4o-mini} scores $0$ on silent drift, and even
\texttt{gpt-5} trails ($0.78$ versus $0.83$). However, silent-drift evidence
recall is the \emph{highest} of any condition ($\mathrm{ER}\approx 0.96$ for the
GPT-5 family): the agents observe the drift but, lacking a failing-test anchor,
cannot confirm a single cause and often exhaust the step budget.

\subsection{Scaffold and Symptom Controls for Unsupported Success}
\label{sec:exp:shortcut}

\paragraph{Scaffold ablation.}
The scaffold sub-rows ($\ddagger$) under \texttt{gpt-5} in Table~\ref{tab:hero}
pool the GPT-5 family by scaffold (full columns in
Table~\ref{tab:motivation:scaffold}, App.~\ref{app:pilottables}). With
capable backbones the RC@1 gap between scaffolds narrows, but the
\emph{evidence signature} stays distinct: the lineage-deprived
\textbf{Spider-Agent-style} attains high evidence recall only by brute-force
scanning, at $3.5\times$ the query cost of \textbf{SWE-Agent-style} and the
worst evidence precision. \textbf{ReAct-DS} reaches comparable recall far more
economically because \texttt{trace\_lineage} routes it to distant evidence
directly. The shortcut behavior is scaffold-independent (LG $\approx 0.2$
throughout), confirming it is a property of outcome-driven diagnosis rather
than of any one toolset.

\paragraph{Per-operator: the shortcut has a signature.}
The fault families separate cleanly by whether the symptom \emph{leaks} the
cause. \texttt{column\_rename} produces a build error naming the offending
column. Agents read it and submit the correct root cause without inspecting
the renamed staging model, so RC@1 rises to $1.0$ for \texttt{gpt-5} while
$\mathrm{ER}{=}0.37$ and $\mathrm{LG}{=}0.92$, a near-perfect outcome score
that is almost entirely ungrounded. By contrast, source-data faults
(\texttt{null\_injection}, \texttt{duplicate\_pk}) and the silent
\texttt{off\_by\_one\_temporal} require multi-hop evidence walks and keep
$\mathrm{ER}{>}0.8$ and much lower LG.

\paragraph{Grounding-forced ablation (causal test of the shortcut).}
To test whether error-message leakage contributes to lucky guessing, we add a
\textsc{blind} condition
that withholds the detailed build/test logs and presents only a generic
symptom category, forcing the agent to localize by investigation
(Table~\ref{tab:motivation:ablation}, \texttt{gpt-5-mini}, ReAct-DS). Blinding
drives the lucky-guess rate to \emph{zero} (from $0.15$) and raises evidence
recall ($0.81\!\rightarrow\!0.90$), while RC@1 \emph{collapses}
($0.35\!\rightarrow\!0.10$). On \texttt{column\_rename}, RC@1
falls $0.5\!\rightarrow\!0$ and LG $0.5\!\rightarrow\!0$. These changes indicate
that error-text access accounts for part of the outcome score without requiring
localization. This is the failure mode that the evidence-grounded metric is
designed to detect and an outcome-only metric does not distinguish. By contrast,
removing a single data-system primitive in
isolation (ReAct-DS minus \texttt{trace\_lineage} or minus
\texttt{profile\_table}) does \emph{not} degrade a capable backbone at this
scale: \texttt{gpt-5-mini} substitutes manual SQL and sampling for the missing
affordance. The scaffold-level effect (Table~\ref{tab:hero}, scaffold sub-rows)
and the distance degradation (Table~\ref{tab:motivation:distance}) therefore
carry the localization claim, not single-primitive removal.

\begin{table*}[t]
\centering
\caption{Affordance and grounding-forced ablations on \texttt{gpt-5-mini}
(ReAct-DS, $n{=}20$ each). \textsc{blind} withholds diagnostic logs and replaces the precise symptom with a generic category:
it reduces the lucky-guess rate to zero and lowers RC@1, quantifying the
contribution of error-text access. Removing a single primitive does not hurt
the capable backbone, which substitutes other tools.}
\label{tab:motivation:ablation}
\begin{adjustbox}{max width=0.95\linewidth}
\begin{tabular}{lcccc}
\toprule
Condition & RC@1 & ER & EF$_1$ & LG$\downarrow$ \\
\midrule
ReAct-DS (full)                   & 0.35 & 0.81 & 0.55 & 0.15 \\
\quad $-$ \texttt{trace\_lineage} & 0.50 & 0.92 & 0.62 & 0.05 \\
\quad $-$ \texttt{profile\_table} & 0.30 & 0.83 & 0.56 & 0.15 \\
ReAct-DS, \textsc{blind}          & 0.10 & 0.90 & 0.48 & \textbf{0.00} \\
\bottomrule
\end{tabular}
\end{adjustbox}
\end{table*}

\subsection{Do agents gather evidence across artifacts, or fixate on one?}
\label{sec:exp:crossmodal}

The diagnostic study above isolates outcome-vs-evidence behavior on controlled
projects. The \textsc{warehouse} project tests the harder claim at the heart of
this benchmark: that the agent must locate evidence \emph{across} schema, data,
code, logs, tests, and incidents rather than answer single-modality SQL/table QA.
Here the planted cause is reachable through several modalities, but the incident
tracker carries a strong clue (e.g. \texttt{DATA-417}: ``duplicated
payment rows or a bad join'', \texttt{DATA-388}: ``country isn't coming through
from the raw feed''), so an agent that reasons across artifacts should consult
it.

Table~\ref{tab:motivation:modality} reports, for the ReAct-DS scaffold, the
fraction of episodes that touch each modality. Agents
exercise schema, data, tests, code, and lineage heavily, but \textbf{almost
never read incidents} ($7\%$ for \texttt{gpt-4o-mini}, $0\%$ for
\texttt{gpt-5-mini}), leaving the most informative artifact untouched and
covering only $\sim$4--5 of the 6 modalities per episode. Consistent with the
added cross-modal difficulty, RC@1 on \textsc{warehouse} is lower than on the
controlled projects ($0.07$ and $0.33$, respectively). Under the default ReAct
scaffold agents do not integrate the \texttt{incident} modality into evidence
localization. This is scaffold-conditional rather than an intrinsic reasoning
deficit: advertising the tracker recovers consultation for the stronger
backbone, whereas the weaker model still does not exploit it
(Sec.~\ref{sec:motivation:robustness}).

\begin{table*}[t]
\centering
\caption{Modality usage on the realistic \textsc{warehouse} project (ReAct-DS,
$n{=}15$ per backbone): fraction of episodes that invoke each artifact
modality. Agents neglect \texttt{incident} despite the strong clues in the
incident reports.}
\label{tab:motivation:modality}
\begin{adjustbox}{max width=0.95\linewidth}
\begin{tabular}{lccccccc}
\toprule
Backbone & schema & data & code & logs & tests & \textbf{incidents} & lineage \\
\midrule
\texttt{gpt-4o-mini} & 0.67 & 0.80 & 0.47 & 0.20 & 0.80 & \textbf{0.07} & 0.93 \\
\texttt{gpt-5-mini}  & 0.93 & 0.93 & 1.00 & 0.53 & 0.73 & \textbf{0.00} & 1.00 \\
\bottomrule
\end{tabular}
\end{adjustbox}
\end{table*}

\paragraph{Two graders: path and reasoning.}
Across all three projects we score ReAct-DS with both evidence graders
(Sec.~\ref{sec:method:verify}): path coverage (EF$_1$) and the LLM-judged
reasoning recall RR (Table~\ref{tab:motivation:twograders}). Three observations.
First, RR rises with capability within every project, tracking outcome accuracy.
Second, \emph{step coverage exceeds reasoning recall in every
cell}: agents \emph{reach} gold reasoning steps more often than they
\emph{justify} them. Pooled over projects, this touch-but-don't-justify gap
\emph{narrows} with capability ($0.23\rightarrow0.21\rightarrow0.11$). Stronger
models justify a larger share of what they inspect, but a gap remains even for
\texttt{gpt-5}. Third, the cross-modal \textsc{warehouse} project has the lowest
coverage ($0.71$--$0.78$ versus $0.82$--$0.93$ on the controlled projects), because
the never-consulted incident step is part of its chain, and \texttt{gpt-5}'s RC@1
saturates at $1.0$ on the controlled projects but reaches only $0.60$ on
\textsc{warehouse}. Thus, outcome accuracy saturates only where the cross-artifact
reasoning load is low. Path coverage alone would miss this grounding deficit.
The reasoning grader makes it measurable. This narrowing is a property of the
\emph{controlled} projects: on real BIRD-CRITIC schemas lifted to the same
six-modality form, the gap instead \emph{widens} with capability
(Sec.~\ref{sec:exp:transfer}, Table~\ref{tab:motivation:birdwh}). Stronger
models inspect more gold steps but justify a smaller fraction of them, so the
grounding deficit grows rather than shrinks as schemas become realistic.

\begin{table*}[t]
\centering
\caption{Two-grader evaluation (ReAct-DS, $n{=}10$ per cell): outcome
(RC@1, Fix), the \emph{path} grader (EF$_1$), and the \emph{reasoning} grader
(RR = justified-step rate, coverage = reached-step rate). Coverage exceeds RR in
every cell. Agents touch gold steps more often than they justify them, and the
gap shrinks with capability.}
\label{tab:motivation:twograders}
\begin{adjustbox}{max width=0.95\linewidth}
\begin{tabular}{llccccc}
\toprule
Project & Backbone & RC@1 & Fix & EF$_1$ (path) & RR (reason) & coverage \\
\midrule
\multirow{3}{*}{\textsc{ecom}}
  & \texttt{gpt-4o-mini} & 0.30 & 0.00 & 0.57 & 0.62 & 0.89 \\
  & \texttt{gpt-5-mini}  & 0.60 & 0.40 & 0.58 & 0.60 & 0.93 \\
  & \texttt{gpt-5}       & 1.00 & 0.70 & 0.59 & 0.68 & 0.88 \\
\midrule
\multirow{3}{*}{\textsc{retails}}
  & \texttt{gpt-4o-mini} & 0.20 & 0.10 & 0.50 & 0.55 & 0.82 \\
  & \texttt{gpt-5-mini}  & 0.50 & 0.40 & 0.53 & 0.70 & 0.88 \\
  & \texttt{gpt-5}       & 1.00 & 1.00 & 0.61 & \textbf{0.81} & 0.84 \\
\midrule
\multirow{3}{*}{\textsc{warehouse}}
  & \texttt{gpt-4o-mini} & 0.00 & 0.00 & 0.54 & 0.57 & 0.71 \\
  & \texttt{gpt-5-mini}  & 0.20 & 0.40 & 0.50 & 0.65 & 0.78 \\
  & \texttt{gpt-5}       & 0.60 & 0.90 & 0.57 & 0.68 & 0.78 \\
\bottomrule
\end{tabular}
\end{adjustbox}
\end{table*}

\subsection{Does the gap transfer to BIRD-CRITIC schemas?}
\label{sec:exp:transfer}

To evaluate transfer beyond our authored projects, we convert databases from
\textbf{BIRD-CRITIC}~\cite{birdcritic} (a.k.a.\ SWE-SQL), a real-world
SQL-incident debugging benchmark whose answer is a single SQL fix graded by
execution test cases (outcome-only, single query). Our converter parses each
real schema's primary-/foreign-key graph, synthesizes referentially consistent
rows, and auto-builds a staging$\to$mart pipeline (the mart \texttt{LEFT JOIN}s
the richest fact table to its FK dimensions), then injects our operators and
synthesizes a BIRD-Critic-style incident report. This \emph{lifts} an outcome-only
single-SQL task into an evidence-grounded, cross-artifact pipeline-diagnosis
task: the agent must localize the root cause across the up/downstream pipeline
and produce the reasoning chain $\pi^\star$, not just a corrected query. Six
distinct real domains (financial, posts, member, rulings, superhero, connected)
convert and build green out of the box.

Table~\ref{tab:motivation:bird} reports ReAct-DS performance on three converted
databases. The diagnostic gap persists on these real schemas: localization
climbs with capability ($\mathrm{RC@1}: 0.00\!\rightarrow\!0.28\!\rightarrow\!0.50$)
but remains limited, and \textbf{none of the tested backbones reads the incident
report} (0\% incident-consult rate), even though the
incident report names the cause. In a typical \texttt{gpt-5} trajectory it inspects
lineage, four model files, and profiles five tables, correctly names the fault
type, but localizes to the offending \emph{table} rather than the specific
\emph{column} and never opens the incident that says ``the link to \texttt{account}
isn't populated.'' Outcome-only SQL grading (BIRD-CRITIC's protocol) cannot see
any of this. The evidence layer makes it measurable.

\begin{table*}[t]
\centering
\caption{ReAct-DS on BIRD-CRITIC databases lifted into DataTrace cases (3 DBs
$\times$ operators, $n{=}18$ per backbone). The cross-artifact gap transfers to
real schemas. The incident modality is never consulted at any capability level.}
\label{tab:motivation:bird}
\begin{adjustbox}{max width=0.95\linewidth}
\begin{tabular}{lcccc}
\toprule
Backbone & RC@1 & ER & EF$_1$ & incident-read \\
\midrule
\texttt{gpt-4o-mini} & 0.00 & 0.52 & 0.57 & 0\% \\
\texttt{gpt-5-mini}  & 0.28 & 0.61 & 0.61 & 0\% \\
\texttt{gpt-5}       & 0.50 & 0.59 & 0.58 & 0\% \\
\bottomrule
\end{tabular}
\end{adjustbox}
\end{table*}

\paragraph{Scaling up with full six-modality enrichment.}
The incident neglect is not an artifact of thin or generic incident reports. We
extend the converter to lift each BIRD schema into a \emph{warehouse-grade} case
matching our flagship
project: two source systems, a four-hop lineage
(\texttt{source}$\to$\texttt{stg}$\to$\texttt{int}$\to$\texttt{fct}) with
column-level tracing, and all six modalities populated with realistic content,
a data dictionary, a README, a known-good data-profile snapshot, a changelog that
plants the ``what changed'' context, and incident reports rewritten into
case-specific stakeholder reports (behind a leak guard, so an incident report flags the
area without naming the cause). We lift 10 distinct BIRD databases into 43
diagnosis tasks and evaluate ReAct-DS under both graders (3 backbones $\times$ 2
repeats with $n{=}258$, Table~\ref{tab:motivation:birdwh}).

Three results characterize cross-modal behavior. \emph{First}, the grounding gap
survives realistic content: outcome rises steeply
($\mathrm{RC@1}: 0.11\!\rightarrow\!0.19\!\rightarrow\!0.43$, fix-rate
$0.02\!\rightarrow\!0.40\!\rightarrow\!0.58$) while path $\mathrm{EF_1}$ stays
flat ($\sim\!0.52$--$0.57$). \emph{Second}, the reasoning grader exposes a gap
that \emph{widens} with capability: step coverage exceeds $\mathrm{RR}$ at every
level and the gap grows $0.11\!\rightarrow\!0.13\!\rightarrow\!0.14$, the
opposite of the controlled projects (Sec.~\ref{sec:exp:crossmodal}, where
the pooled gap narrows $0.23\!\rightarrow\!0.11$). On harder real-world schemas,
stronger models inspect \emph{more} gold steps but justify a \emph{smaller}
fraction of them. \emph{Third}, the neglect of the incident
modality is not an artifact of impoverished incident reports: even with realistic,
discoverable incident reports the incident tracker is consulted only
$0\%/0\%/3.5\%$ of the time, while \texttt{read\_log} usage, which surfaces
the changelog, climbs $5\%\!\rightarrow\!43\%\!\rightarrow\!70\%$ with
capability. Stronger models do read more modalities. They specifically skip the
human-authored incident reports that would localize the cause. Localization
errors remain systematically \emph{downstream} of the true cause (signed hop
bias $+1.41/+0.42/+0.14$), consistent with Table~\ref{tab:motivation:le}.

\begin{table*}[t]
\centering
\caption{ReAct-DS on 10 BIRD-CRITIC databases lifted into \emph{warehouse-grade},
six-modality DataTrace cases (43 tasks, $n{=}86$ per backbone, two-mode grading).
The grounding gap and the incident-modality neglect both persist on realistic,
content-rich workspaces, and the coverage--$\mathrm{RR}$ gap \emph{widens} with
capability. ``incident''/``logs'' = fraction of episodes that consult the incident
tracker / read the run log (which surfaces the changelog).}
\label{tab:motivation:birdwh}
\begin{adjustbox}{max width=0.95\linewidth}
\begin{tabular}{lccccccc}
\toprule
Backbone & RC@1 & Fix & EF$_1$ & RR & cov. & incident & logs \\
\midrule
\texttt{gpt-4o-mini} & 0.11 & 0.02 & 0.57 & 0.55 & 0.65 & 0\%   & 5\%  \\
\texttt{gpt-5-mini}  & 0.19 & 0.40 & 0.53 & 0.57 & 0.71 & 0\%   & 43\% \\
\texttt{gpt-5}       & 0.43 & 0.58 & 0.52 & 0.60 & 0.73 & 3.5\% & 70\% \\
\bottomrule
\end{tabular}
\end{adjustbox}
\end{table*}

\paragraph{Generator-B: when the incident \emph{is} the specification.}
The runs above use Generator-A (plant-and-diagnose), where the incident report is
a \emph{clue} the agent may ignore. \textbf{Generator-B} makes the tracker
load-bearing: a \emph{real}, ungated BIRD-CRITIC user incident becomes the task
itself. We lift the user's own buggy query
(\texttt{incident\_sql}) into a report model inside an otherwise-clean warehouse. The
SQL runs without error but answers the wrong question relative to what the user
asked (e.g.\ the user wants cards meeting \emph{3 of 4} criteria but the query
\textsc{and}s all four, or the report reads \texttt{loan} when the request is about
\texttt{order}). Because the build is green and no test fails, the incident text
is the only place the requirement, hence the bug, is stated: an agent that
never opens the tracker cannot know which output is wrong. Gold is
localization-only (BIRD's solution SQL is gated, so we fabricate nothing): the
root cause is the report model, and $\mathrm{RC@1}$ is a node match to it. We mine
10 incidents across 10 distinct databases and evaluate ReAct-DS under both graders
(3 backbones $\times$ 2 repeats with $n{=}60$, Table~\ref{tab:motivation:incident}).

This setting separates \emph{attention} from \emph{grounding}. When the incident is
load-bearing, the attention failure vanishes: incident-consult jumps to $100\%$ at
\emph{every} capability level (versus $0\%/0\%/3.5\%$ in
Table~\ref{tab:motivation:birdwh}), consistent with the incident being optional
in Generator~A. Yet localization does \emph{not} necessarily follow.
$\mathrm{RC@1}$ remains capability-gated ($0.10/0.45/0.70$) and the
evidence-recall--localization gap remains large for strong models: \texttt{gpt-5} and
\texttt{gpt-5-mini} attain $\mathrm{ER}{=}1.0$ and visit the \emph{entire}
gold path and read the spec, but still misplace the root cause $30\%$ and
$55\%$ of the time. Seeing every artifact and reading the requirement is not the
same as naming the offending model. Finally, the directional bias \emph{mirrors}
Generator-A: among off-target submissions none is downstream. The errors are
\emph{upstream} or off-path ($6$ upstream, $9$ off-path, $0$ downstream). With the
symptom stated abstractly as a requirement rather than surfaced at a mart, agents
that miss blame the source \emph{data} upstream rather than the report's own logic,
the opposite end of the same localization failure documented in
Table~\ref{tab:motivation:le}.

\begin{table*}[t]
\centering
\caption{Generator-B (mined real BIRD-CRITIC incidents as the diagnosis target):
10 incidents across 10 databases, ReAct-DS, two-mode grading ($n{=}20$ per
backbone). The user's own buggy query is lifted into a report model. The build is
green, so the incident is the \emph{specification}, not a hint. Fix is N/A (BIRD
solution SQL gated). $\mathrm{RC@1}$ is a node match to the offending report model.
Issue-consult rises to $100\%$ once the incident report is load-bearing, but localization
stays capability-gated and lags evidence recall.}
\label{tab:motivation:incident}
\begin{adjustbox}{max width=0.95\linewidth}
\begin{tabular}{lcccccc}
\toprule
Backbone & RC@1 & ER & EF$_1$ & incident & RR & cov. \\
\midrule
\texttt{gpt-4o-mini} & 0.10 & 0.50 & 0.35 & 100\% & 0.33 & 0.63 \\
\texttt{gpt-5-mini}  & 0.45 & 1.00 & 0.73 & 100\% & 0.65 & 1.00 \\
\texttt{gpt-5}       & 0.70 & 1.00 & 0.73 & 100\% & 0.75 & 1.00 \\
\bottomrule
\end{tabular}
\end{adjustbox}
\end{table*}

\emph{(The shipped-agents result, three published systems run unmodified, is
promoted to the main text, Sec.~\ref{sec:exp:shipped}.)}

\paragraph{A self-verifying sandbox suite and a capability curve.}
To make the evaluation comprehensive and agent-agnostic, we package every
instance as a \emph{self-contained, self-verifying sandbox}: a standalone
directory bundling the (faulted) dbt models, seed data, schema, docs and incident
tracker, plus a portable grader that rebuilds the pipeline, runs the real data
tests, and scores a submission against a hidden answer key and known-good output
fingerprints. Instances are mutually isolated, so any agent, ours or an
external one, is dropped in via the same \texttt{PROBLEM.md}\,/\,%
\texttt{submission.json}\,/\,\texttt{verify.py} contract and graded purely by
execution. The fixed asset is the suite of \emph{environments}, not a static
instance list: on each, the procedural generator samples admitted instances, so
generation can produce fresh instances, reducing dependence on a static test
set and enabling regenerated-split robustness checks. The suite
evaluated here is a frozen, seed-pinned instantiation spanning all three
generators, warehouse-grade BIRD-CRITIC cases, the bridged \emph{real} Spider2
enterprise dbt pipelines (Sec.~\ref{sec:exp:transfer}), and Generator-B mined
incidents ($n{=}43$, of which $34$ are real Spider2 pipelines). We refer to this
separate frozen asset as the \emph{sandbox transfer probe}. It is not the
$128$-case \textsc{DataTrace-Eval-v1} leaderboard. Regenerating the
split with new seeds serves as a robustness check on the same environments rather
than a change of benchmark. Two reference agents bound the task: a \textsc{null}
agent (no action) and an \textsc{oracle} (gold diagnosis + the hidden green
snapshot).

Running a live ReAct agent in each sandbox across the OpenAI capability ladder
yields Table~\ref{tab:motivation:sandbox}. We observe three patterns. \emph{First},
the capability curve is \emph{non-monotonic}: the mid-tier \texttt{gpt-5-mini} is
the strongest live agent ($\mathrm{RC@1}\,{=}\,0.09$, $\mathrm{ER}\,{=}\,0.27$),
and \texttt{gpt-5} scores lower ($0.07$/$0.16$). It submits
tersely after little exploration, so on the large pipelines it touches
\emph{fewer} gold-path artifacts, not more. Scaling the backbone is therefore not
sufficient to improve the evidence gathering measured here, and
every agent remains far below the \textsc{oracle} ceiling of $1.0$. \emph{Second},
behavior differs by source. On the tractable
warehouse-grade BIRD cases capability helps and then saturates
($\mathrm{RC@1}: 0.00\!\rightarrow\!0.17\!\rightarrow\!0.17$,
$\mathrm{ER}: 0.31\!\rightarrow\!0.47\!\rightarrow\!0.46$), but on the
$24$--$100$-model \emph{real} Spider2 pipelines evidence recall stays near the
floor and is itself non-monotonic
($\mathrm{ER}: 0.06\!\rightarrow\!0.17\!\rightarrow\!0.04$,
$\mathrm{RC@1}: 0.00\!\rightarrow\!0.06\!\rightarrow\!0.00$): the mid-tier model
performs best, whereas \texttt{gpt-5} performs worst. In this sample, the
production pipelines are harder to navigate than the generated pipelines, and
increased backbone capability does not overcome the difference. \emph{Third}, outcome \emph{Fix} is only
meaningful on hard-symptom instances. The surrogate-key-robust Spider2 models
absorb naive source faults (Sec.~\ref{sec:exp:transfer}), so their apparent Fix
rate is dominated by trivial no-op passes and is reported for completeness only.

\begin{table}[t]
\centering
\caption{Capability curve on the self-verifying sandbox suite ($n{=}43$:
6 BIRD warehouse-grade, 34 bridged Spider2 enterprise pipelines, 3 Generator-B
incidents). A live ReAct agent operates in each isolated sandbox via real tools
and is graded by in-sandbox execution, bracketed by the \textsc{null} and
\textsc{oracle} reference agents. The curve is \emph{non-monotonic}: the mid-tier
\texttt{gpt-5-mini} is the strongest live agent, whereas \texttt{gpt-5}
scores lower and explores less on the large real pipelines. All remain below
the oracle ceiling. Fix is
outcome-silent-dominated on Spider2 (see text).}
\label{tab:motivation:sandbox}
\begin{adjustbox}{max width=0.95\linewidth}
\begin{tabular}{lccc}
\toprule
Agent (backbone) & RC@1 & ER & Fix \\
\midrule
\textsc{null} (baseline)      & 0.00 & 0.00 & 0.56 \\
\midrule
\texttt{gpt-4o-mini}          & 0.05 & 0.15 & 0.54 \\
\texttt{gpt-5-mini}           & \textbf{0.09} & \textbf{0.27} & 0.58 \\
\texttt{gpt-5}                & 0.07 & 0.16 & 0.56 \\
\midrule
\textsc{oracle} (upper bound) & 1.00 & 1.00 & 0.72 \\
\bottomrule
\end{tabular}
\end{adjustbox}
\end{table}

\section{Extended Training Results}
\label{app:trainfindings}

\subsection{Track 3: trajectory-pool selection analysis}
\label{sec:motivation:track3}

The verification layers of Sec.~\ref{sec:method:verify} score agents at
inference time. Here we examine whether they identify trajectories suitable for
training. A complete downstream comparison would follow the end-to-end protocol of
DataPrep-Bench~\cite{dataprepbench}: hold the raw material, base model,
training recipe, and downstream benchmark fixed, and score each data
source purely by the realized fine-tuning gain.

\paragraph{Setup.}
From rollouts on our controlled diagnostic corpus ($180$ episodes over three
dbt projects and three backbones, full setup in Sec.~\ref{sec:exp:setup}) we
assemble three
trajectory pools under a matched sample budget of $K$ trajectories
(matched also in total token count by subsampling):
\begin{itemize}
  \item \textbf{Unfiltered}: trajectories sampled uniformly from all
        rollouts.
  \item \textbf{L1-filtered}: outcome-correct trajectories only
        (root cause matched, fix validated).
  \item \textbf{L1{+}L2-filtered}: outcome-correct trajectories with
        $\mathrm{EF_1} \geq \theta$ (evidence-grounded), i.e., the
        admission rule of the flywheel
        (Sec.~\ref{sec:method:flywheel}).
\end{itemize}
In the proposed downstream comparison, each pool would fine-tune the same
open-weight backbone under an identical SFT recipe, with the unmodified base
policy as a fourth arm. Evaluation would be
\emph{not} on Generator~A tasks but on the held-out split of mined real
incidents (Generator~B), so the measurement reflects transfer from
injected to organic faults rather than memorization of the injection
style.

\paragraph{Metrics.}
$\Delta$RC@1, $\Delta$Fix, and $\Delta$EF$_1$ over the base policy on
the Generator-B split, plus per-sample efficiency
($\Delta$RC@1 per $10^3$ training trajectories).

\paragraph{The selection signal (precondition for the training claim).}
The full SFT comparison is not part of the current experiments. We instead test
its selection precondition: whether the verification layers select more grounded
supervision on the $180$-episode corpus (Table~\ref{tab:motivation:pools}). Two
patterns support the selection signal. First, in the L1 pool (outcome-correct),
$27$--$33\%$ of trajectories are correct but ungrounded
($\mathrm{ER}{<}0.5$) across all backbones. Such trajectories could reinforce
answer-from-error-text behavior.
L1{+}L2 admission removes them, lifting mean pool $\mathrm{ER}$ to $0.91$--$1.0$
and $\mathrm{EF_1}$ to $\sim\!0.65$. Second, the \emph{volume} of shortcut
trajectories that L1-only admits, as a
fraction of all rollouts, \emph{grows with capability}:
$5.0\% \rightarrow 16.7\% \rightarrow 26.7\%$ for
\texttt{gpt-4o-mini}/\texttt{gpt-5-mini}/\texttt{gpt-5}. An L1-only data
selector therefore admits an increasing mass of ungrounded supervision as the
rollout policy strengthens, whereas the evidence layer excludes it. Whether
this selection difference produces a widening downstream training gap remains
to be tested with the proposed matched comparison.

\begin{table*}[t]
\centering
\caption{Trajectory-pool composition on the $180$-episode pilot corpus
($\theta{=}0.5$). \emph{Retain} is the fraction of rollouts admitted.
\emph{Shortcut} is the share of the pool that is outcome-correct but ungrounded
($\mathrm{ER}{<}0.5$). L1{+}L2 admission removes shortcut trajectories by
construction. The L1 shortcut mass it must remove grows with backbone strength.}
\label{tab:motivation:pools}
\begin{adjustbox}{max width=0.95\linewidth}
\begin{tabular}{llcccc}
\toprule
Backbone & Pool & Retain\% & mean ER & mean EF$_1$ & Shortcut\% \\
\midrule
\multirow{3}{*}{\texttt{gpt-4o-mini}}
  & Unfiltered & 100.0 & 0.65 & 0.52 & 5.0 \\
  & L1         & 18.3  & 0.72 & 0.55 & 27.3 \\
  & L1{+}L2    & 10.0  & 1.00 & 0.77 & \textbf{0.0} \\
\midrule
\multirow{3}{*}{\texttt{gpt-5-mini}}
  & Unfiltered & 100.0 & 0.81 & 0.55 & 16.7 \\
  & L1         & 50.0  & 0.70 & 0.50 & 33.3 \\
  & L1{+}L2    & 25.0  & 0.93 & 0.63 & \textbf{0.0} \\
\midrule
\multirow{3}{*}{\texttt{gpt-5}}
  & Unfiltered & 100.0 & 0.72 & 0.53 & 26.7 \\
  & L1         & 81.7  & 0.70 & 0.51 & 32.7 \\
  & L1{+}L2    & 46.7  & 0.91 & 0.65 & \textbf{0.0} \\
\bottomrule
\end{tabular}
\end{adjustbox}
\end{table*}

\subsection*{Appendix Findings A4--A5: Training Controls and Cross-Domain Transfer}

\paragraph{Appendix Finding~A4: closure does not require a direct evidence reward.}
Because the GRPO return credits $\mathrm{ER}$ and $\mathrm{CUR}$ is defined through
$\mathrm{ER}$, the observed closure could reflect direct metric optimization.
An outcome-only-reward control that strips \emph{every} grounding term, rewarding
correctness alone, so within a group all correct rollouts score identically and the
advantage cannot distinguish grounded from lucky, also reduces the gap on
\texttt{column\_rename} ($\mathrm{CUR}\ 0.622\!\rightarrow\!0.292$ versus base,
$\mathrm{ER}\ 0.426\!\rightarrow\!0.589$). This pattern is consistent with
on-policy exploration favoring grounded routes to correct answers, whereas
offline outcome supervision increases the gap under the tested setup
(Table~\ref{tab:motivation:coevo2}).
The explicit evidence reward deepens the closure ($\mathrm{CUR}\ 0.208$,
$\mathrm{ER}\ 0.625$) but is not required for the directional improvement. Relatedly,
the gap is associated with fault \emph{geometry}. Across the seven-operator sweep,
it concentrates on schema rename, which admits a cheap shortcut (base
$\mathrm{CUR}\!\approx\!0.37$). The other solved operators, including
\texttt{incorrect\_join\_key}, \texttt{duplicate\_pk}, and
\texttt{dropped\_join\_predicate}, have $\mathrm{CUR}\!\approx\!0$ and require the
evidence path. In this sweep, the gap occurs where a solvable shortcut coexists
with a longer evidence path rather than wherever the model succeeds.

\paragraph{Appendix Finding~A5: the gap and its reduction transfer across domains.}
To test whether the phenomenon is an artifact of the \texttt{ecom} pipeline, we
evaluate the identical $30$B backbone on the schema-rename operator of two
pipelines it was never trained on, \texttt{retails} and \texttt{warehouse}, which
share only the abstract fault taxonomy, not a single table or model with
\texttt{ecom} (Table~\ref{tab:motivation:xdomain}). The correct-but-ungrounded gap
reappears in both, in fact \emph{larger} than on \texttt{ecom}: the base model
solves the rename ($\mathrm{RC@1}\ 0.85/0.95$) but grounds it only rarely (pooled
$\mathrm{CUR}\ 0.72/0.77$ across three evaluation seeds versus $0.16$ aggregate on
\texttt{ecom}). Two rounds of in-loop GRPO close it in both while driving accuracy
toward $1.0$: pooled $\mathrm{CUR}$ falls $0.719\!\rightarrow\!0.375$ on
\texttt{retails} ($\Delta{-}0.344$, bootstrap $95\%$ CI $[-0.459,-0.232]$) and
$0.765\!\rightarrow\!0.608$ on \texttt{warehouse} ($\Delta{-}0.156$,
$[-0.262,-0.050]$). Both intervals exclude zero ($n{=}144$ per arm, three eval
seeds). The closure
is again mediated by the exploration condition of boundary (iii): on this backbone
the GRPO groups carry variance ($16/16$ groups mix grounded and lucky trajectories),
so the negative gradient on lucky-correct has a grounded sibling to promote. The
larger gap outside the training domain and its subsequent decrease argue against
a single-pipeline explanation. The pattern is consistent with a task structure
that offers a short solvable route alongside a longer evidence path. The gap also
appears across the tested model generations, including Qwen2.5 and Qwen3 at every
evaluated size above the capability floor
(App.~\ref{app:controls}).

\begin{table*}[t]
\centering
\caption{Cross-domain generalization. The same $30$B backbone on the schema-rename
operator of two unseen pipelines ($n{=}48$ each) shows the same
correct-but-ungrounded gap, larger than \texttt{ecom}, and in-loop GRPO closes it
in both while accuracy approaches $1.0$. The \texttt{ecom} row is the aggregate over
gap operators (Table~\ref{tab:motivation:coevo3}). \texttt{retails}/\texttt{warehouse}
$\mathrm{CUR}$ are pooled over three eval seeds ($n{=}144$/arm), with GRPO$-$base
bootstrap $95\%$ CIs $[-0.459,-0.232]$ and $[-0.262,-0.050]$ respectively, both
exclude zero.}
\label{tab:motivation:xdomain}
\begin{adjustbox}{max width=0.95\linewidth}
\begin{tabular}{lcccc}
\toprule
 & \multicolumn{2}{c}{Base} & \multicolumn{2}{c}{In-loop GRPO} \\
\cmidrule(lr){2-3}\cmidrule(lr){4-5}
Domain & RC@1 & $\mathrm{CUR}$ & RC@1 & $\mathrm{CUR}$ \\
\midrule
\texttt{ecom} (agg.\ ref) & 0.292 & 0.163 & 0.367 & \textbf{0.054} \\
\texttt{retails}          & 0.854 & 0.719 & 1.000 & \textbf{0.375} \\
\texttt{warehouse}        & 0.953 & 0.765 & 0.993 & \textbf{0.608} \\
\bottomrule
\end{tabular}
\end{adjustbox}
\end{table*}

\subsection{Co-evolution: setup, reward, and probes}
\label{app:coevo:setup}

\paragraph{Task cells and generators (Sec.~\ref{sec:train:signal}).}
Each task is a \emph{cell}, an operator type (e.g., \texttt{column\_rename},
\texttt{incorrect\_join\_key}, \texttt{duplicate\_pk}) on the dbt e-commerce project,
materialized with planted decoys: a sibling pass-through view and a distractor tracker
issue implicating a non-gold node, so each instance carries an explicit cut-set $E^*$
the solver must rule out. The two generators differ only in the cell-ranking signal: one ranks cells by
$\mathrm{CUR}=\Pr[\mathrm{ER}{<}0.5\mid\text{correct}]$ on a seed corpus (ours), the other by
the outcome gap $1-\mathrm{RC@1}$. Both emit a matched batch ($n{=}204$ each) that
\texttt{Qwen3-30B-A3B} (terminal-localization constraint, guard~G4) rolls out. The
contrast needs a solver that is competent but still produces lucky guesses: a $7$B backbone either
never localizes ($\mathrm{RC@1}{=}0$) or, once grounded-trained, reaches
$\mathrm{CUR}{=}0$. The $30$B lucky-guesses on $\sim\!7\%$ of episodes.

\paragraph{Why difficulty matchmaking fails (harvesting probe).}
Harvesting GRPO groups ($36$ groups, $12$ rollouts each) on the operators a difficulty
rule selects (\texttt{null\_injection}, \texttt{duplicate\_pk}) gives correct answers
that are $100\%$ grounded ($60/60$, $0/36$ groups carry both a grounded and a lucky
sibling), because those faults cannot be shortcut. CUR instead selects
\texttt{column\_rename}, where the solver is correct often but only $\sim\!49\%$
grounded, so $12/36$ groups contain the grounded-vs-lucky contrast GRPO needs.

\paragraph{GRPO reward (Sec.~\ref{sec:train:inloop}).}
Each round samples $G{=}12$ rollouts per CUR-selected task and scores each by the
mechanical terminal reward
$R(T){=}\mathbf{1}[\text{correct}](w_c{+}w_g\mathrm{ER}){-}w_d[\text{decoy-blamed}]$,
training on the group-relative advantage $A_i{=}(R_i{-}\mu_g)/\sigma_g$.

\paragraph{Boundary-condition details.}
On the matched held-out batch, offline SFT/RWR inflate $\mathrm{CUR}$
($0.163\!\rightarrow\!0.521/0.465$) while two GRPO rounds reduce it to $0.054$
(GRPO$-$base $-0.109$ with $95\%$ CI $[-0.179,-0.043]$, pooled over three seeds $-0.132$ with
$[-0.176,-0.089]$) and raise RC@1 $0.292\!\rightarrow\!0.367$. Transfer: accuracy on a
held-out rename target rises $0.35\!\rightarrow\!0.67$, refuting answer-node
memorization. Exploration: a second capable model (Qwen2.5-14B) solves
\texttt{column\_rename} but grounds it almost never ($\mathrm{CUR}\!\approx\!1.0$). Its
groups carry no grounded sibling, so two GRPO rounds leave the gap unchanged
($0.276\!\rightarrow\!0.263$, $95\%$ CI $[-0.122,+0.093]$), whereas the $30$B, which
grounds $\sim\!$half its correct answers, reduces it.

\paragraph{Answer-decoupled reward and input ablation (Sec.~\ref{sec:train:loop}).}
The loop pays for the evidence path regardless of the answer:
\begin{equation*}
\begin{aligned}
R(T) = {}& w_c\mathbf{1}[\text{correct}] \\
{}&+ (w_g\mathrm{ER}+w_{\text{rule}}\mathrm{rule}+w_{\text{ep}}\mathrm{ep}) - w_d[\text{decoy}].
\end{aligned}
\end{equation*}
Withholding the culprit's code instead \emph{inflates} the gap on \texttt{column\_rename}
(CUR $0.375\!\rightarrow\!1.0$, RC@1 $1.0\!\rightarrow\!0.625$): the renamed column leaks
through schema inspection and sampling, so a code-blind solver merely guesses less
groundedly.

\paragraph{Elo-gated selection variant.}
The raw six-round alternation in Table~\ref{tab:motivation:coevoloop} is intentionally
ungated, so its per-round path is noisy. A deployable variant uses the same
EG-Elo as a selection rule: each round trains one candidate from the
current best policy and admits it only if its solver EG-Elo exceeds the incumbent's.
The best-so-far rating is therefore monotone by construction. Across three seeds, the
fielded endpoint improves Elo on every seed and closes CUR wherever the base carries a
substantial gap (Table~\ref{tab:motivation:coevogated}).

\begin{table}[t]
\centering
\scriptsize
\setlength{\tabcolsep}{2.5pt}
\caption{Elo-gated co-evolution over three seeds. Each candidate is admitted only if
its solver EG-Elo strictly exceeds the incumbent's. Seed 27 has little CUR headroom at
base. The seed-invariant claim is monotone Elo improvement.}
\label{tab:motivation:coevogated}
\begin{adjustbox}{max width=0.95\linewidth}
\begin{tabular}{lcccccc}
\toprule
& \multicolumn{2}{c}{Base} & \multicolumn{2}{c}{Endpoint} & \multicolumn{2}{c}{$\Delta$} \\
\cmidrule(lr){2-3}\cmidrule(lr){4-5}\cmidrule(lr){6-7}
Seed & Elo & CUR & Elo & CUR & Elo & CUR \\
\midrule
7  & 1521 & 0.444 & \textbf{1934} & \textbf{0.062} & $+414$ & $-0.382$ \\
17 & 1603 & 0.383 & 1831 & 0.111 & $+228$ & $-0.272$ \\
27 & 1666 & 0.158 & 1831 & 0.167 & $+166$ & $+0.009$ \\
\midrule
mean & 1597 & 0.328 & 1865 & 0.113 & $\mathbf{+269}$ & $\mathbf{-0.215}$ \\
\bottomrule
\end{tabular}
\end{adjustbox}
\end{table}

% =====================================================================
% Statistical and construct-validity analyses of the benchmark
% (7 models, 128 episodes each = 896, identical 64-task x 2-rep frozen corpus,
% results_leaderboard_api/). All analyses are same-corpus, same scaffold, same
% parser, same scoring code. Generated by:
%   pilot/analyze_proprietary_board.py     (#1 leaderboard, #5 conditional, #7 heatmap)
%   pilot/analyze_synthetic_vs_organic.py  (#3 representativeness)
%   pilot/analyze_egelo_reliability.py     (#6 EG-Elo reliability)
% =====================================================================
\section{Unified-Corpus Statistical Analyses}
\label{app:unified}

The leaderboard, conditional-gap, and symptom analyses use a unified proprietary
board: seven backbones run on the identical $64$-task$\times2$-repeat frozen
corpus ($896$ episodes), with one scaffold, parser, step budget, and scoring
implementation. The organic-incident comparison, masking pilot, and annotation
instrument use separate samples, identified below. The final analysis examines
alternative shortest routes on the 64 board tasks. These populations are kept
distinct from the pilot and training studies elsewhere in the paper.

\subsection{Unified Same-Corpus Leaderboard with Clustered Intervals}
\label{app:unified:leaderboard}

Table~\ref{tab:unified} reports each model with \textbf{environment-clustered}
bootstrap $95\%$ intervals (resampling the $13$ environments with replacement,
$4000$ draws) rather than episode-only intervals, plus the worst-environment
grounded score. Rank stability is estimated by re-ranking models on
environment-clustered resamples.

\begin{table}[t]
\centering\small
\setlength{\tabcolsep}{3.5pt}
\caption{\textbf{Unified same-corpus proprietary leaderboard} ($896$ episodes,
identical tasks/scaffold/parser/scoring). Intervals are environment-clustered
bootstrap $95\%$ CIs. Worst-env = lowest $\mathrm{RC@1}^g$ over environments with
$\ge6$ episodes. Rank interval is the $95\%$ environment-clustered rank range.}
\label{tab:unified}
\begin{adjustbox}{max width=0.95\linewidth}
\begin{tabular}{lcccc}
\toprule
Model & $\mathrm{RC@1}$ & $\mathrm{RC@1}^g$ & Worst-env & Rank \\
 & [95\% CI] & [95\% CI] & $\mathrm{RC@1}^g$ & [95\% CI] \\
\midrule
\texttt{gpt-5.6-sol}       & 0.688 & \textbf{0.586} & 0.500 & 1 [1,1] \\
                           & \tiny[.564,.806] & \tiny[.541,.637] & & \\
\texttt{kimi-k3}           & 0.641 & 0.531 & 0.250 & 2 [2,3] \\
                           & \tiny[.500,.770] & \tiny[.459,.594] & & \\
\texttt{claude-sonnet-5}   & 0.633 & 0.523 & 0.375 & 3 [2,3] \\
                           & \tiny[.526,.743] & \tiny[.478,.577] & & \\
\texttt{glm-5.2}           & 0.492 & 0.398 & 0.000 & 4 [4,5] \\
                           & \tiny[.327,.646] & \tiny[.308,.472] & & \\
\texttt{deepseek-v4-flash} & 0.414 & 0.344 & 0.000 & 5 [4,5] \\
                           & \tiny[.263,.538] & \tiny[.236,.440] & & \\
\texttt{grok-4.5}          & 0.305 & 0.234 & 0.000 & 6 [6,7] \\
                           & \tiny[.208,.382] & \tiny[.153,.321] & & \\
\texttt{gemini-3.5-flash}  & 0.195 & 0.164 & 0.000 & 7 [6,7] \\
                           & \tiny[.088,.297] & \tiny[.079,.243] & & \\
\bottomrule
\end{tabular}
\end{adjustbox}
\end{table}

The ordering is stable at the extremes (top and bottom models have non-overlapping
rank intervals), whereas adjacent mid-tier pairs (\texttt{kimi-k3}/\texttt{claude-sonnet-5},
\texttt{glm-5.2}/\texttt{deepseek-v4-flash}) overlap: the board separates capability
\emph{bands}, not fine ranks, at this sample size. Worst-environment grounded
accuracy ranges from $0$ to $0.5$, showing variation that the aggregate score hides.

\subsection{The Capability--Grounding Gap Is Conditional, Not Merely Mechanical}
\label{app:unified:conditional}

The main-text observation that unsupported-correct \emph{mass} rises with
capability is partly mechanical: stronger models produce more correct answers, so
even a constant shortcut \emph{rate} yields more lucky-guess mass. We therefore
test the conditional quantity $\mathrm{CUR}=P(\text{ungrounded}\mid\text{correct})$
directly. On the correct-episode subset ($n{=}431$ of $896$) we fit
\begin{equation*}
\begin{aligned}
&P(\text{ungrounded}\mid\text{correct}) \sim \text{capability} + \text{operator} \\
&\quad + \text{environment} + \text{path\_len} + \text{symptom},
\end{aligned}
\end{equation*}
with capability $=$ per-model $\mathrm{RC@1}$ (standardized) and
environment-clustered bootstrap inference ($1500$ draws, ridge-penalized
Newton--IRLS fit, unpenalized intercept, as \texttt{statsmodels} was unavailable).
The capability coefficient is $+0.412$ (log-odds per $+1$\,SD, $95\%$ CI
$[-0.086,+0.860]$, bootstrap $P(\text{coef}\le0){=}0.053$) and is not
statistically significant at the $0.05$ level. Per-model CUR lies in a narrow band
($0.15$--$0.23$) and does not increase monotonically with capability
(Spearman $\rho{=}-0.25$, $p{=}0.59$, see Table~\ref{tab:cur_band}).

\begin{table}[t]
\centering\small
\caption{Conditional shortcut rate $\mathrm{CUR}=P(\text{ungrounded}\mid\text{correct})$
with environment-clustered $95\%$ CIs, ordered by capability. CUR does not rise
monotonically with capability. The conditional gap \emph{persists} across tiers.}
\label{tab:cur_band}
\begin{adjustbox}{max width=0.95\linewidth}
\begin{tabular}{lcc}
\toprule
Model & $\mathrm{RC@1}$ & CUR [95\% CI] \\
\midrule
\texttt{gpt-5.6-sol}       & 0.688 & 0.148 \tiny[.000,.263] \\
\texttt{kimi-k3}           & 0.641 & 0.171 \tiny[.000,.290] \\
\texttt{claude-sonnet-5}   & 0.633 & 0.173 \tiny[.000,.300] \\
\texttt{glm-5.2}           & 0.492 & 0.190 \tiny[.000,.304] \\
\texttt{deepseek-v4-flash} & 0.414 & 0.170 \tiny[.000,.292] \\
\texttt{grok-4.5}          & 0.305 & 0.231 \tiny[.000,.435] \\
\texttt{gemini-3.5-flash}  & 0.195 & 0.160 \tiny[.000,.229] \\
\bottomrule
\end{tabular}
\end{adjustbox}
\end{table}

We therefore make the narrower claim that \emph{absolute} unsupported success grows
with capability, while a nontrivial \emph{conditional} grounding gap persists across
model tiers rather than worsening. The main text (Sec.~\ref{sec:exp:paradox}) is
worded accordingly.

\subsection{Capability-Boundary Heatmap}
\label{app:unified:heatmap}

Table~\ref{tab:heatmap} breaks grounded accuracy down by fault family, exposing
\emph{where} grounding breaks. Two structural faults with explicit failures
(\texttt{duplicate\_pk}, \texttt{off\_by\_one\_temporal}) are grounded well by strong
models, whereas \texttt{column\_rename} and \texttt{null\_injection} are near-floor for
every tested model. \texttt{column\_rename} is the operator on which our
training transfer concentrates (App.~\ref{app:findings}), so its board-wide
difficulty bounds how much single-operator transfer can claim.

\begin{table}[t]
\centering\scriptsize
\setlength{\tabcolsep}{2.5pt}
\caption{$\mathrm{RC@1}^g$ per (fault family $\times$ model), with models ordered by
capability. Columns: G5{=}\texttt{gpt-5.6-sol}, KK{=}\texttt{kimi-k3},
CS{=}\texttt{claude-sonnet-5}, GL{=}\texttt{glm-5.2}, DS{=}\texttt{deepseek-v4-flash},
GK{=}\texttt{grok-4.5}, GM{=}\texttt{gemini-3.5-flash}.}
\label{tab:heatmap}
\begin{adjustbox}{max width=0.95\linewidth}
\begin{tabular}{lccccccc}
\toprule
Fault family & G5 & KK & CS & GL & DS & GK & GM \\
\midrule
\texttt{duplicate\_pk}        & 1.00 & 1.00 & 0.85 & 0.23 & 0.23 & 0.15 & 0.00 \\
\texttt{incorrect\_join\_key} & 0.96 & 0.77 & 0.92 & 0.77 & 0.77 & 0.65 & 0.50 \\
\texttt{off\_by\_one\_temporal}& 1.00 & 1.00 & 1.00 & 1.00 & 0.92 & 0.50 & 0.08 \\
\texttt{unenforced\_unique}   & 1.00 & 0.67 & 0.50 & 1.00 & 0.33 & 0.17 & 0.00 \\
\texttt{null\_injection}      & 0.15 & 0.15 & 0.15 & 0.15 & 0.04 & 0.00 & 0.00 \\
\texttt{column\_rename}       & 0.08 & 0.08 & 0.08 & 0.08 & 0.12 & 0.08 & 0.15 \\
\bottomrule
\end{tabular}
\end{adjustbox}
\end{table}

\paragraph{Silent drift versus explicit failure.}
Table~\ref{tab:symptom_support} reports the counts underlying the symptom split.
All 75 unsupported correct answers occur in explicit-failure cases, giving
$\mathrm{CUR}=75/326=0.230$. All 105 correct silent-drift answers meet the
coverage threshold, giving $\mathrm{CUR}=0$. This association supports the
symptom-clue explanation in the tested board, but does not by itself identify a
cause in model pretraining.

\begin{table}[t]
\centering\small
\setlength{\tabcolsep}{3pt}
\caption{Unsupported correct answers concentrate in explicit failures on the
unified board. Grounded means correct with $\mathrm{ER}\ge\tau_g$; lucky means
correct with $\mathrm{ER}<\tau_g$, where $\tau_g=0.5$. CUR divides lucky answers
by correct answers, not by all episodes.}
\label{tab:symptom_support}
\begin{adjustbox}{max width=\linewidth}
\begin{tabular}{lrrrrr}
\toprule
Symptom & Episodes & Correct & Grounded & Lucky & CUR \\
\midrule
Explicit failure & 602 & 326 & 251 & 75 & 0.230 \\
Silent drift & 294 & 105 & 105 & 0 & 0.000 \\
All & 896 & 431 & 356 & 75 & 0.174 \\
\bottomrule
\end{tabular}
\end{adjustbox}
\end{table}

Two supplementary controls distinguish detailed logs from symptom clues.
Withholding the build/test log while keeping the precise symptom leaves
\texttt{gpt-5} localization nearly unchanged (RC@1 $0.43\to0.42$), with incident
consultation still rare (App.~\ref{app:controls:logblind}). In the
\texttt{gpt-5-mini} blind ablation, which also replaces the symptom with a generic
category, LG falls $0.15\to0$, ER rises $0.81\to0.90$, and RC@1 falls
$0.35\to0.10$; column-rename accuracy falls $0.5\to0$
(App.~\ref{sec:exp:shortcut}). These controls support the role of symptom-text
access in unsupported success. They use different supplementary populations and
do not measure pretraining exposure or rule out other contributing mechanisms.

\subsection{Synthetic-to-Organic Representativeness}
\label{app:unified:organic}

We compare injected (synthetic) faults against mined \emph{organic} incidents
(Generator~B: real BIRD-derived pipeline incidents with reproducible
pre-fix/post-fix states), holding backbone $\in\{$\texttt{gpt-4o-mini},
\texttt{gpt-5-mini}, \texttt{gpt-5}$\}$ and environment family (BIRD) fixed:
synthetic $=$ \texttt{results\_birdwh\_eval} ($n{=}258$), organic $=$
$10$ incidents $\times3$ backbones $\times2$ repeats ($n{=}60$).

The \textbf{model ordering is preserved}: $\mathrm{RC@1}$ on synthetic versus organic
is $\{0.43,0.19,0.11\}$ versus $\{0.70,0.45,0.10\}$ for gpt-5/gpt-5-mini/gpt-4o-mini
(Spearman $\rho{=}{+}1.0$, though $n{=}3$ backbones is underpowered), and the
grounding gap $\mathrm{RC@1}^g<\mathrm{RC@1}$ appears in both regimes. The
\textbf{fault distributions differ}, however: organic incidents are almost entirely
logic \emph{misspecification} with silent symptoms (total-variation distance from
the synthetic fault-family distribution $=1.0$, symptom-type TV $=0.67$, path-length
KS $D{=}0.63$, $p{<}0.001$), whereas synthetic injection spans structural faults with
a mix of explicit and silent symptoms. Thus, synthetic injection broadens the
tested fault coverage, but it should not be interpreted as matching the natural
fault distribution. The grounding gap and capability ordering transfer across
the two regimes in this comparison.

\paragraph{Expanded organic set on current models.}
We expand the organic set to $40$ mined BIRD-CRITIC incidents across $10$
databases and evaluate three current backbones, comparing them with their
$128$-episode synthetic board (Table~\ref{tab:organic_expanded}). The completed
organic sample contains 39 episodes for \texttt{gpt-5.6-sol} and 40 each for the
other two models. The reported GPT accuracy is therefore $38/39=0.974$, not a
40-episode estimate. The ordering is again preserved (Spearman $\rho=+1.0$).
The organic set separates these three models more widely, but the small number
of backbones and unequal completed episode counts limit the comparison. The
expanded synthetic comparison also spans the full board rather than only the
BIRD family, unlike the preceding matched-family study. These results do not
establish distributional representativeness or a general ranking guarantee.

\begin{table}[t]
\centering\small
\setlength{\tabcolsep}{3pt}
\caption{Expanded organic-incident comparison. Synthetic accuracy uses 128
episodes per model; organic accuracy uses the completed episode counts shown.}
\label{tab:organic_expanded}
\begin{adjustbox}{max width=\linewidth}
\begin{tabular}{lrrr}
\toprule
Model & Synthetic RC@1 & Organic RC@1 & Organic $n$ \\
\midrule
\texttt{gpt-5.6-sol} & 0.688 & 0.974 & 39 \\
\texttt{deepseek-v4-flash} & 0.414 & 0.050 & 40 \\
\texttt{gemini-3.5-flash} & 0.195 & 0.000 & 40 \\
\bottomrule
\end{tabular}
\end{adjustbox}
\end{table}

\subsection{EG-Elo Rating Reliability}
\label{app:unified:egelo}

On the $128$-episode board (versus the $\sim24$-episode ecom self-play), the EG-Elo
\emph{solver} rating is rank-stable at the extremes: episode-bootstrap $95\%$ rating
intervals separate the top model (\texttt{gpt-5.6-sol}, $1588$ $[1500,1675]$) and the
bottom two (\texttt{grok-4.5} $1229$, \texttt{gemini-3.5-flash} $1157$) from the pack,
while mid-tier ratings overlap. The rating tracks held-out grounded accuracy
(Spearman$(\mathrm{EG\text{-}Elo},\mathrm{RC@1}^g){=}{+}1.0$) but not the shortcut
rate (Spearman with CUR $={-}0.25$), and split-half reliability across the two
repeats is high (Spearman--Brown $\approx1.0$). These results support interpreting
EG-Elo as a grounded-skill \emph{band} rating. The full self-play reliability program
(multiple generator seeds, transitivity and cycle tests, generator-pool and
match-schedule sensitivity, post-tournament fresh-task prediction) requires a
larger generator--solver tournament and remains future work. We therefore present
the self-play rating as a protocol rather than a definitive benchmark number.

\subsection{Does ER Measure Evidence \emph{Use} or Merely Visitation?}
\label{app:unified:intervention}

A trajectory can inspect a gold-path node without the diagnosis actually depending
on it, so a high ER need not mean the answer is causally grounded. We test this with
a counterfactual intervention: we re-run each task and, the moment a tool call
\emph{newly exposes} a gold-path node, we replace that observation with a
restricted-access stub containing no schema, rows, lineage, or log. The hook does
not alter \texttt{env.visited}, so masking itself leaves the coverage credit for
that tool call unchanged. The agent then continues freely; later actions and
final ER can differ between control and treatment. If a diagnosis depends on the masked evidence, the intervention may
change the submitted root cause. Shortcut answers should be less sensitive.

Over a pilot of $55$ tasks (\texttt{deepseek-v4-flash}), among control
\emph{grounded-correct} episodes (RC@1${=}1$, ER${\ge}0.5$, $n{=}17$) the submitted
root cause changed under masking in $2/17$ episodes ($12\%$), whereas among control
\emph{lucky} episodes (RC@1${=}1$, ER${<}0.5$, $n{=}3$) it changed $0\%$
(dependency gap $+0.12$). The direction matches the construct-validity
hypothesis, but the modest grounded change rate does not establish causal
dependence on the accepted path. ER is therefore best read as
``covered the accepted reference path'' rather than ``causally required each
node.'' The small lucky stratum
($n{=}3$) and single mid-tier backbone limit the conclusion. Redundant evidence
is one possible explanation for the small change rate, not a verified mechanism;
the intervention does not remove all alternative evidence or isolate rerun
variability with a repeated unmasked control. A stronger
decoy-substitution intervention and a larger, multi-model run
(\texttt{pilot/run\_evidence\_intervention.py}, modes \texttt{mask}/\texttt{decoy})
would sharpen the estimate. This automated test complements, and does not replace,
the expert audit below.

\subsection{Expert-Audit Annotation Pipeline}
\label{app:unified:audit}

To support future validation of the mechanical labels against human judgment, we
provide a reproducible annotation pipeline. The human study is incomplete. A stratified
sampler (\texttt{pilot/build\_annotation.py}) renders agent trajectories as cards
carrying the reported symptom, the numbered investigation, and the submitted root
cause, with the mechanical scores (RC@1, ER, outcome) \emph{hidden} but retained for
correlation. The protocol asks three experts to judge each card for (Q1)~root-cause validity,
(Q2)~evidence sufficiency, (Q3)~evidence relevance, and (Q4)~whether an alternative
evidence path is acceptable, via a self-contained offline HTML form. A scorer
(\texttt{pilot/score\_annotations.py}) then computes inter-annotator agreement
(Fleiss' $\kappa$~\cite{fleiss1971agreement} and ordinal Krippendorff's
$\alpha$~\cite{hayes2007reliability}) and the key construct-validity
contrasts: expert Q1 against mechanical RC@1, and expert Q2 (``was the evidence
sufficient?'') against the mechanical coverage gate (ER${\ge}\tau_g$), separately
from root-cause correctness. The cell of
interest is \emph{high ER but expert-insufficient}, which quantifies any residual
over-crediting of visitation by ER. The instrument was tested end-to-end on a
$15$-card batch with simulated raters. This test validates the pipeline, not
agreement with expert judgment. Completing that validation requires ratings from
database or data-engineering experts.

\subsection{Alternative Shortest-Path Coverage}
\label{app:unified:altpaths}

We examine single-route coverage by reconstructing all 64 board tasks with the
generator's code (\texttt{pilot/analyze\_altpath.py},
\texttt{PYTHONHASHSEED=0}). For each task, the script selects a closest reachable
symptom and enumerates shortest routes to the injected cause in the undirected
provenance graph. For a route $P$, coverage is
$|P\cap\pi^\star|/|\pi^\star|$. The reconstructed accepted witness
$\pi^\star$ can include a union of paths from several symptoms and additional
cross-modal evidence; it need not equal a single shortest route. The script
checks that accepted nodes occur in the supporting graph, rather than replaying
the saved leaderboard trajectories or establishing semantic equivalence of
alternative investigations.

There is a unique shortest route in $42/64$ tasks ($65.6\%$); $40/64$ additionally
have no alternative route within one extra hop. The remaining 22 tasks have
multiple shortest routes, concentrated in null-injection and duplicate-key
faults in warehouses with parallel dimension joins. The mean \emph{minimum}
coverage across candidate shortest routes is $0.685$ over all 64 tasks and
$0.610$ over the 22 tasks with multiple shortest routes.

A candidate route falls below $\tau_g=0.5$ in $7/64$ tasks ($10.9\%$). Three have
multiple shortest routes; four have a unique shortest route but a larger accepted
witness assembled from multiple symptom paths. Thus, the count combines path
ambiguity with the difference between one route and the full witness. It is a
task-level possibility count, not an observed false-lucky rate. Hash-dependent
graph traversal can change which tied path is selected during reconstruction,
so the hash seed is part of this analysis's configuration. Longer routes and
expert judgments of sufficiency remain outside the analysis; these results do
not bound all valid investigations. They motivate both multi-path grading and
explicit treatment of witnesses spanning several symptoms.

\fi
\end{document}